\documentclass[aip,jcp,amsmath,amssymb,reprint,numeric]{revtex4-1}

\title{pinned system}
\usepackage{graphicx}
\usepackage{subfigure}
\usepackage{dcolumn}
\usepackage{bm}
\usepackage{mathrsfs}
\usepackage[dvipsnames]{xcolor}
\usepackage[normalem]{ulem}
\usepackage{caption}
\begin{document}
\title{Dynamic heterogeneity in polydisperse systems: A comparative study of the role of local structural order parameter and particle size}

\author{Palak Patel}
\address{\textit{Polymer Science and Engineering Division, CSIR-National Chemical Laboratory, Pune-411008, India}}
\affiliation{\textit{Academy of Scientific and Innovative Research (AcSIR), Ghaziabad 201002, India}}
\author{Mohit Sharma}
\address{\textit{Polymer Science and Engineering Division, CSIR-National Chemical Laboratory, Pune-411008, India}}
\affiliation{\textit{Academy of Scientific and Innovative Research (AcSIR), Ghaziabad 201002, India}}
\author{Sarika Maitra Bhattacharyya}
\email{mb.sarika@ncl.res.in}
\address{\textit{Polymer Science and Engineering Division, CSIR-National Chemical Laboratory, Pune-411008, India}}
\affiliation{\textit{Academy of Scientific and Innovative Research (AcSIR), Ghaziabad 201002, India}}

\begin{abstract}

In polydisperse systems, describing the structure and any structural order parameter (SOP) is not trivial as it varies with the number of species we use to describe the system, $M$. Depending on the degree of polydispersity, there is an optimum value of $M=M_{0}$ where we show that the mutual information of the system increases.  However, surprisingly the correlation between a recently proposed SOP  and the dynamics is highest for $M=1$. This effect increases with polydispersity. We find that the SOP at $M=1$ is coupled with the particle size, $\sigma$, and this coupling increases with polydispersity and decreases with an increase in $M$. Careful analysis shows that at lower polydispersities, the SOP is a good predictor of the dynamics. However, at higher polydispersity, the dynamics is strongly dependent on $\sigma$. Since the coupling between the SOP and $\sigma$ is higher for $M=1$, it appears to be a better predictor of the dynamics. We also study the Vibrality an order parameter independent of structural information. Compared to SOP, at high polydispersity we find Vibrality to be a marginally better predictor of the dynamics. However, this high predictive power of Vibrality, which is not there at lower polydispersity, appears to be due to its stronger coupling with $\sigma$. Therefore our study suggests that for systems with high polydispersity, the correlation of any order parameter and $\sigma$ will affect the correlation between the order parameter and dynamics and need not project a generic predictive power of the order parameter.

\end{abstract}

\maketitle

\section{INTRODUCTION}
\label{introduction_sec}
When a liquid is cooled fast enough, it enters the supercooled liquid regime, where the properties of the liquid are very different from those of the normal liquid regime. When the supercooled liquid approaches the glass transition, its dynamics increases by orders of magnitude \cite{stillinger, ANGELL}, with the structure showing marginal changes. This observation questioned the role of structure in dynamics and the application of liquid state theories \cite{W_Gotze_1992,annu_rev.physchem, Hansen_and_McDonald} in the supercooled regime. However, studies have shown that although the structure does not change drastically, static properties that depend on the structure can change enough to affect the dynamics \cite{Atreyee_PRL, Kenneth_PRL_2015,manoj_PRL_2021}. One of the key signatures of supercooled liquids is their dynamical heterogeneity which increases with a decrease in temperature \cite{Ediger, Hurley_1995,walter_PRL_1997}. There have been a large number of studies attempting to causally connect this dynamical heterogeneity and local order parameters, some of which are purely structural in origin.\cite{Harowell, Andrea_liu_nature, Paddy_PRL, Tanaka_PRL_2020, Richard_PRM,coslovich_JCP_2020, Andrea_liu_PRE, Tanaka_nature,manoj_PRL_2021,mohit_PRE} In recent studies, we have defined a structural order parameter (SOP) that is connected to the depth of the mean-field caging potential.\cite{manoj_PRL_2021,mohit_PRE} Our study has shown that for a large number of systems, the SOP is a good parameter to describe the relaxation process in the systems.\cite{manoj_PRL_2021}  We have also shown that this causality persists even at the local level .\cite{mohit_PRE} The distribution of the particle level SOP becomes wider at lower temperatures, thus suggesting an increase in local structural heterogeneity. The correlation between the SOP and the dynamics at the particle level is observed only below the onset of the glassy dynamics, $T_{onset}$, and increases as the temperature is decreased. Therefore, according to this study, the structural heterogeneity and the coupling between the SOP and dynamics increase at lower temperatures.\cite{mohit_PRE} 

Given the good predictive power of this new structural order parameter, it should be tested for other glass-forming liquids. Among systems that are good glass formers, polydisperse systems with size polydispersity come high in the order.\cite{tanka_PDI, Devid_PRE, Tanaka_PRL_2007, Daan_JCP, Tanaka_JCP_2013, Tanaka_NAS_2015, Ran_2019, Quirke_1990} Polydisperse systems beyond some degree of polydispersity can be easily supercooled, \cite{Devid_PRE,lacks1999,williams2001,pinaki2005,solid_liquid_transition_by_bagchi, Bidisperse_simulation, PDI_bagchi_sneha_sarika} and most experimental colloidal systems are polydisperse.\cite{Bitsanis_2,colloidal_particle_1994,Ryan_JCIS, Andreas_2020,Janne_2020,Andreas_NAS_2021, Andreas_NAS_2021,Daniel_JCP} Moreover, the swap Monte Carlo algorithm, which allows the system to be cooled to unprecedentedly low temperatures, is best applied to polydisperse systems with continuous size polydispersity. \cite{paris_PRE,Pastore_CPL,Coslovich_2018,ozawa,ozawa_JCP_2018,Ozawa_PRL_2016,Berthier_NAS_2017}

However, for a system with continuous size polydispersity, describing the structure and any parameter that depends on the structure is a challenge. For these systems, the number of species, $M$, equals the total number of particles. Many a time, these systems are treated like a monodisperse system ($M=1$), and the average structure/radial distribution function (rdf) shows an artificial softening.\cite{palak_polydisperse,peter_poole_2001,PDI_bagchi_sneha_sarika,Trond_Dyre_2023} Therefore any property calculated using the rdf does not show the correct value. Depending on the diameter of the particles, we can always approximate the system in terms of a certain number of species. However, what is the optimum number of species, $M=M_{o}$, needed to describe the properties of the system ? This is a question often asked.\cite{peter_poole_2001,ozawa,main_paper_Truskett,palak_polydisperse} In earlier work, we used the correlation between the total excess entropy of the system and its two-body counterpart, which needs the information of the rdf, to obtain the optimum number of species, $M=M_{o}$.\cite{palak_polydisperse} The method is quite simple and much less computer intensive, but it provides similar results as those obtained from the study of configurational entropy using diameter permutation.\cite{ozawa}

Here, we first show that our method of describing the system into multiple species increases the mutual information (MI) of the system. We then show that the SOP and its correlation with the dynamics depend on $M$. It was earlier shown that the correlation between SOP and dynamics helps us identify.$T_{onset}$ \cite{mohit_PRE} Since the SOP and its correlation vary with $M$, so does the $T_{onset}$. Similar to our earlier study,\cite{palak_polydisperse} the $T_{onset}$ first changes with $M$ and then saturates. This clearly suggests that for a polydisperse system, for the calculation of the SOP, the system needs to be described in terms of multiple species.  However, to our surprise, we find that the correlation between the SOP and the dynamics is at its maximum for $M=1$. Furthermore the study reveals that at low polydispersity, the SOP is a good predictor of the dynamics, but at high polydispersity, the size of the particle plays a dominant role in determining the dynamics. Moreover, the SOP and the size are also correlated, and this correlation increases with an increase in polydispersity and decreases with an increase in $M$. Therefore at high polydispersity and for $M=1$, where the SOP and the particle size are strongly correlated, the SOP appears to be strongly correlated with the dynamics. However, this does not depict the true correlation, which is mediated by particle size. We also study Vibrality, another order parameter independent of the system's structure. We find that at high polydispersity, compared to the SOP, the Vibrality has an even stronger correlation with the particle size. Therefore it appears to be a better predictor of dynamics. These results clearly suggest that for systems with high polydispersity, any local order parameter correlated with the particle size might appear to be a good predictor of the dynamics, and these results should be cautiously interpreted and not assumed to be a generic result.

The organization of the rest of the paper is as follow. Section \ref{simulation_details_section} contains the simulation details. In Section \ref{MI_AND_rdf_section}, we discuss mutual information as a function of the radial distribution function. In section \ref{computing_Local_caging_potential_section}, we present the calculation of the caging potential in a polydisperse system. In section \ref{SPECIES_DEPENDENCE_OF_THE_CAGING_POTENTIAL}, we discuss the species dependence on the caging potential. In section \ref{onset_caluclation_PRS}, we discuss the species dependence of the correlation between the SOP and the particle dynamics. In section \ref{ANALYSIS_OF_DYNAMICS_OF_SOFT_AND_HARD_PARTICLES}, we analyze the dynamics of the particles with soft and hard SOP. In section \ref{COMPARATIVE_STUDY_OF_THE_ROLE_OF_PARTICLE_SIZE_AND_LOCAL_STRUCTURE_DYNAMICS}, we do a comparative analysis of the role of particle size and SOP in the dynamics. The paper ends with a brief conclusion in Section \ref{CONCLUSION}. This paper contains three Appendix sections at the end.

\section{SIMULATION DETAILS}
\label{simulation_details_section}
For this study, we have performed three-dimensional MD simulations [using the Large-scale Atomic/Molecular Massively Parallel Simulator (LAMMPS) package\cite{lammps}] for polydisperse systems in a canonical (NVT) ensemble. N = 4000 particles are present in a cubic box with volume V and density $\rho = \frac{N}{V} = 1.0$. We have used periodic boundary conditions for the simulation. In this simulation, a Nos\'{e}-Hoover thermostat with an integration timestep 0.001$\tau$ and 100 timesteps as time constants is used. The study involves the Gaussian type of size distribution for continuous size polydispersity. This means each of the N particles has a different radius. The Gaussian distribution is given by 

\begin{equation}
P(\sigma) = \frac{1}{\sqrt{2\pi\Lambda^2}}\exp^{\frac{-(\sigma-<\sigma>)^2}{2\Lambda^2}} ,
\label{gaussian}
\end{equation}
\noindent
where $\Lambda$ is the standard deviation. In this distribution, we consider $\sigma_{max/min}=<\sigma> \pm 3\Lambda$. The degree of polydispersity is quantified by,

\begin{center}
PDI =$\frac{\sqrt{<\sigma^2>-<\sigma>^2}}{<\sigma>}$ =$\frac{\Lambda}{<\sigma>}$    
\end{center}

For all the polydisperse systems, the particle sizes are chosen such that $<\sigma>=\int P(\sigma)\sigma d\sigma$ = 1.

In this study, particles i and j interact via inverse power law (IPL) potential. The form of the potential is given by,\cite{Bidisperse_simulation,tridisperse_simulation}

\begin{equation}
 U(r_{ij}) =
 \begin{cases}
 \epsilon_{ij} (\frac{\sigma_{ij}} {r_{ij}})^{12} + \sum_{l=0}^{2} c_{2l} (\frac{r_{ij}} {\sigma_{ij}})^{2l}, & (\frac{r_{ij}} {\sigma_{ij}}) \leq x_{c}\\
 0, & (\frac{r_{ij}} {\sigma_{ij}}) > x_{c} 
\end{cases} 
\label{potential} 
\end{equation}
\noindent
where $c_0, c_2$, and $c_4 $ are constants, and they are selected such that the potential becomes continuous up to its second derivative at the cutoff $x_c=1.25\sigma_{ij}$. 

The interaction strength between two particles i and j is  $\epsilon_{ij}$ = 1.0. $\sigma_{ij}=\frac{(\sigma_i+\sigma_j)}{2}$, where $\sigma_{i}$ is the diameter of particle i. Length, temperature, and time are given in units of $<\sigma>, \epsilon_{ij}$, and $\big(\frac{m<\sigma>^2}{\epsilon_{ij}}\big)^\frac{1}{2}$, respectively. For all state points, the equilibration is performed for $100 \tau_{\alpha}$ ($\tau_{\alpha}$ is the $\alpha$-relaxation time; details are given in Appendix A).\cite{palak_polydisperse}

During the analysis, when the system is described in terms of $M$ species, the particles in the diameter range $(\sigma_{max}-\sigma_{min})/M$ are treated as singles species where M = 1,2,3,.... . Therefore for $M=1$, all particles are assumed to have the same average diameter.

Since these systems are not that well known, we provide information on the different characteristic temperatures of the systems in Appendix - A

\section{MUTUAL INFORMATION AND RADIAL DISTRIBUTION FUNCTION}
\label{MI_AND_rdf_section}
As mentioned in Sec. \ref{introduction_sec}, in earlier work, we used the correlation between the total excess entropy and its two body counterpart to determine the optimum number of species required to describe the system.\cite{palak_polydisperse} We will now show that this method is similar in spirit to the calculation of mutual information (MI) between species and their structures.\cite{coslovich_JCP_2020}

 The excess entropy $S_{ex}$ of a system is the loss of entropy due to correlation. This is usually calculated via the thermodynamic integration method.  \cite{sastry_2000,palak_ujjwal_JCP} It can also be expressed as $S_{ex}=S_{2}+S_{3}+....$, where $S_{n}$ denotes the entropy due to the $n$ body correlation.\cite{Green_N_body_correlation_1958} $80\%$ of the total entropy comes from the two body excess entropy and is given by\cite{Goel_2008,Evans_1989}
\begin{equation}
    S_2 = -2\pi\rho \int_{0}^{\infty} \{g(r)\ln g(r) - g(r)+1 \}r^2 dr
    \label{total_s2_from_tot_gr_eq}
\end{equation}
\noindent 
where g(r) is the radial distribution function. For a large number of systems, it was shown that the two body excess entropy calculated using the rdf crosses the total excess entropy, and this crossover temperature $T_{cross}$ is similar to the onset temperature of the glassy dynamics.\cite{Atreyee_2017,Atreyee_PRL} Above this temperature, the two-body excess entropy is larger than the total excess entropy, and below this temperature, the reverse happens. This concept was then used to describe systems with continuous polydispersity in terms of an optimum number of species, $M_{0}$.\cite{palak_polydisperse} The two body entropy for $M$ species can be written as
\begin{equation}
\small\frac{S_2}{k_B}=-2\pi \rho \sum_{u,v = 1}^{M} \chi_{u} \chi_{v} \int_{0}^{\infty}\{g_{uv}(r)\ln g_{uv}(r) - g_{uv}(r) + 1 \} r^2 dr ,
\label{S_2}  
\end{equation}
\noindent 
where $\chi_{u}$ is the fraction of particles in $u$ species and $g_{uv}(r)$ is the partial radial distribution function of species $u$ and $v$, and it is expressed as,\cite{Hansen_and_McDonald} $g_{uv}(r) = \frac{V}{N_{u} N_{v}} \Big< \sum_{i=1}^{N_{u}} \sum_{j=1, j \neq i}^{N_v} \delta (r - r_{i} + r_{j}) \Big>$   

The temperature dependence of $S_{2}$ changes with $M$, and this also changes the $T_{cross}$ value (shown in Appendix A). As shown in Fig. \ref{MI_global_gr_fig}, as a function of the number of species $M$, $T_{cross}$ first increases and then almost saturates at a certain value. This saturation value is similar to the onset temperature obtained from other methods.\cite{Atreyee_2017,sastry_inherent_1998,palak_polydisperse} The saturation of $T_{cross}$  implies that the structure/rdf does not change considerably when an even larger number of species is used to describe the system. Therefore this provides us with information on the optimum number of species, $M=M_{0}$, required to describe the system.

Interestingly, our formalism is similar in spirit to the formalism suggested recently using mutual information (MI) theory; the difference in two body entropy when the system is expressed as a single species and as $M$ species can be approximately written as,\cite{coslovich_JCP_2020}
\begin{equation}
    \Delta S_2 \simeq \sum_{k=1}^{M} \int_{0}^{R} 4\pi r^2 \rho \chi_k g_k (r) \ln \Big(\frac{g_k(r)}{g(r)} \Big) dr ,
\label{entropy_diff}
\end{equation}
\noindent 
 where $g_{k}(r)$ is the rdf of the $k^{th}$ species and the total rdf $g(r) = \sum_k \chi_k g_k (r)$.  Coslovich $\it et al.$ beautifully argued that this difference in two body entropy is similar to the MI.\cite{coslovich_JCP_2020} From Eq. \ref{entropy_diff}, we notice that in this formalism, the probability distribution in the MI is replaced by the radial distribution function, which is the probability of finding a particle at a distance $r$ from a central particle over and above the ideal gas prediction. Therefore this formalism, instead of using the bare probability of finding a particle, is based on the probability of finding particles at certain interparticle distances.
 
In Fig. \ref{MI_global_gr_fig}, along with $T_{cross}$, we also plot $\Delta S_{2}$ as a function of $M$ for systems with $7\%$ and $15\%$ polydispersities. Both quantities are scaled by their respective saturation values at high $M$. We find that both show similar $M$ dependence. Ideally, the peak in the $\Delta S_{2}$ vs $M$ plot should describe the optimum value of $M$. However, there is no such peak, but just like $T_{cross}$, the $\Delta S_{2}$ value increases sharply with $M$ and then shows saturation. Note that MI is large when the distribution between two species is well segregated. However, the rdf of two consecutive species overlaps. This may be the reason the entropy difference does not show a peak. Results shown in Fig. \ref{MI_global_gr_fig} clearly suggest that for these systems, the structure and any quantity that needs structure as an input must be described by dividing the particles into a certain optimum number of species, and this division is going to increase the MI. 

\begin{figure}[h!]
  \centering
    \subfigure{\includegraphics[width=.45\textwidth]{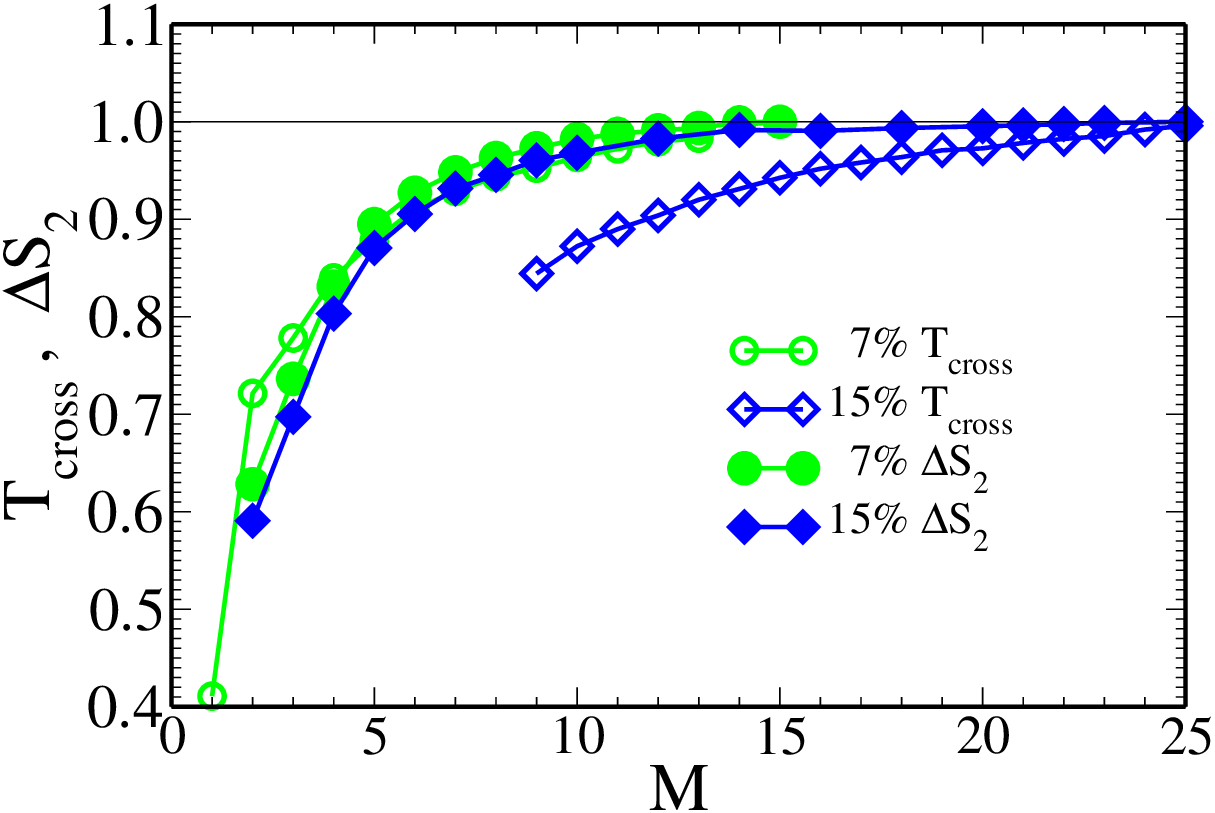}}
    \caption{\emph{Scaled MI of $\Delta S_2$ and $T_{cross}$ \cite{palak_polydisperse} saturated at same M. Scaling is performed by dividing the $\Delta S_2$ and $T_{cross}$ value by their respective saturation values at high $M$. }}
    \label{MI_global_gr_fig}
\end{figure}

\section{COMPUTING LOCAL CAGING POTENTIAL}
\label{computing_Local_caging_potential_section}
In a recent study, we described a structural parameter that describes the local caging potential \cite{manoj_PRL_2021,mohit_PRE}. We have also shown that for the KA model, the softness of this potential and the short-time dynamics are causal \cite{mohit_PRE}. The computation of the local caging potential requires information on the radial distribution function. As discussed in Sec. \ref{MI_AND_rdf_section}, the radial distribution function of a system with continuous polydispersity depends on the number of species we divide the particles of the system into. Extending our earlier work, the average depth of the mean-field caging potential for a system with $M$ species can be written as,\cite{mohit_PRE}

\begin{equation}
     \beta\Phi_{r}^{av}(\Delta r = 0) =
      -\rho \int dr \sum_{u=1}^M\sum_{v=1 }^M C_{uv}(r) \chi_{u}\chi_{v}g_{uv}(r) ,
      \label{potential_at_zero_displacement_eq}
\end{equation}
\noindent
where $\beta = \frac{1}{k_B T}$, $k_B = 1$, $\rho$ is a density, and $r$ denotes the distance between the central tagged particle and its neighbors. $\Delta r$ is the distance of the tagged particle from its equilibrium position. $C_{uv}(r)$ is the direct correlation function and, according to the Hypernetted chain approximation, can be written as,\cite{Hansen_and_McDonald}

\begin{equation}
    C_{uv}(r) = -\beta U_{uv}(r) + [g_{uv}(r)-1] - \ln[g_{uv}(r)]
    \label{c_mu_nu__with_all_term_equation}
\end{equation}
\noindent where $U_{uv}$ is the interaction potential. It was shown that the depth of the potential is inversely proportional to the curvature and, therefore the softness parameter \cite{manoj_PRL_2021,mohit_PRE}. Please note that we consider the depth of the caging potential as an energy barrier and, thus, we work with the absolute magnitude of the caging potential [given by Eq. \ref{potential_at_zero_displacement_eq}]

For the microscopic analysis, we need to calculate $\beta\Phi_{r}(\Delta r = 0)$ for every snapshot at a single particle level. This is given by Eq. \ref{potential_at_zero_displacement_eq}, where the rdf and the direct correlation function are now obtained at the single particle level. The single particle partial rdf in a single frame can be expressed as a sum of Gaussian, and it is calculated as,\cite{piggi_PRL}

 \begin{equation}
    {g_{uv}}^{i}(r) = \frac{1}{4\pi\rho r^2} \sum_{j} \frac{1}{\sqrt{2\pi\delta^{2}}} \exp^{-\frac{(r-r_{ij})^2}{2\delta^2}} ,
    \label{partial_rdf_eq}
 \end{equation}
\noindent
where  $\delta$ is the variance of the Gaussian distribution used to make the discontinuous function a continuous one. In this work, we assume $\delta = 0.09 <\sigma>$. The direct correlation function can also be calculated at the single particle level using Eq. \ref{c_mu_nu__with_all_term_equation} but with single particle rdf. At higher polydispersity index (PDI), when the system is described by one species, the rdf shows a large softening and is non-zero at very small values of 'r' compared to the interaction potential. Therefore any function that calculates the product of the potential and rdf incurs a large error.\cite{palak_polydisperse} This error is higher for repulsive potential and increases with PDI (as shown in Appendix B). In our calculation of the potential depth, such products lead to unphysically large positive values of the caging potential. This implies an unstable potential and a negative curvature /softness parameter. Note that this is an artificial effect. To overcome this problem, we have made one approximation. We assume that the potential of mean force is the same as the interaction potential, i.e., $-\beta U_{uv}(r) = \ln[g_{uv}(r)]$ and $C^{approx}_{uv}(r) \approx [g_{uv}(r)-1]$. For smaller polydispersity, where the error due to softening of the rdf is less and we can compute physically meaningful caging potential by assuming all three terms in the direct correlation function, we have compared our theoretical prediction with total and approximate direct correlation functions. As discussed in Appendix B, although the absolute value of the caging potential is different, the prediction of the correlation of the dynamics and softness parameters remains the same. Therefore, in this work, we use the approximate direct correlation function $C^{approx}(r)$ at the single particle level to avoid the unphysical results of the caging potential at higher PDI.

  The inverse of the depth of the caging potential is related to the softness, but they are not the same \cite{mohit_PRE}. There are some system dependent but temperature independent constants that are needed for the calculation of the absolute value of softness but not its correlation with dynamics.\cite{mohit_wca_lj} In this work, we will seamlessly use the terms "inverse" of the depth of the caging potential and "softness" of the caging potential, as they are qualitatively the same.

\section{SPECIES DEPENDENCE OF THE CAGING POTENTIAL}
\label{SPECIES_DEPENDENCE_OF_THE_CAGING_POTENTIAL}
\begin{figure*}
  \centering
    \subfigure{\includegraphics[width=.29\textwidth]{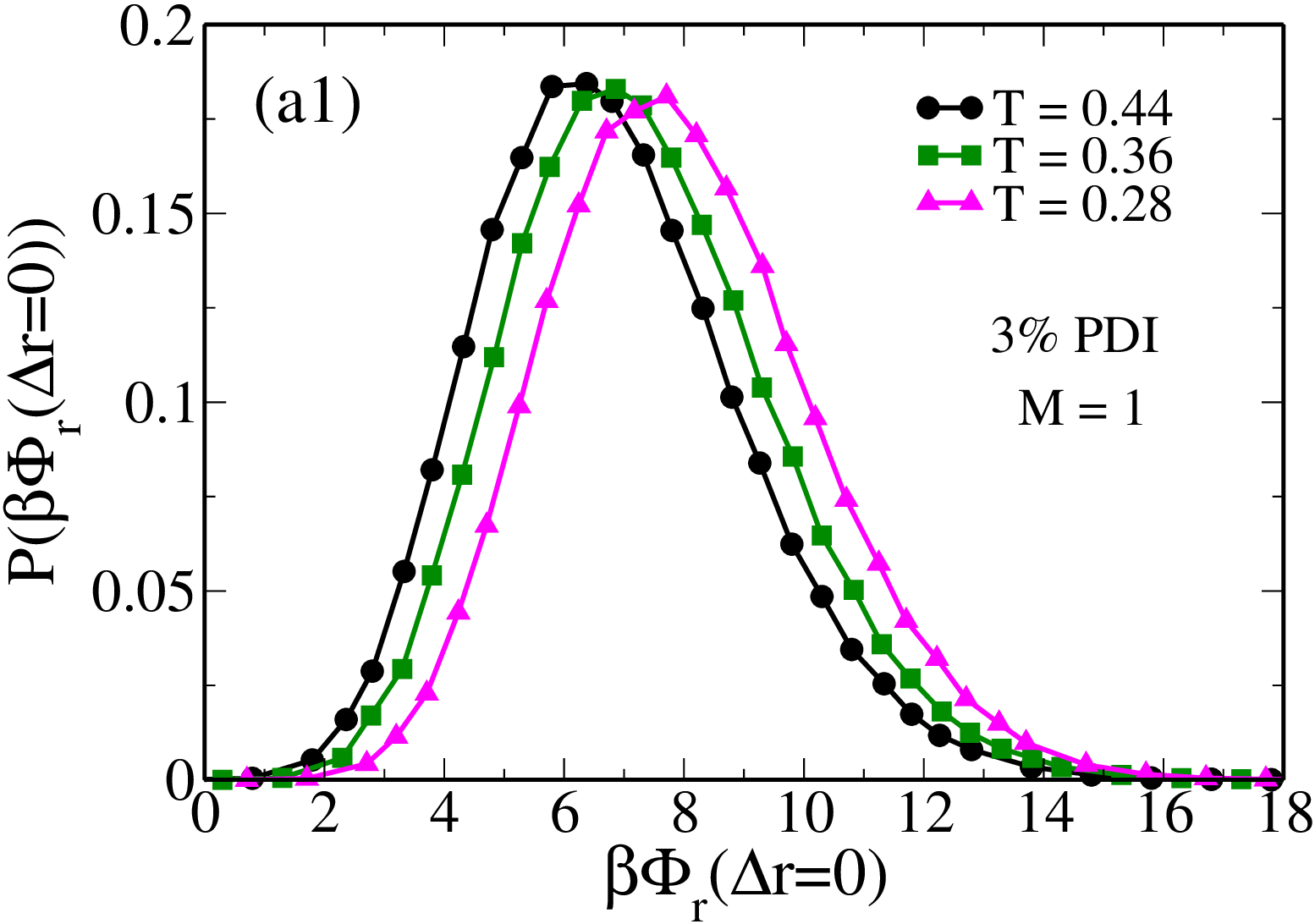}}
     \hspace{0.1cm}
      \vspace{0.2cm}
      \subfigure{\includegraphics[width=.29\textwidth]{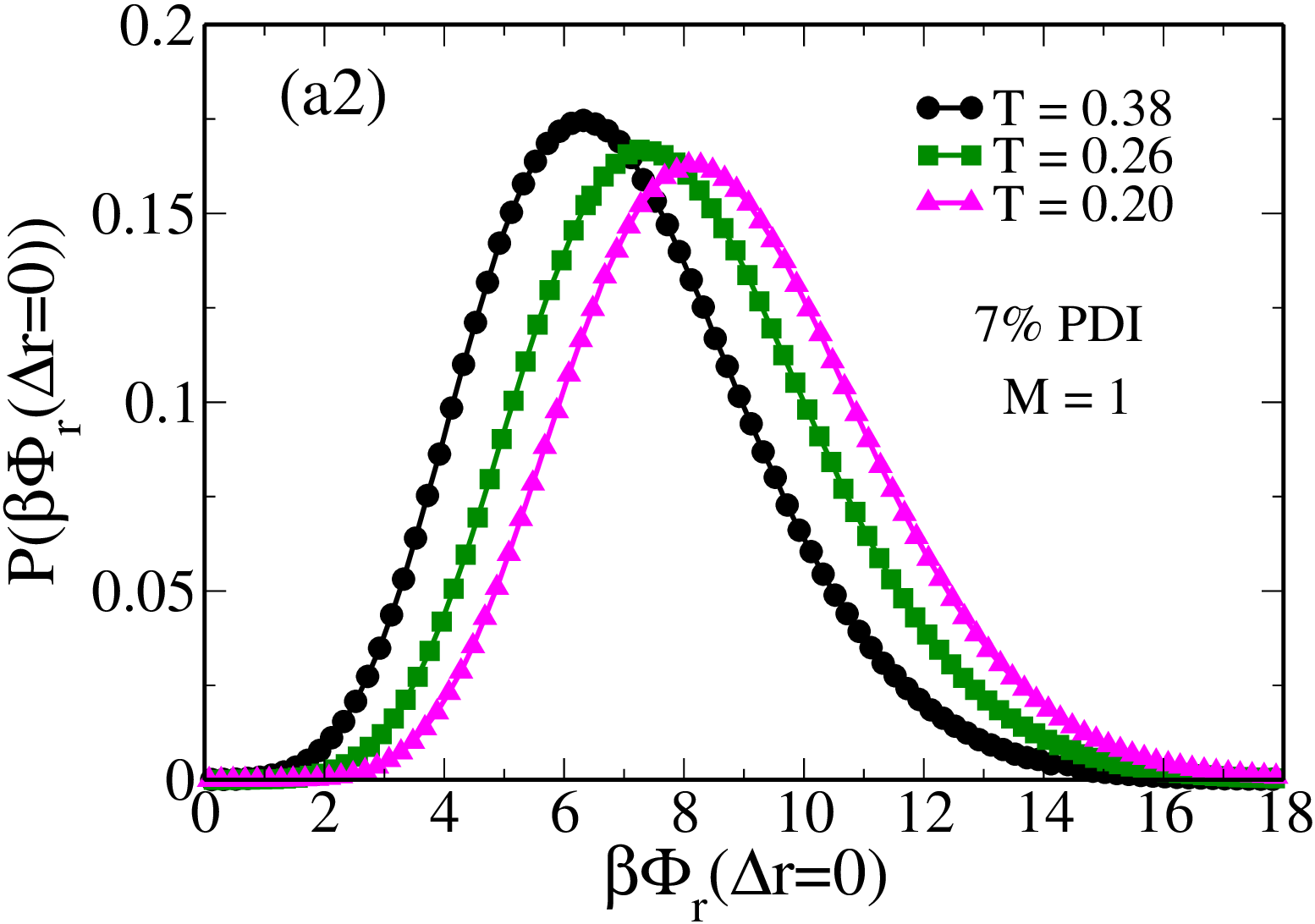}}
      \hspace{0.1cm}
       \vspace{0.2cm}
      \subfigure{\includegraphics[width=.29\textwidth]{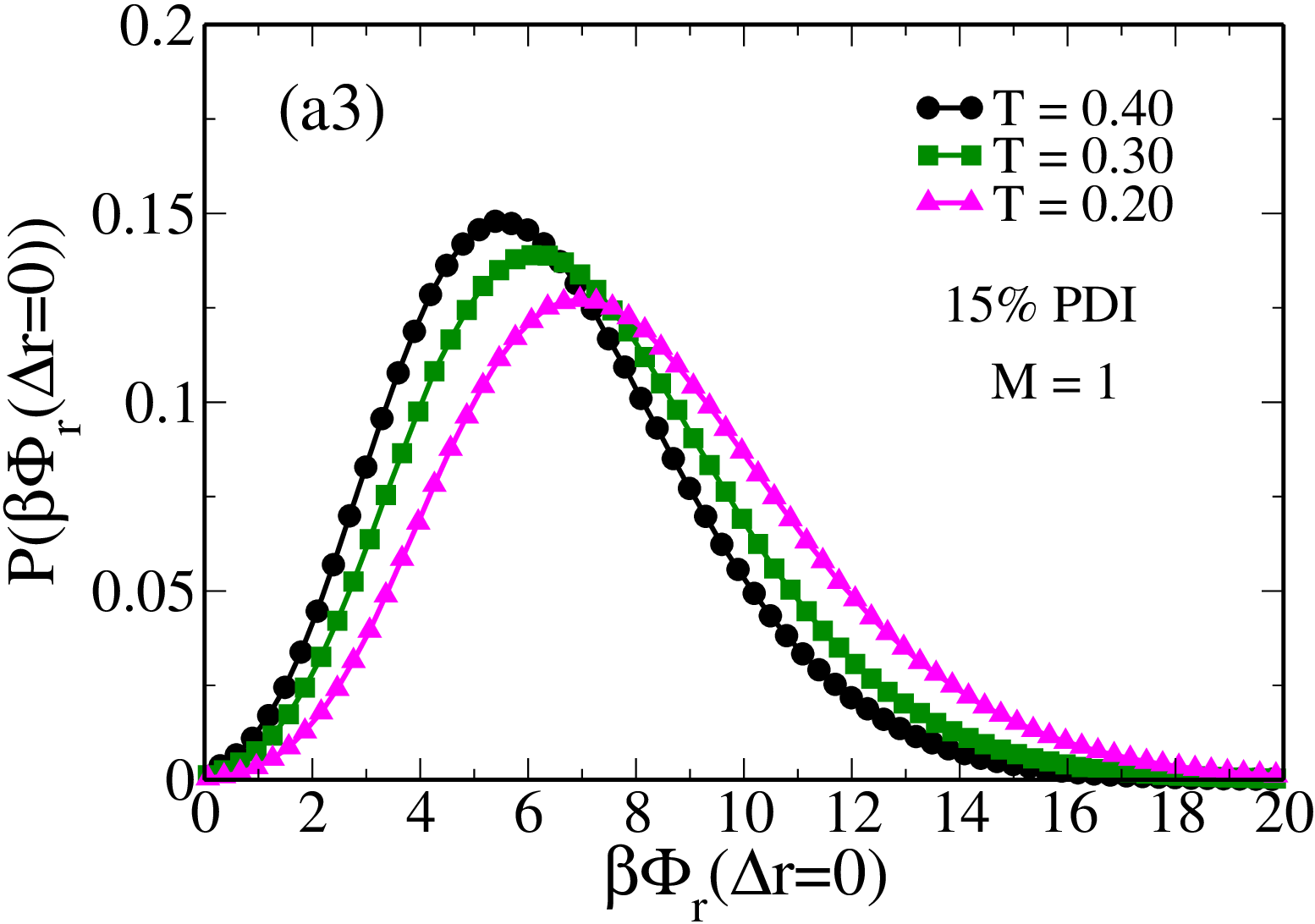}}
      \hspace{0.1cm}
       \vspace{0.2cm}
        \subfigure{\includegraphics[width=.29\textwidth]{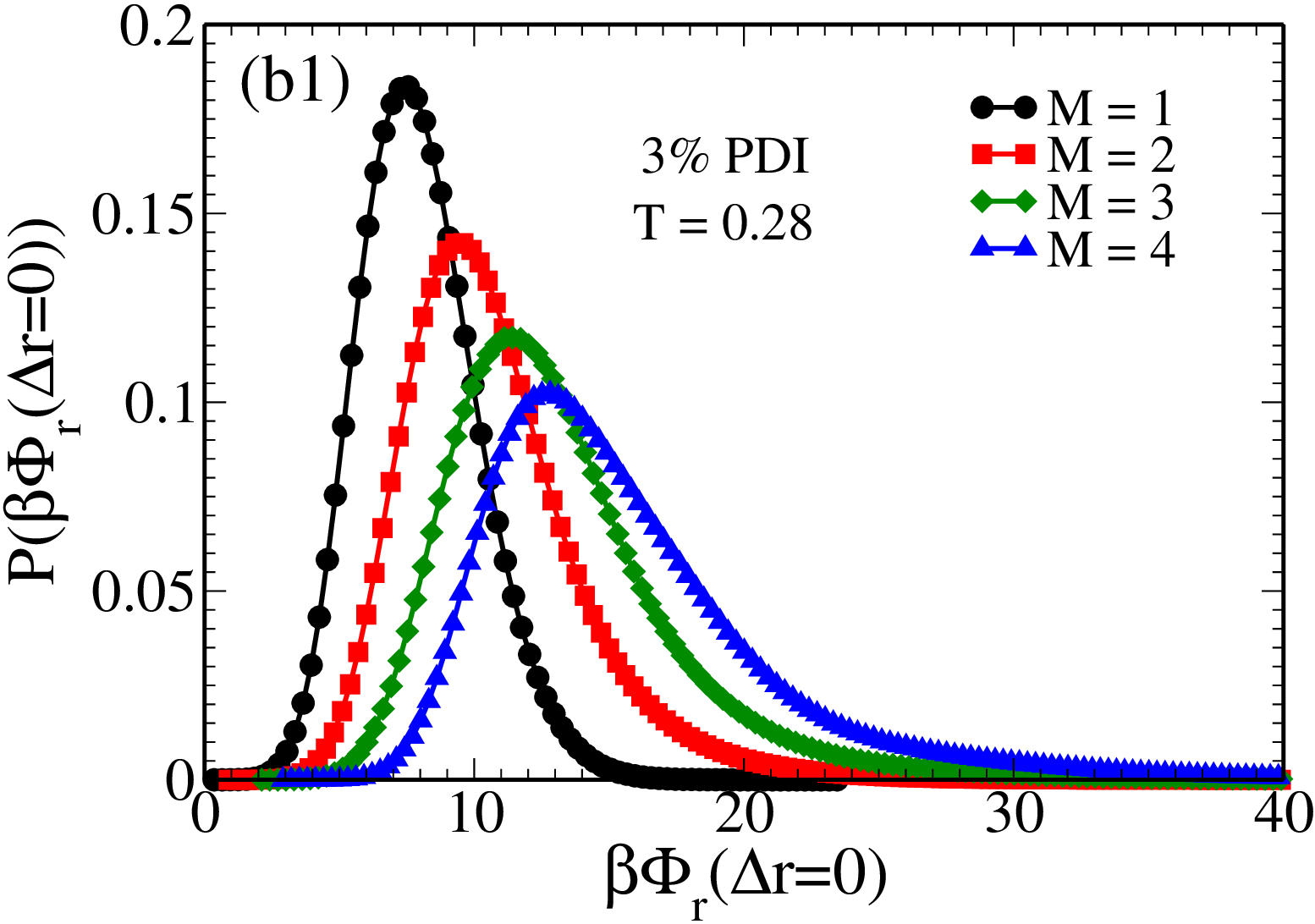}}
        \hspace{0.1cm}
     \subfigure{\includegraphics[width=.29\textwidth]{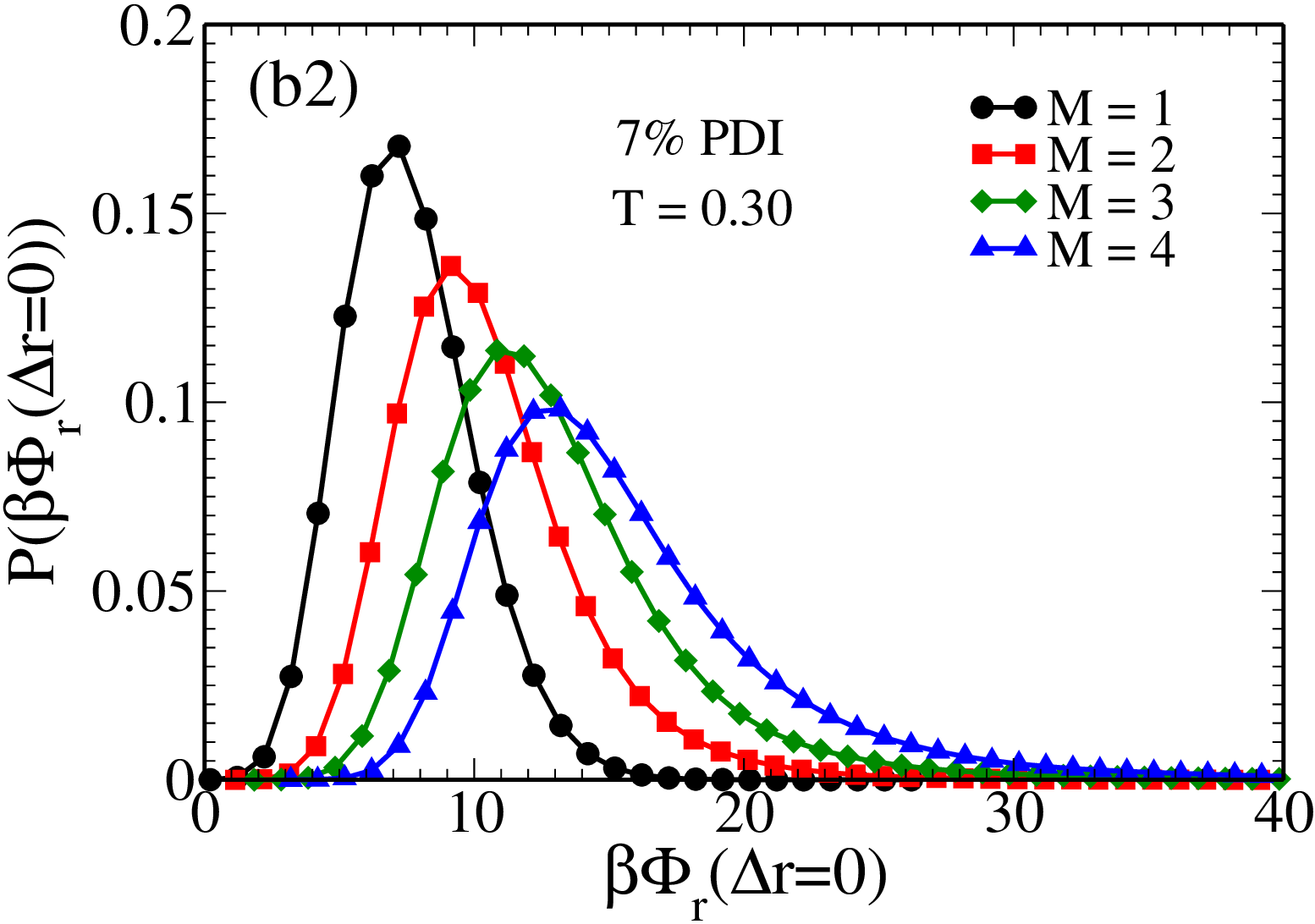}}
     \hspace{0.1cm}
      \subfigure{\includegraphics[width=.29\textwidth]{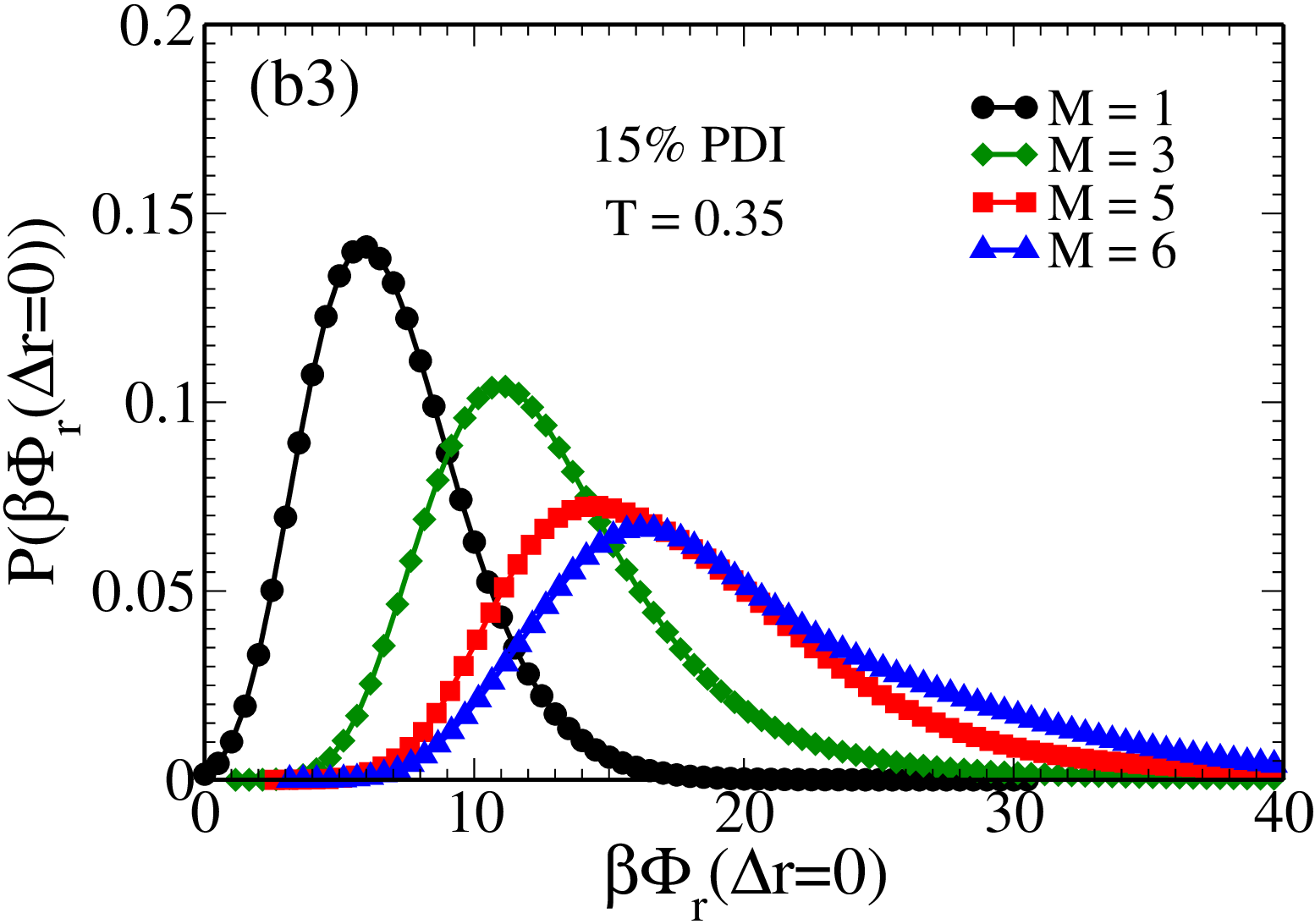}}
    \caption{\emph{Distribution of caging potential. Top panel:- Different temperatures for a fixed M =1. Bottom panel:- Different M values. (a1) and (b1) - 3\% PDI and T = 0.28, (a2) and (b2) - 7\% PDI and T = 0.30, (a3) and (b3)- 15\% PDI and T = 0.35. As expected, the caging potential increases with decreasing T. The caging potential also increases with increasing M. The temperatures are chosen such that the relaxation times are similar for all the systems.}}
    \label{potential_softness_entropy_M1}
\end{figure*}

\begin{figure*}
  \centering
    \subfigure{\includegraphics[width=.31\textwidth]{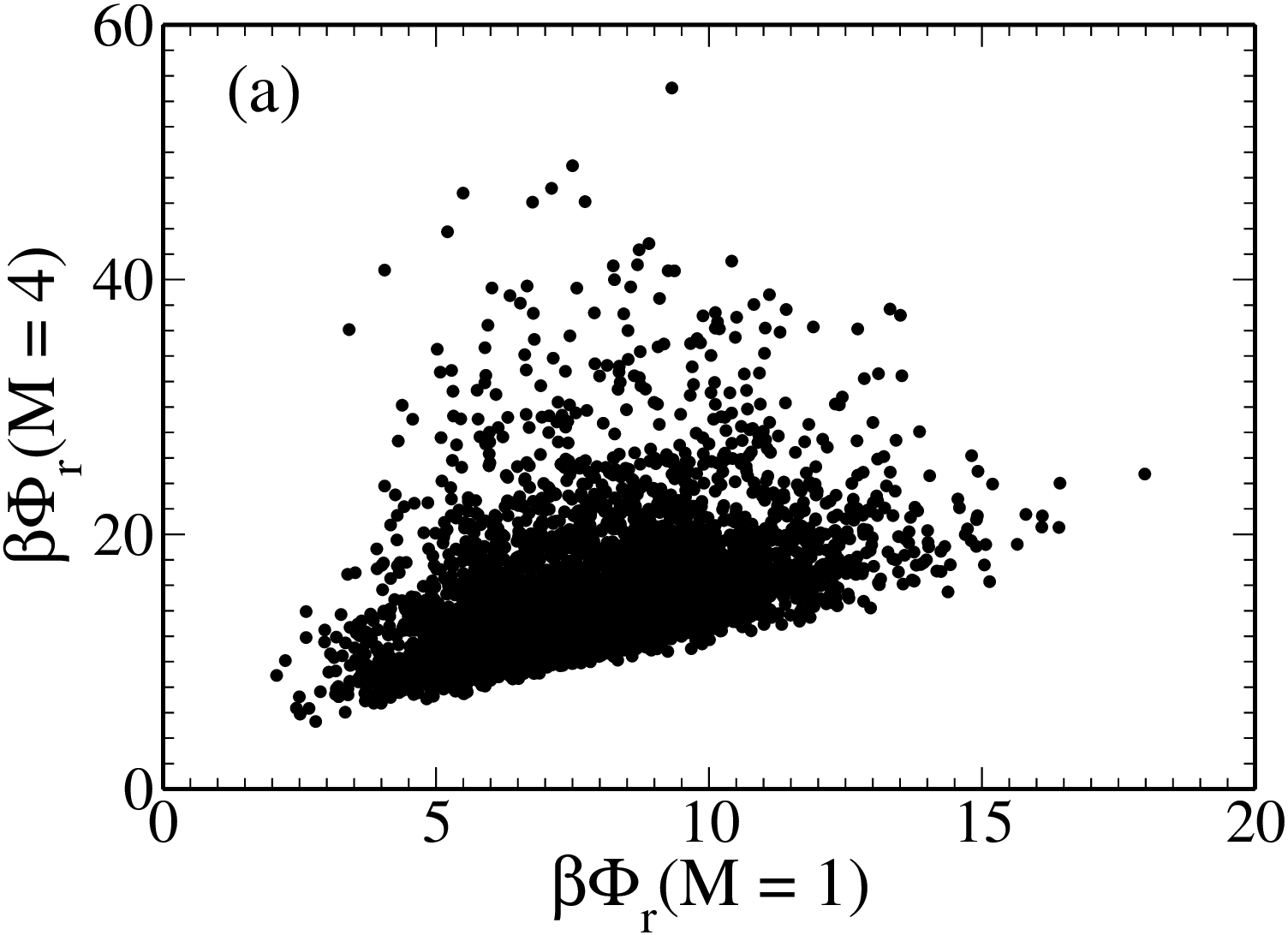}}
    \subfigure{\includegraphics[width=.31\textwidth]{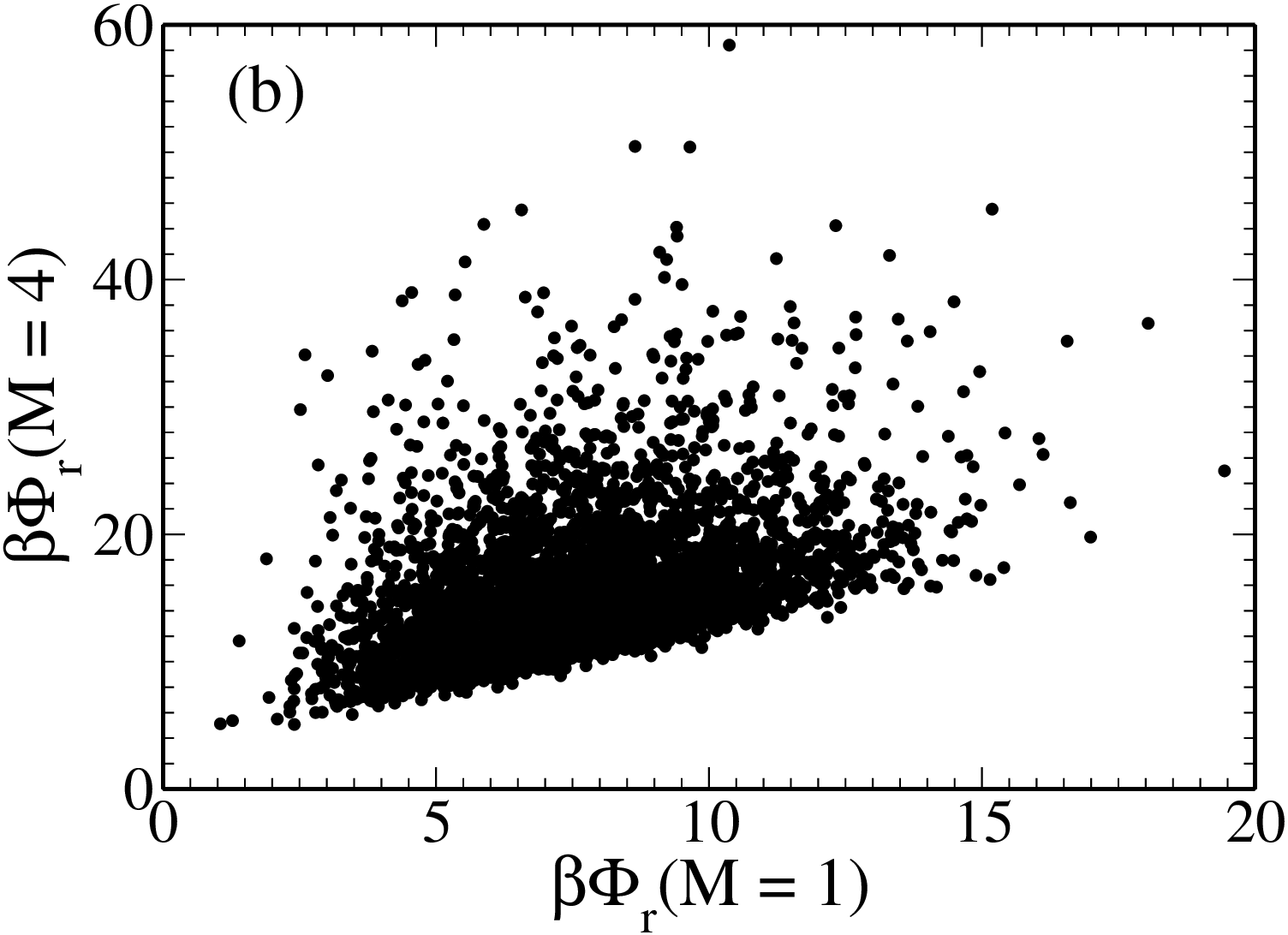}}
    \subfigure{\includegraphics[width=.31\textwidth]{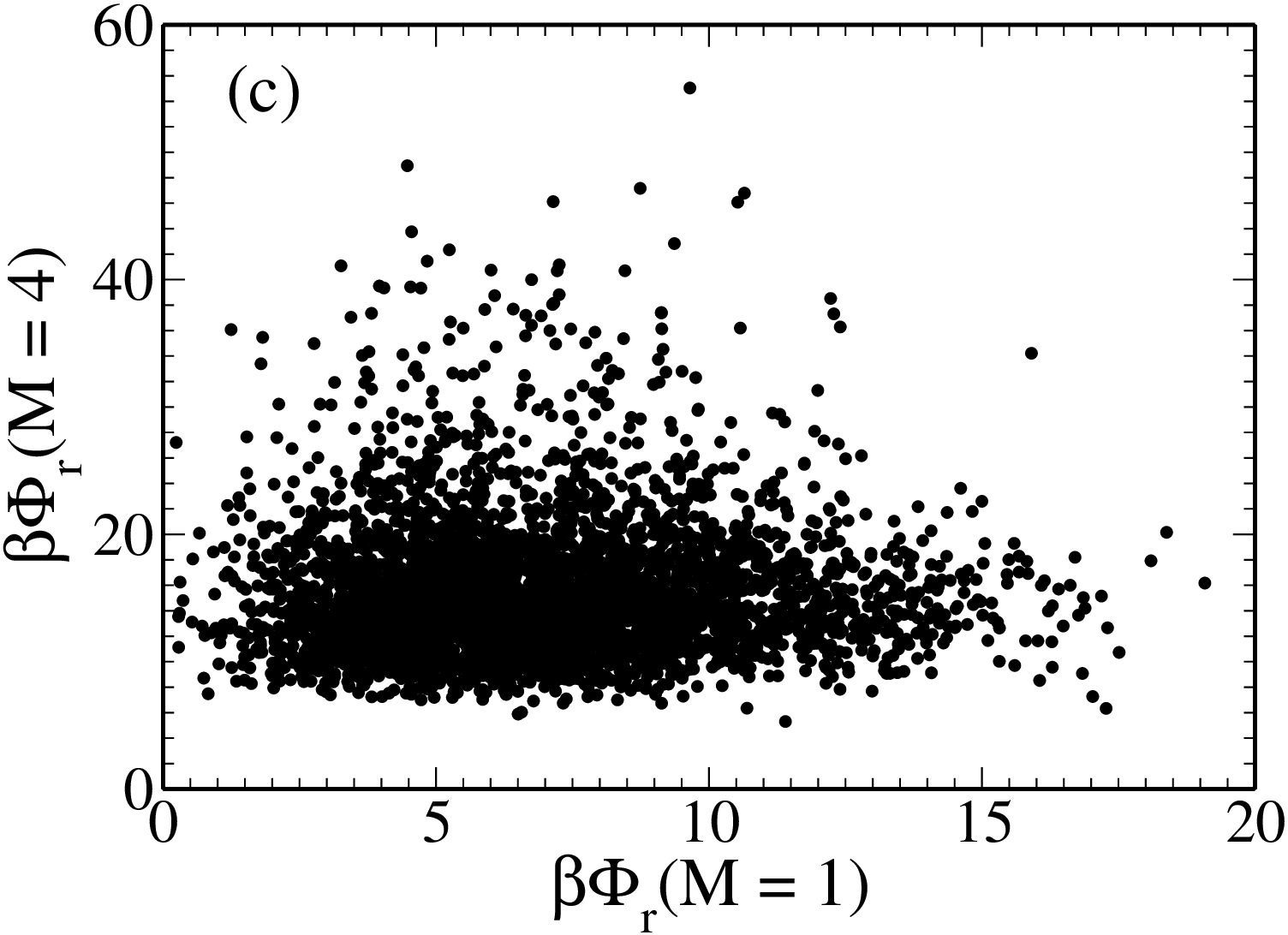}}
    \caption{\emph{Scatter plot between $\beta\Phi_r$ at M = 1 and M = 4 at different PDIs (a) 3\% PDI at T=0.28; (b) 7\% PDI at T=0.30; (c) 15\% PDI at T=0.35. The temperatures are chosen such that the relaxation times are similar for all the systems. It clearly shows that the rank of the structural order parameter of a particle changes with M. The effect increases with PDI.}}
    \label{phir_at_M1_M4_fig}
\end{figure*}

First, we assume the systems to be monodisperse i.e., $M=1$, and obtain the per-particle depth of the caging potential from the microscopic version of Eq.\ref{potential_at_zero_displacement_eq}. As shown in Fig. \ref{potential_softness_entropy_M1}, for all the systems, with a decrease in temperature, there is a shift of the probability distribution of $\beta \Phi_{r}(\Delta r=0)$ to higher values. This implies that, as expected, the cage structure is more well-defined at lower temperatures and the particles sit at a deeper potential minimum. In Fig. \ref{potential_softness_entropy_M1}, we also plot the probability distribution of $\beta \Phi_{r}(\Delta r=0)$ as a function of $M$. We find that for all the systems with an increase in $M$, the probability distribution of $\beta \Phi_{r}(\Delta r=0)$ moves to higher values of $\beta \Phi_{r}(\Delta r=0)$. This shift is concurrent with the fact that when a polydisperse system is treated as a monodisperse system, the RDF shows artificial softening \cite{palak_polydisperse}. However, when the polydisperse system is divided into an $M$ number of species, the inter- and intra-species RDFs become sharper than the RDF obtained assuming a single species. Therefore the cage is better defined by the multispecies system. This gives rise to an increase in the depth of the minima. This increase in the depth of the caging potential with an increase in $M$ is similar to the decrease in the two-body pair entropy obtained in our earlier study \cite{palak_polydisperse}.  Furthermore, to understand if this shift in the distribution of the caging potential with $M$ is just an increase in the depth of the particle level caging potential affecting all particles equally, in Fig. \ref{phir_at_M1_M4_fig}, as a representative plot, we show a scatter plot of the particle level caging potential obtained for M = 1 and M = 4. This clearly shows that this shift in the distribution is not just a shift in the value of the particle level caging potential and affects each particle differently. As expected, the M dependence is more at a higher PDI.

\section{SPECIES DEPENDENCE OF THE CORRELATION OF CAGING POTENTIAL WITH PARTICLE DYNAMICS}
\label{onset_caluclation_PRS}
\begin{figure}[h!]
  \centering
  \subfigure{\includegraphics[width=.24\textwidth]{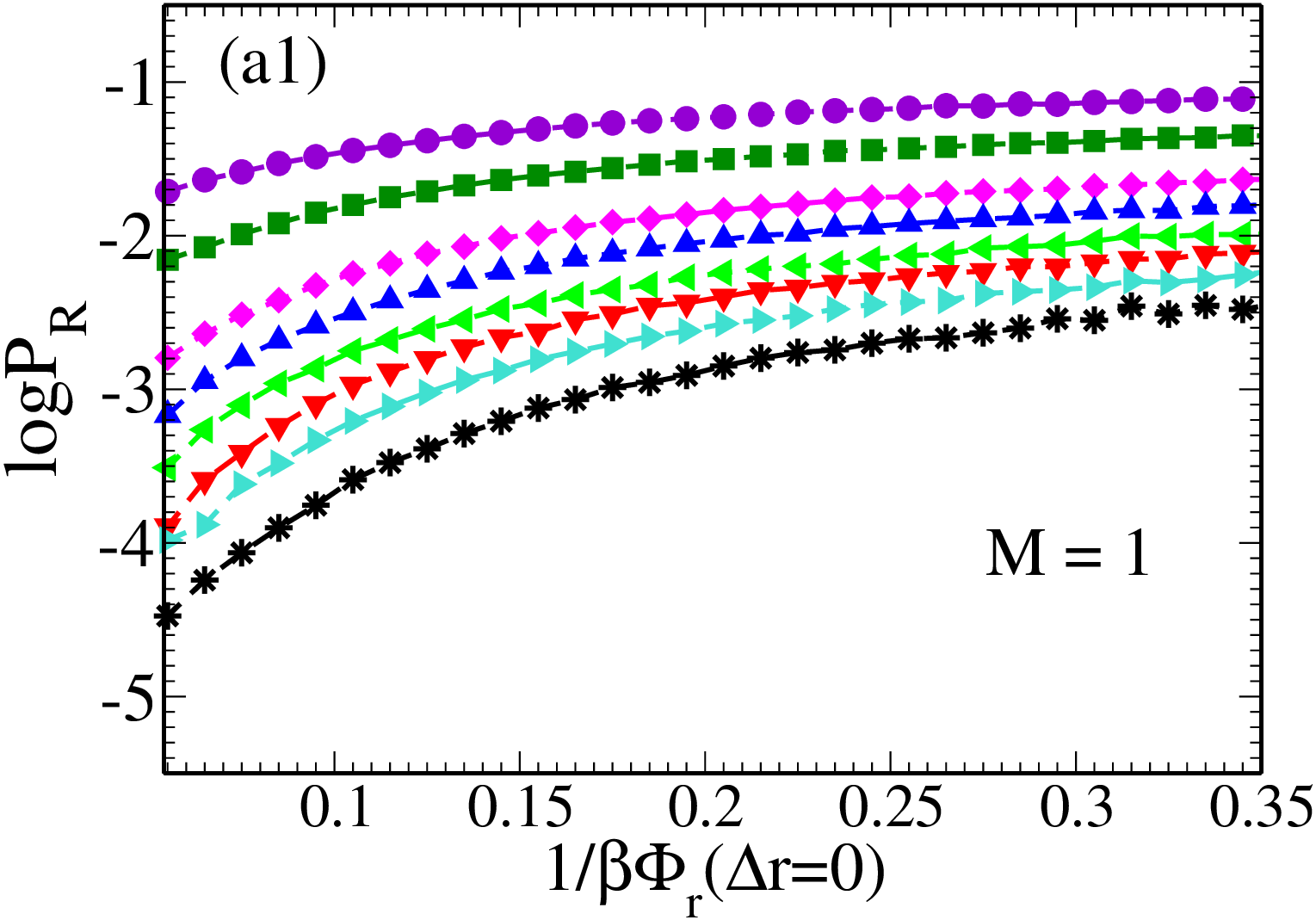}}
   \vspace{0.1cm}
    \subfigure{\includegraphics[width=.235\textwidth]{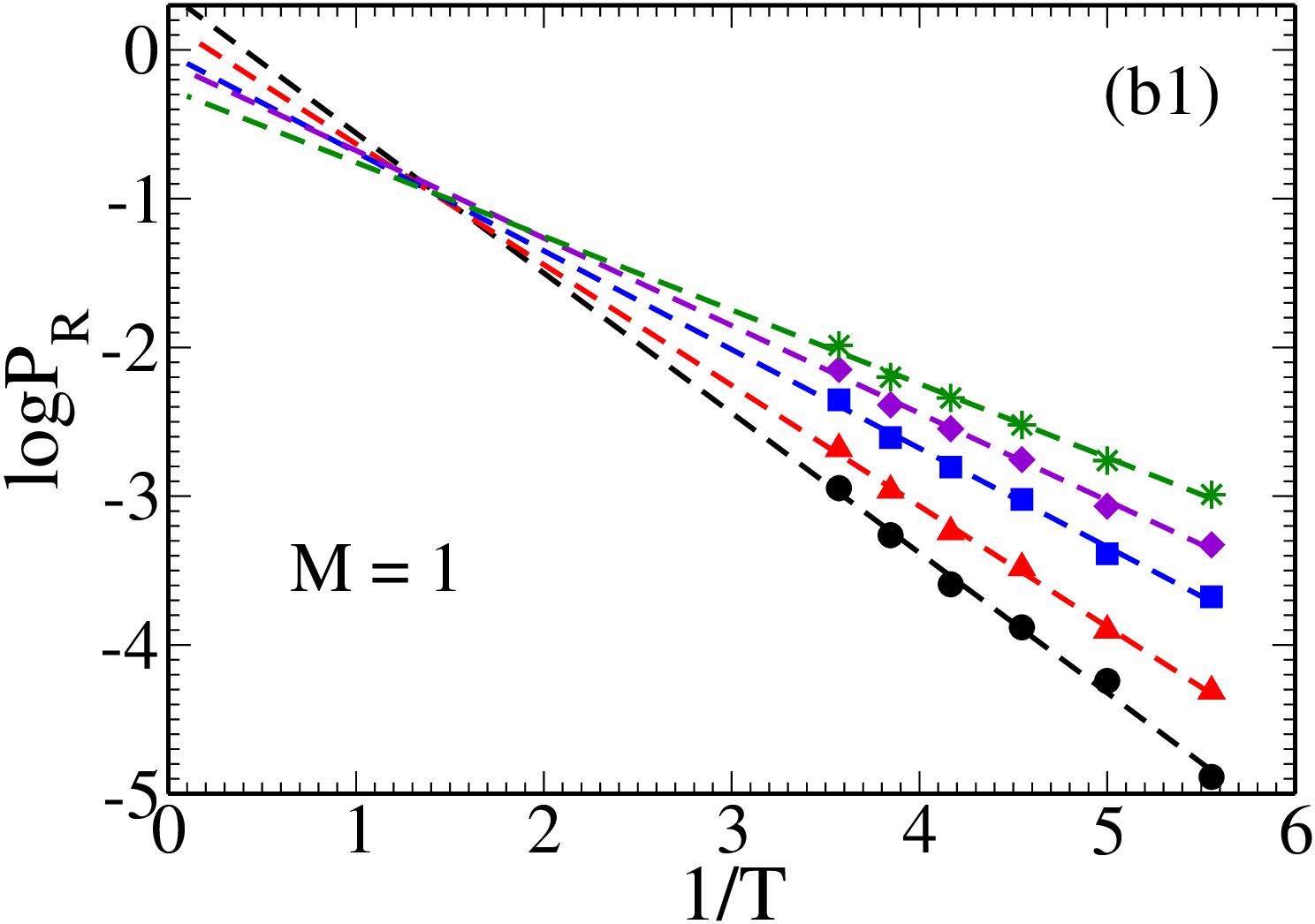}}
     \vspace{0.1cm}
    \subfigure{\includegraphics[width=.24\textwidth]{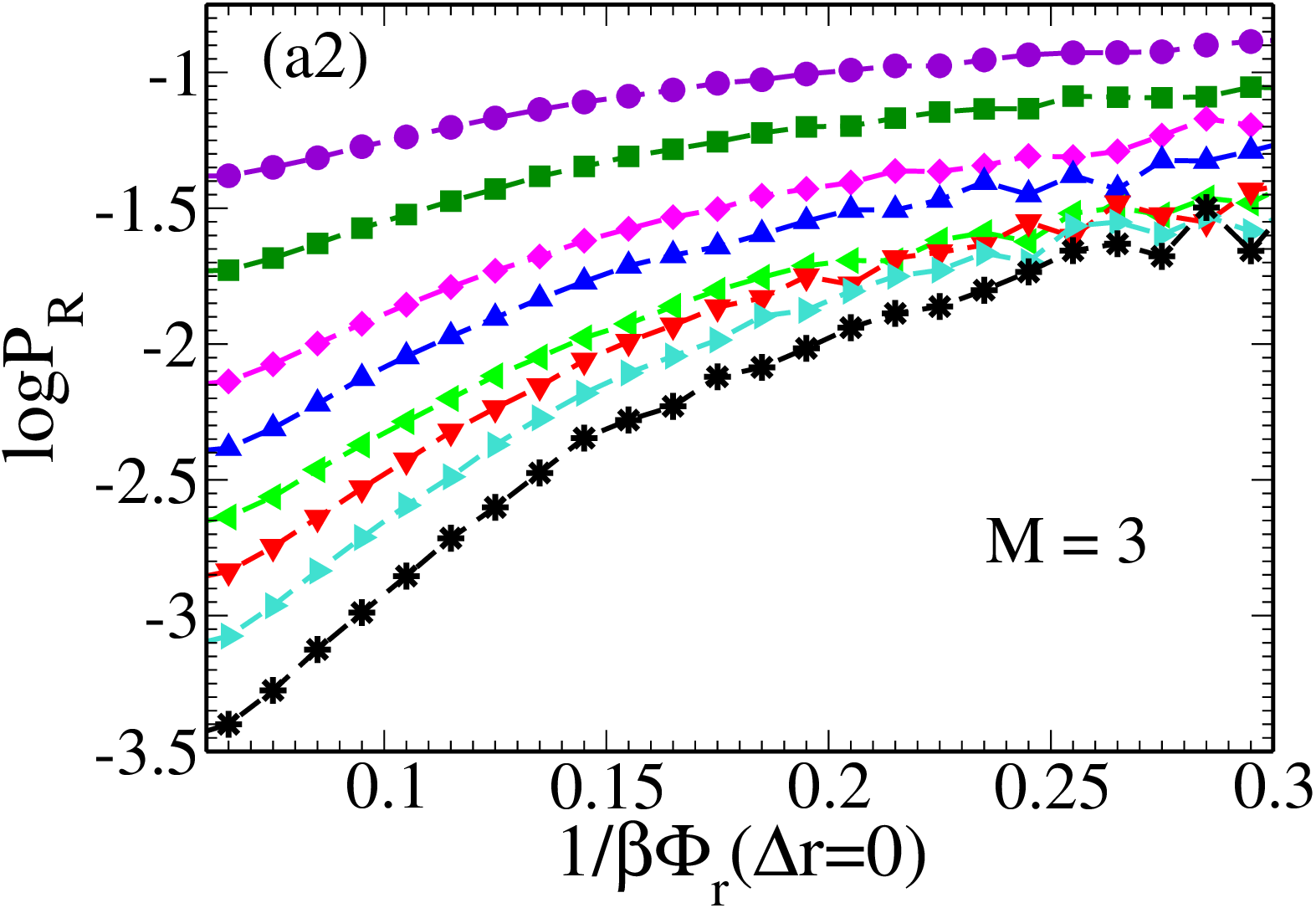}}
     \vspace{0.1cm}
    \subfigure{\includegraphics[width=.235\textwidth]{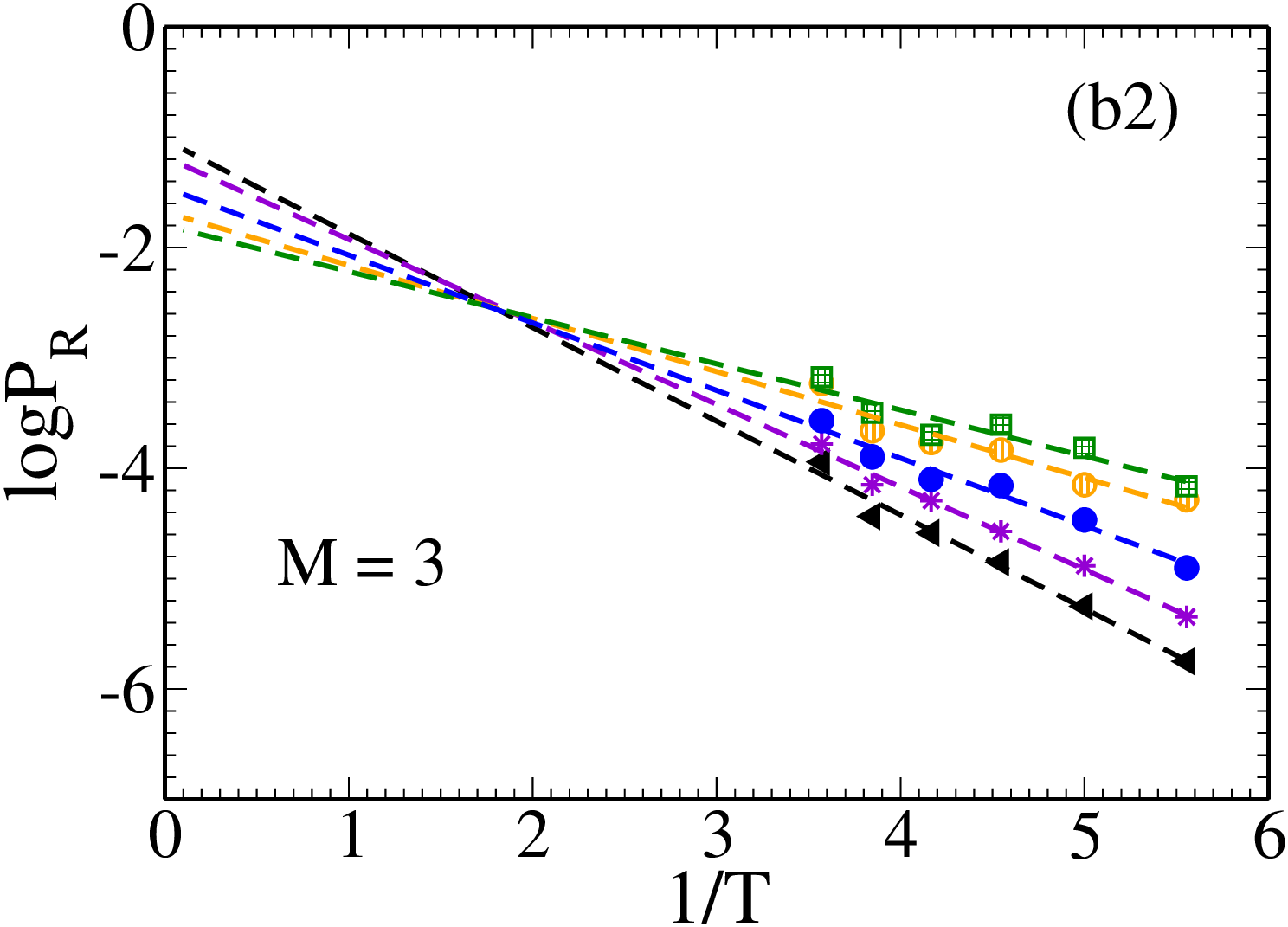}}
     \vspace{0.1cm}
    \subfigure{\includegraphics[width=.234\textwidth]{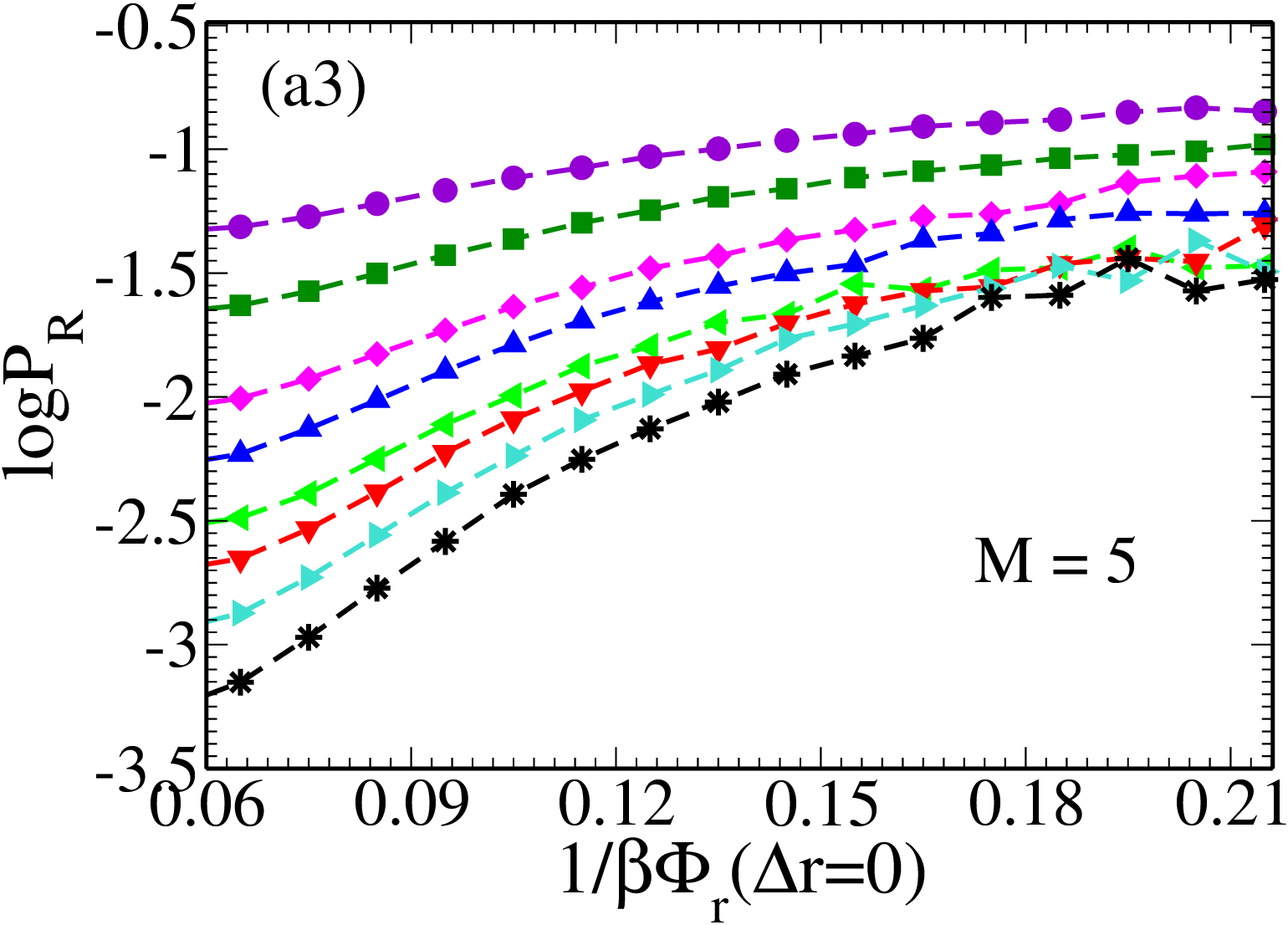}}
     \vspace{0.1cm}
    \hspace{0.01cm}
    \subfigure{\includegraphics[width=.225\textwidth]{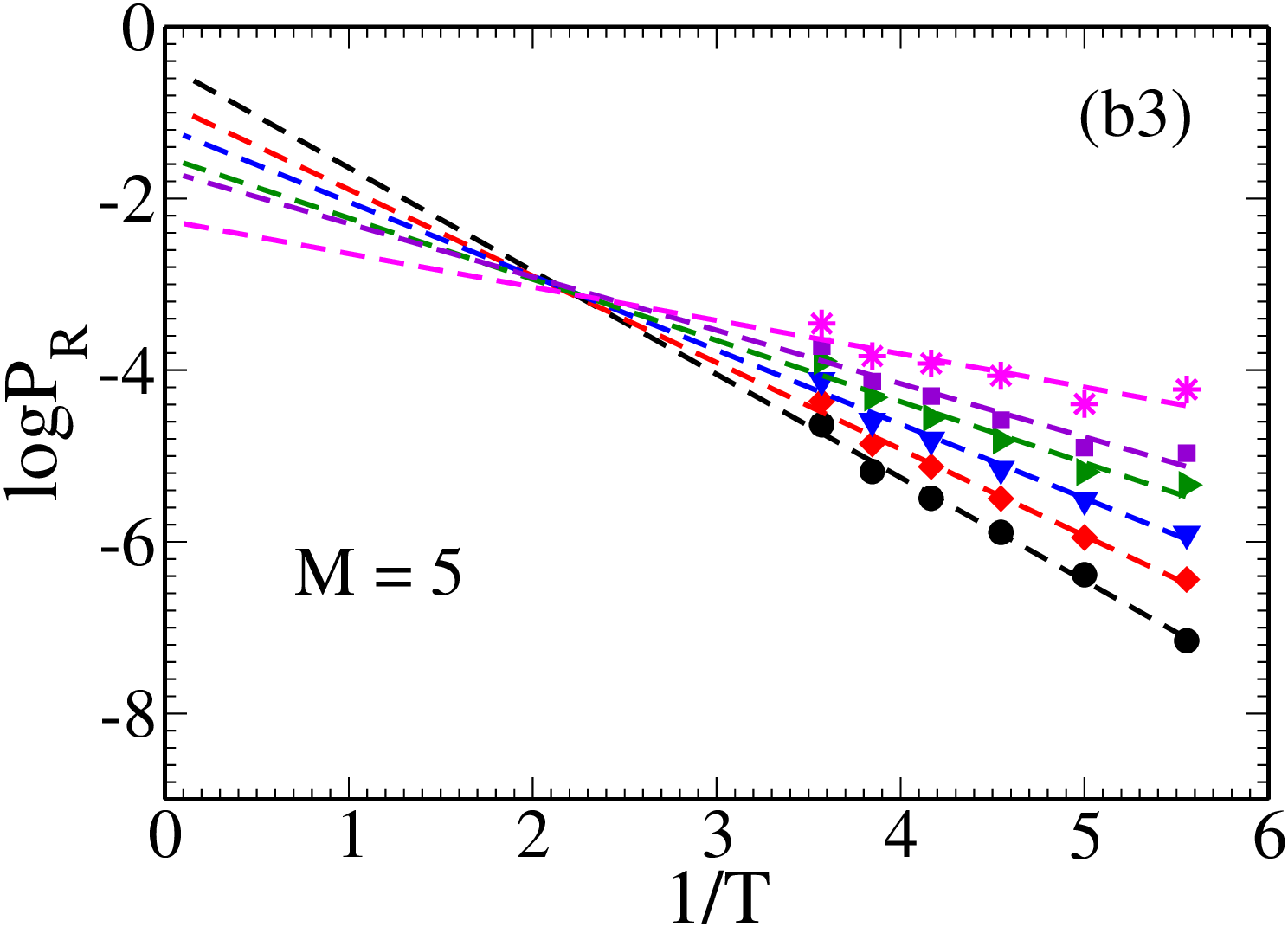}}
    \caption{\emph{Correlation between structure and dynamics as a function of $M$. Left panel: (a1)-(a3) The fraction of particles that underwent rearrangement $P_R(1/\beta\Phi_r)$ vs the depth of the caging potential, $1/\beta\Phi_r$ at different T [0.4 (violet circle) - 0.2(black star)], Right panel: (b1)-(b3) $P_R(1/\beta\Phi_r)$ as a function of 1/T at different values of the inverse of the caging potential. The results are for a 15\% PDI.}}
    \label{log_prs_fig_M1}
\end{figure}

\begin{figure}[h!]
  \centering
  \subfigure{\includegraphics[width=.35\textwidth]{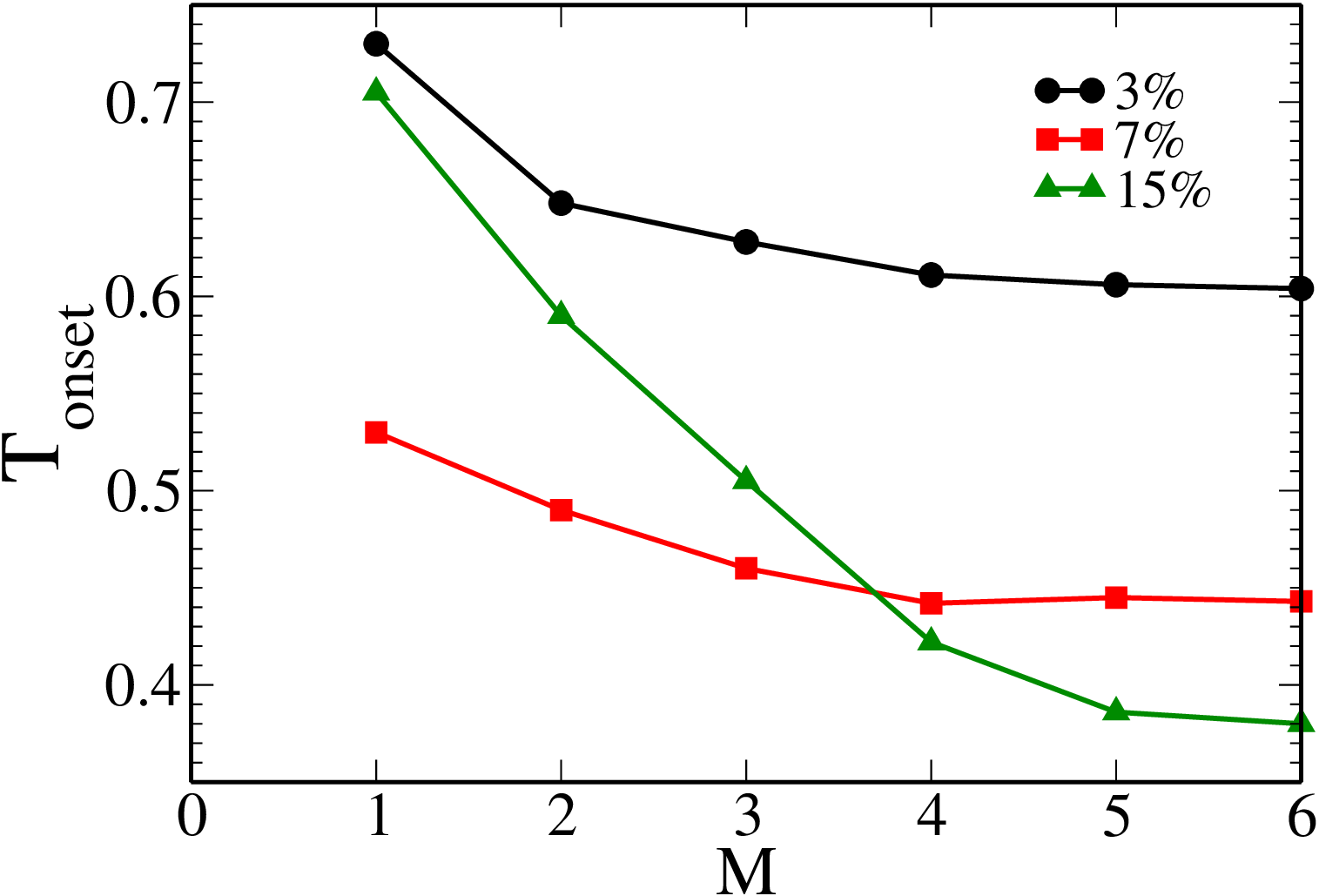}}
    \caption{\emph{ (a) $T_{onset}$ obtained from the crossing of $P_R(1/\beta\Phi_r)$ vs 1/T plots (Fig. \ref{log_prs_fig_M1}) as a function of $M$. The $T_{onset}$ initially decreases with $M$ before saturating at higher values of $M$. With an increase in polydispersity, the saturation increases to higher $M$ values.}}
    \label{T_onset_diff_M_fig}
\end{figure}

In Sec \ref{SPECIES_DEPENDENCE_OF_THE_CAGING_POTENTIAL}, we have shown that the distribution of the local caging potential varies with $M$. Suppose this variation was just a shift in the value of the caging potential of each particle. In that case, we do not expect the correlation between the caging potential and the dynamics to be affected by $M$. However, as shown, that is not the case. Therefore in this section, we study the correlation between the dynamics and the structure obtained via the local caging potential as a function of $M$. To understand the correlation between the dynamics and the structure, we follow the methodology used in earlier works.\cite{Andrea_liu_nature,mohit_PRE} We identify fast particles using a well-documented method\cite{Candelier_PRL,smessaert_PRE,mohit_PRE} also given in Appendix C. After identifying the fast particles, we correlate them with the local SOP. We calculate the fraction of particles having a specific value of $1/\beta\Phi_r$ that undergo rearrangement, $P_R(1/\beta\Phi_r)$, and plot it as a function of $1/\beta\Phi_r$ at different T and M values. The plots for the system with 15$\%$ PDI, where the effect is maximal, are shown here in Fig. \ref{log_prs_fig_M1}. The results are similar for other systems. We find that $P_R(1/\beta\Phi_r)$ has a dependence on the SOP that becomes stronger at lower temperatures. At lower temperatures, particles with a higher value of softness (sitting in a shallow caging potential) have a higher probability of moving. Apparently, the behavior appears to be $M$ independent.  Following our earlier work, we plot the $P_{R}(1/\beta\Phi_r)$ as a function of temperature for different $1/\beta\Phi_r$ values. We find that for all the cases, it can be expressed in an Arrhenius form: $P_R(1/\beta\Phi_r) = P_0(1/\beta\Phi_r)\exp [\Delta E(1/\beta\Phi_r)/T]$, where $\Delta E$ is the activation energy. These plots also appear to be similar for all $M$ values. It was earlier shown that the temperature at which these $P_{R}(1/\beta\Phi_r)$ vs $T$ plots for different softness values intercept marks the onset temperature of glassy dynamics \cite{Andrea_liu_nature,mohit_PRE}. The origin of this observation was explained by the microscopic mean field theory \cite{mohit_PRE,manoj_PRL_2021}. According to the theory, we can correlate softness and dynamics only when the cage around the particles is well-defined. It is well known in the supercooled liquid literature that only below the onset temperature where there is a separation between the short and long time dynamics the particles in the short time feel caged by their neighbors and this cage becomes longer lived at lower temperatures. Therefore the crossing of the plots marks the highest temperature where this theory is valid, and beyond that, due to the absence of any well-defined cage, the theoretical formulation breaks down. In addition, at lower temperatures, where the lifetime of the cage increases, the structure becomes a better predictor of the dynamics. We extract the onset predicted by the crossing of the $P_{R}(1/\beta\Phi_r)$ vs $T$ plots. They are plotted in Fig. \ref{T_onset_diff_M_fig}. It clearly shows that the $T_{onset}$ values have a $M$ dependence. The value of $M$, where it saturates, increases with the percentage of polydispersity. The saturation temperature is similar to the onset temperature obtained using other methods.\cite{palak_polydisperse} This result is similar in spirit to that obtained in our earlier work using two-body excess entropy.\cite{palak_polydisperse}

\section{ANALYSIS OF DYNAMICS OF SOFT AND HARD PARTICLES}
\label{ANALYSIS_OF_DYNAMICS_OF_SOFT_AND_HARD_PARTICLES}
\begin{figure*}
  \centering
  \subfigure{\includegraphics[width=.295\textwidth]{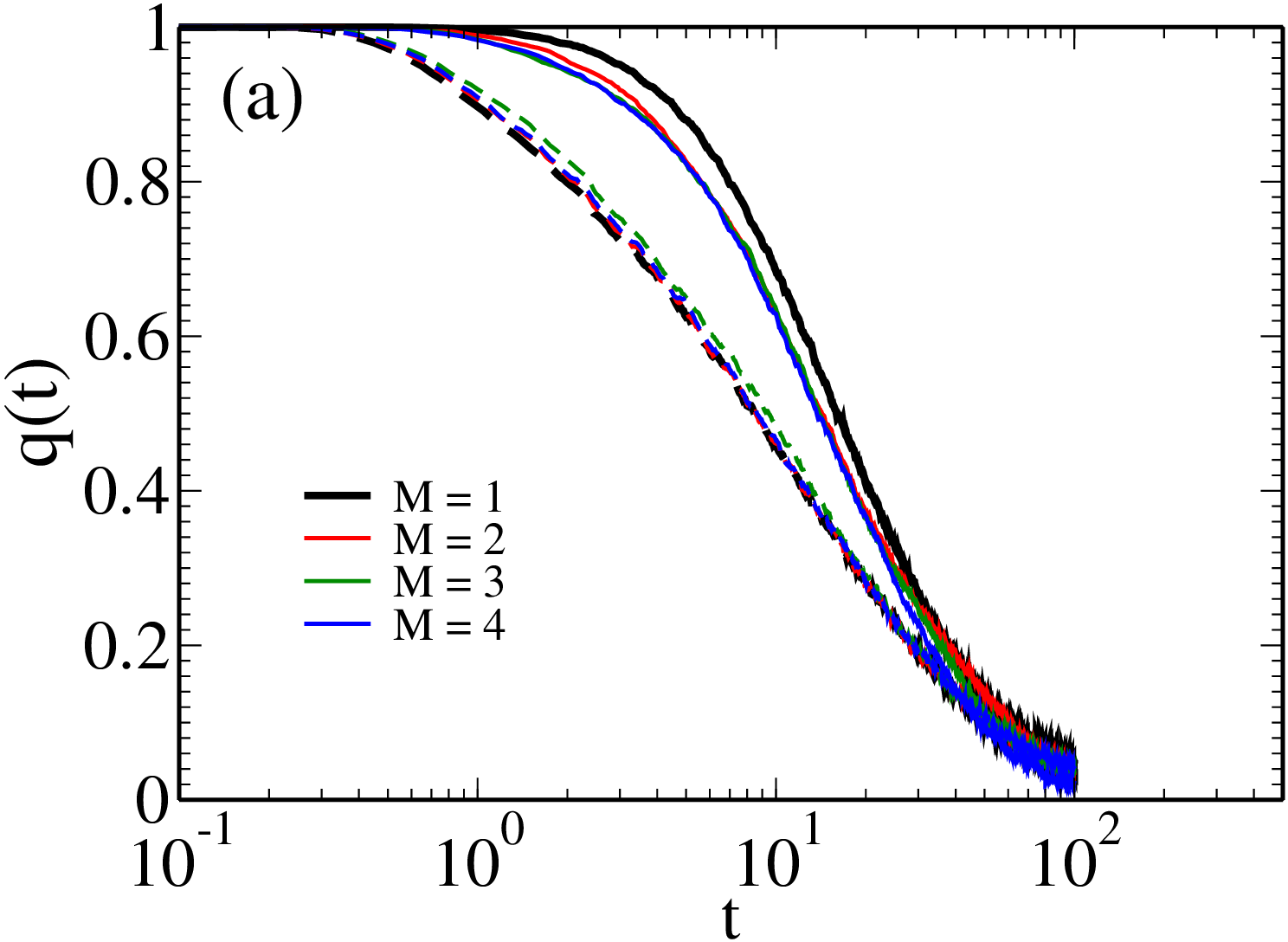}}
   \hspace{0.1cm}
  \subfigure{\includegraphics[width=.29\textwidth]{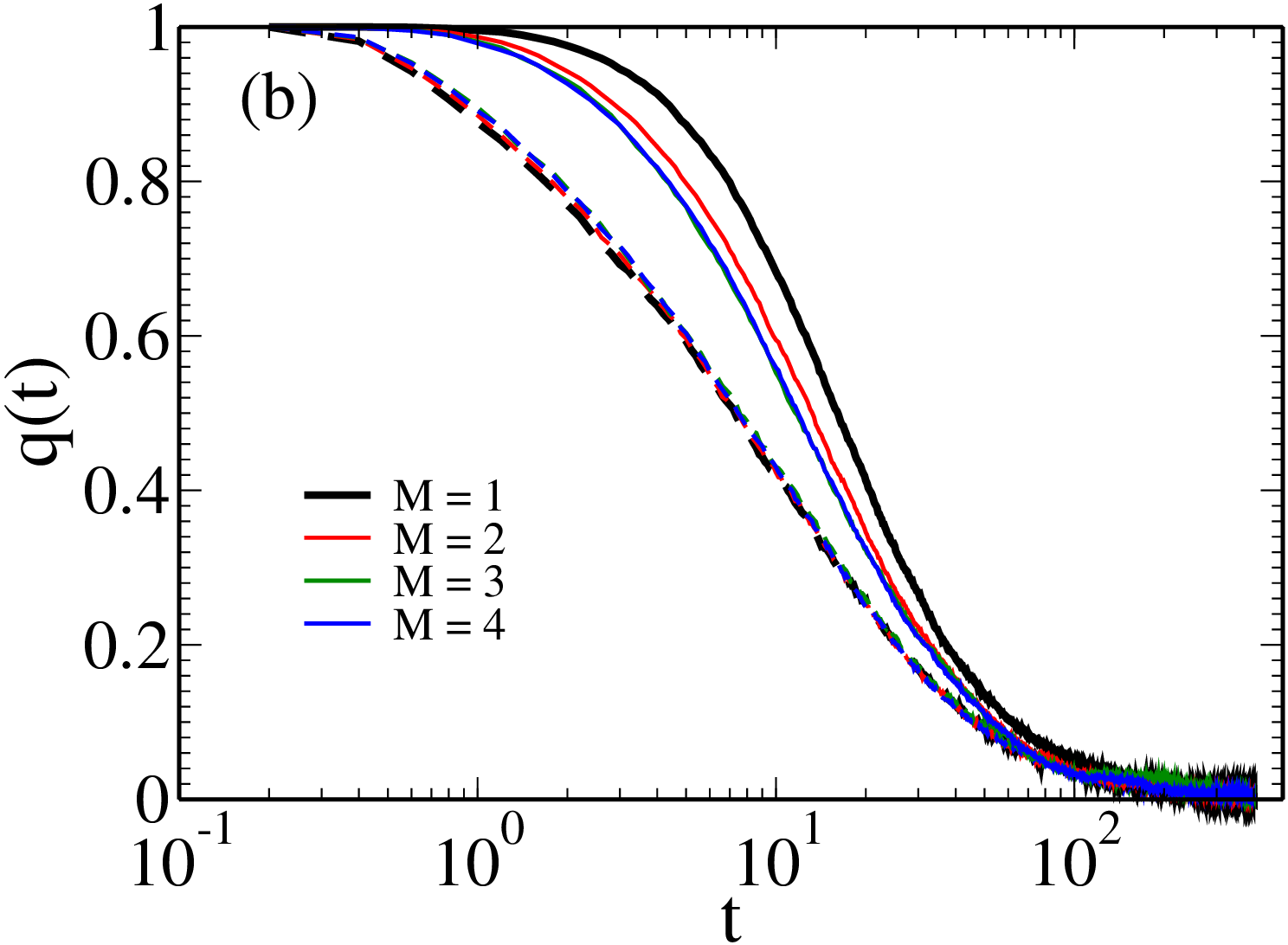}}
   \hspace{0.1cm}
  \subfigure{\includegraphics[width=.29\textwidth]{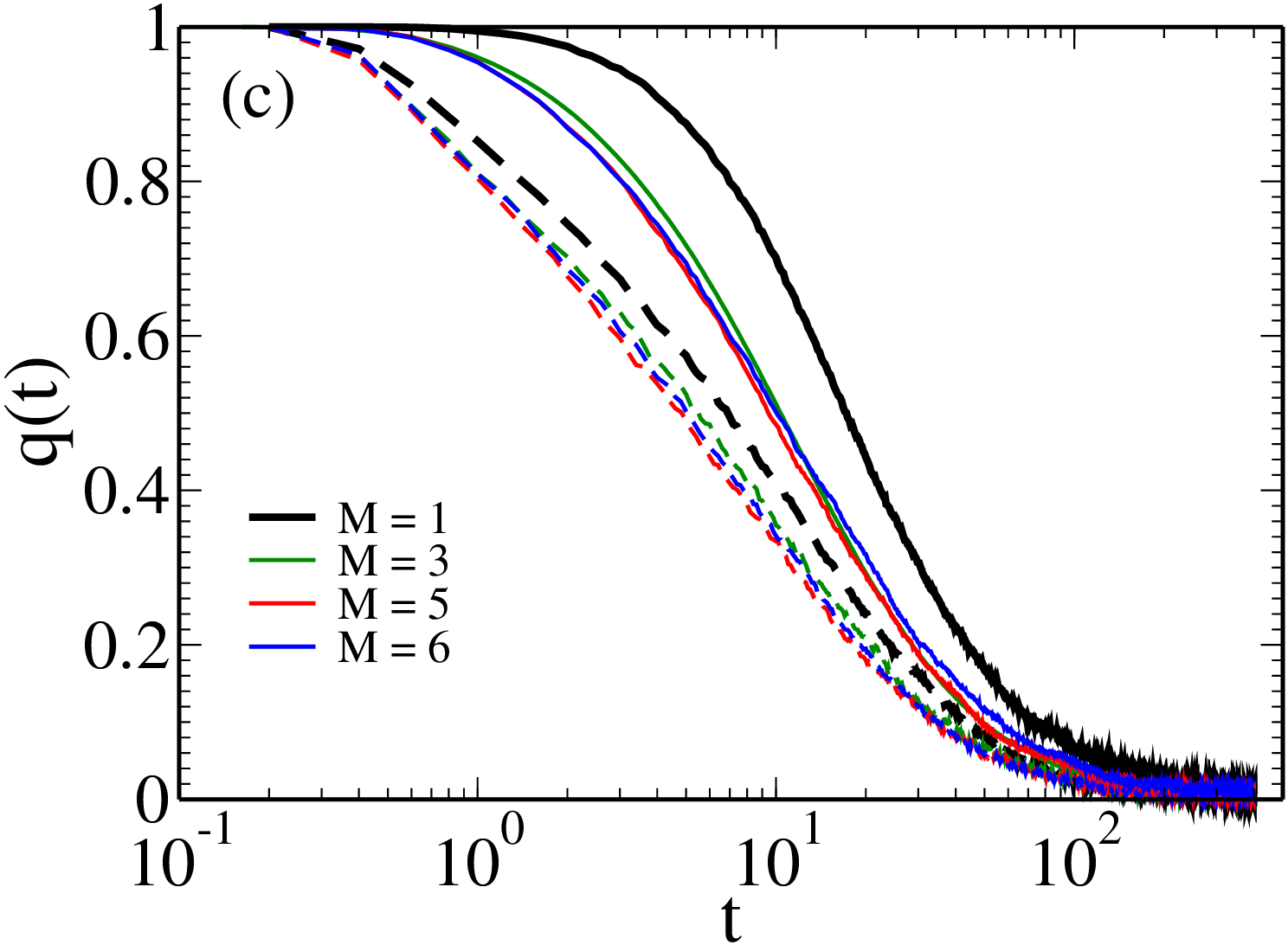}}
    \caption{\emph{Dynamics of a few softest (high $1/\beta\Phi_r$ value)(dotted line) and few hardest (low $1/\beta\Phi_r$ value)(solid line) particles at different M values. (a) 3\% PDI (T = 0.28), (b) 7\% PDI (T = 0.30), and (c) 15\% PDI (T = 0.35). T is chosen such that the relaxation times of each system are in the same range. The dark lines are for M=1, and with an increase in $M$, the plots shift.}}
    \label{overlap_hard_soft_region_at_same_t_fig}
        \end{figure*}

Since we have established that, on average, the particles with higher softness have a higher probability of moving, we can expect that if we compare the dynamics (via the overlap function) of a few of the hardest and softest particles, then at short times they will show a large difference, and eventually due to the evolution of the cage and its softness around the particle they will decay at the same time.\cite{mohit_PRE} Dynamics of particles via the overlap function [q(t)] can be calculated as: 

\begin{equation}
    q(t)= \frac{1}{N} \sum_{i=1}^{N} \omega(|r_{i}(t) - r_{i}(0)|) ,
    \label{overlap_equation}
\end{equation}
\noindent
where function $\omega(x) = 1$ when $0 \leq x \leq a$ and $\omega (x) = 0$ otherwise. The cutoff of the overlap parameter a = 0.5 is chosen such that particle positions separated due to small amplitude vibrational motion are treated as the same.\cite{overlap_shiladitya} Here, we restrict our study to one temperature for each system. For the $3\%$ PDI system, we choose T = 0.28, the lowest temperature at which we can run the system before it undergoes crystallization. For the other two systems, we study them at temperatures where the relaxation times are similar to those of the $3\%$ PDI system at T=0.28. We pick a few (around 2) of the hardest and softest particles, and the softness parameter is calculated for the same system at different values of $M$ (Fig. \ref{overlap_hard_soft_region_at_same_t_fig}). 

We find that the difference in the overlap of the few hardest and softest particles changes with M. However, beyond a certain value of $M$, the overlap functions of the hardest and the softest particles do not change with $M$. This suggests that beyond this $M$ value, the identification of the hardest and softest particles becomes independent of $M$. We consider this the optimum value of the species, $M_{0}$, needed to describe the system. For 3$\%$ PDI, $M_{0}=3$, for 7$\%$ PDI, $M_{0}=4$, and for 15$\%$ PDI, $M_{0}=6$. Note that the $T_{onset}$ values for different PDIs (Fig. \ref{T_onset_diff_M_fig}) also show saturation at similar values of $M$. Therefore the results are consistent. The results obtained also agree with our previous study, where we showed that parameters that need structural input are better determined when the system is described in terms of multiple species,\cite{palak_polydisperse} and the optimum number of species increases with polydispersity. Therefore we can say that the structural order parameter of a system should be calculated by describing the system in terms of $M_{0}$ species. This structural order parameter will provide a true description of the local caging potential and will correlate with the dynamics.

\begin{figure*}
  \centering
  \subfigure{\includegraphics[width=.285\textwidth]{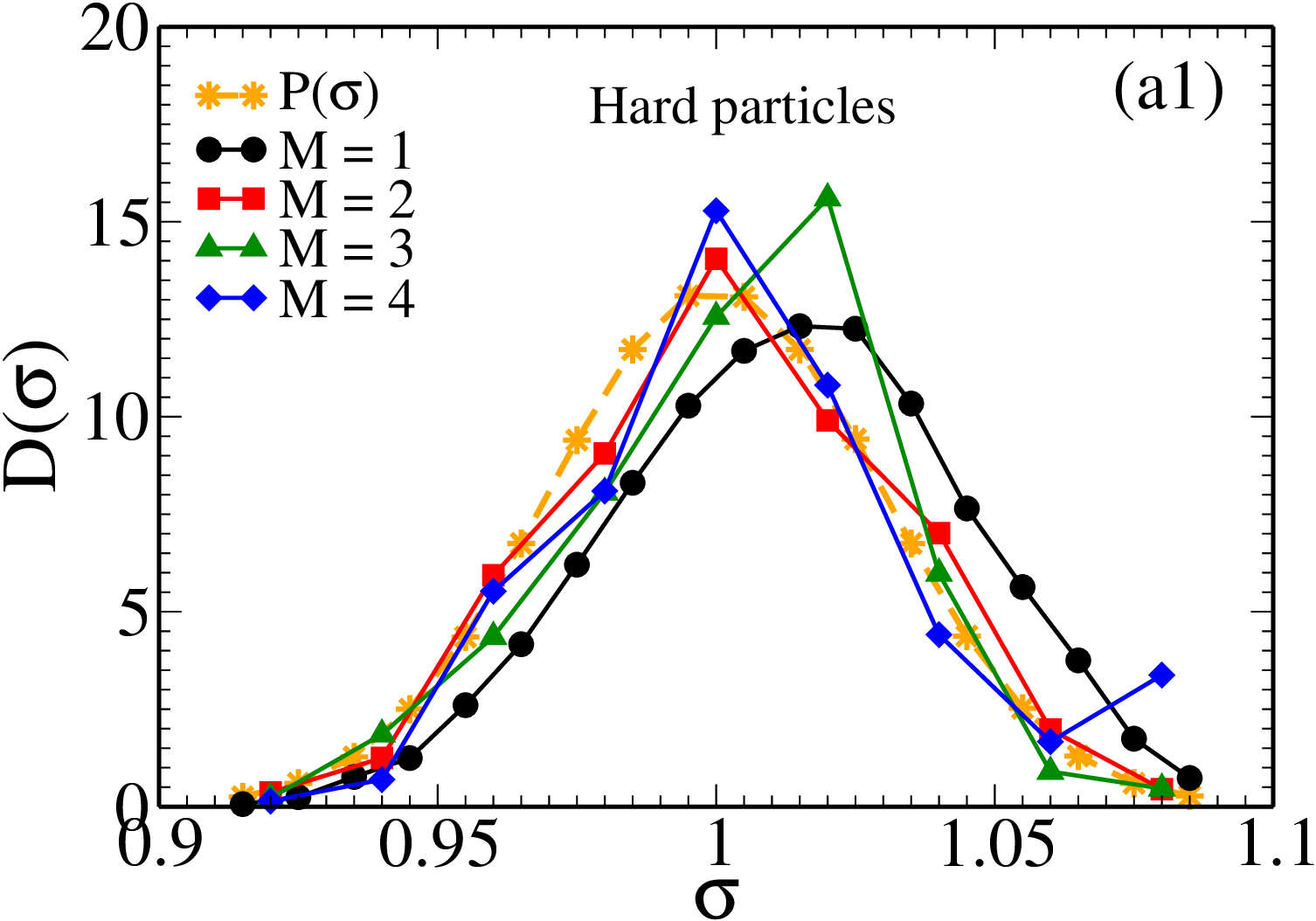}}
   \hspace{0.2cm}
         \vspace{0.1cm}
    \subfigure{\includegraphics[width=.27\textwidth]{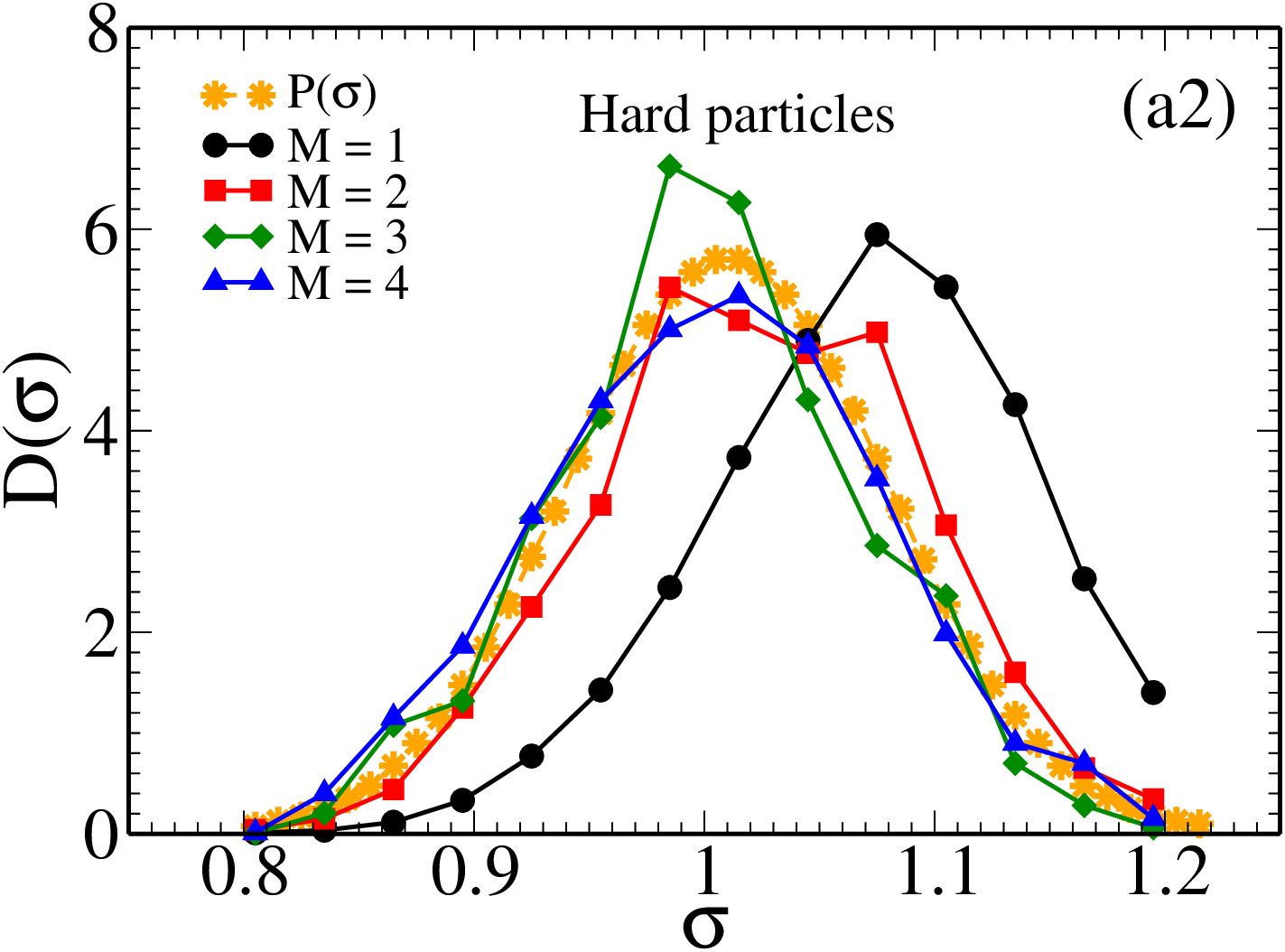}}
     \hspace{0.2cm}
         \vspace{0.1cm}
    \subfigure{\includegraphics[width=.275\textwidth]{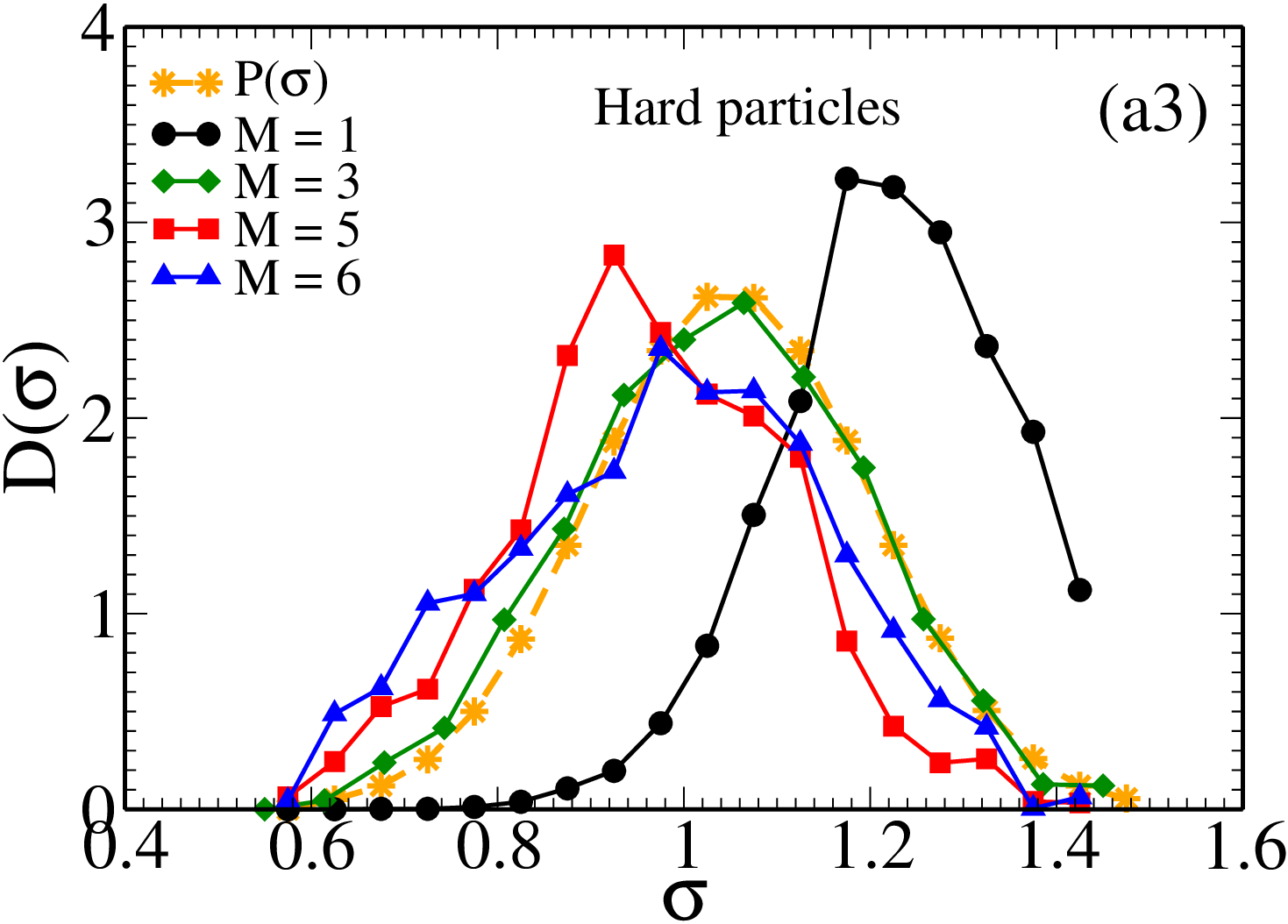}}
     \hspace{0.2cm}
         \vspace{0.1cm}
  \subfigure{\includegraphics[width=.285\textwidth]{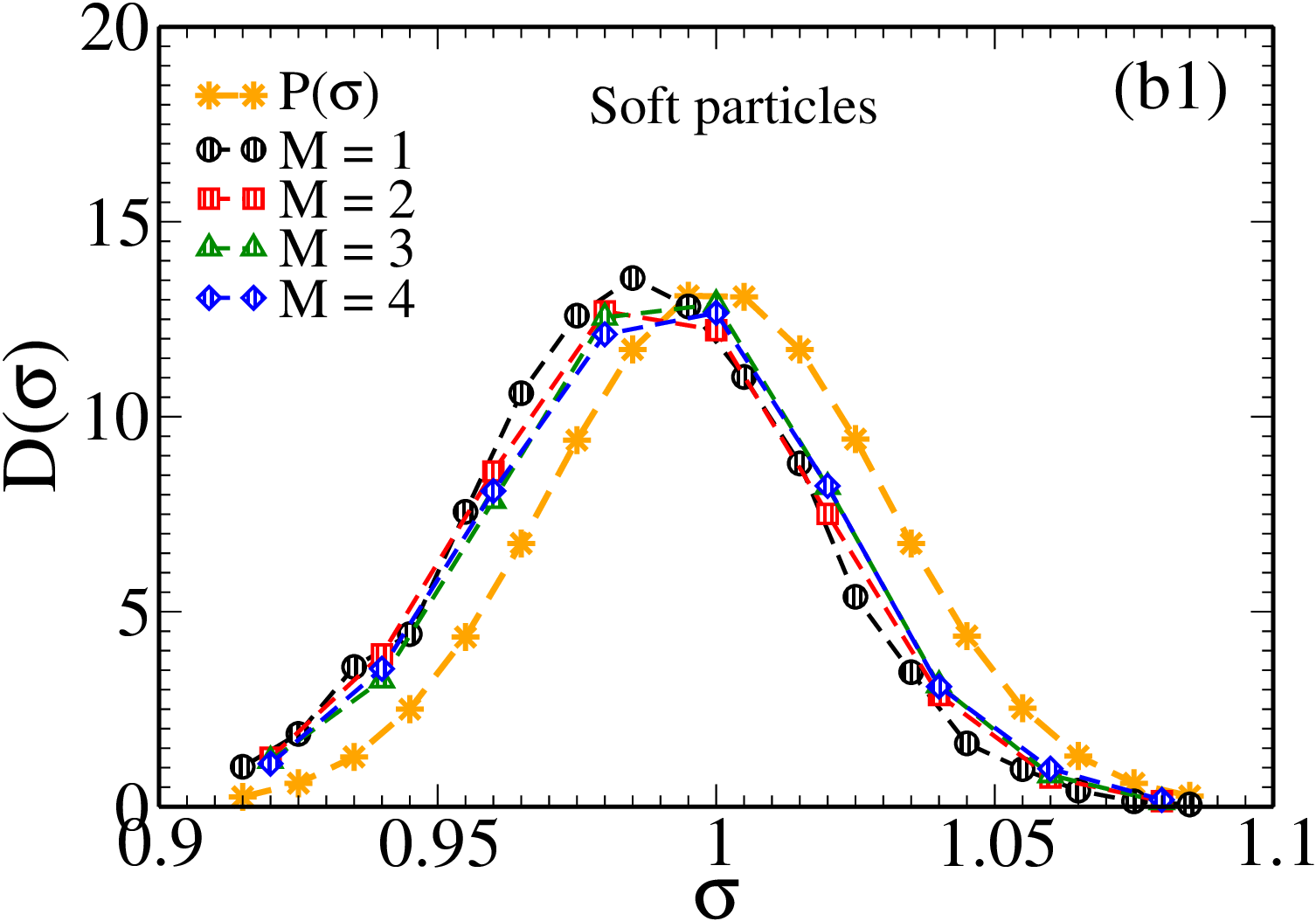}}
  \hspace{0.2cm}
  \subfigure{\includegraphics[width=.27\textwidth]{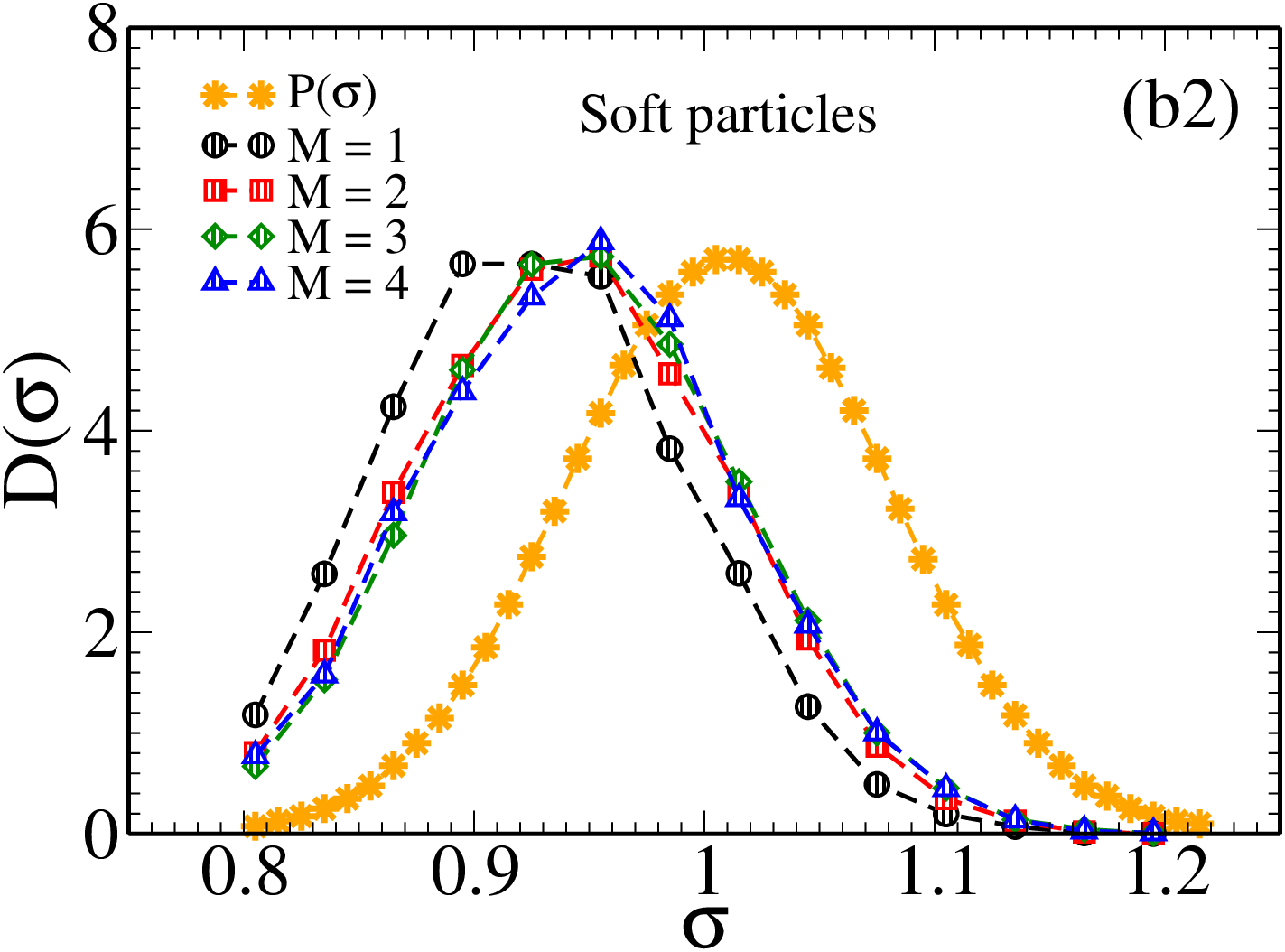}}
   \hspace{0.2cm}
   \subfigure{\includegraphics[width=.27\textwidth]{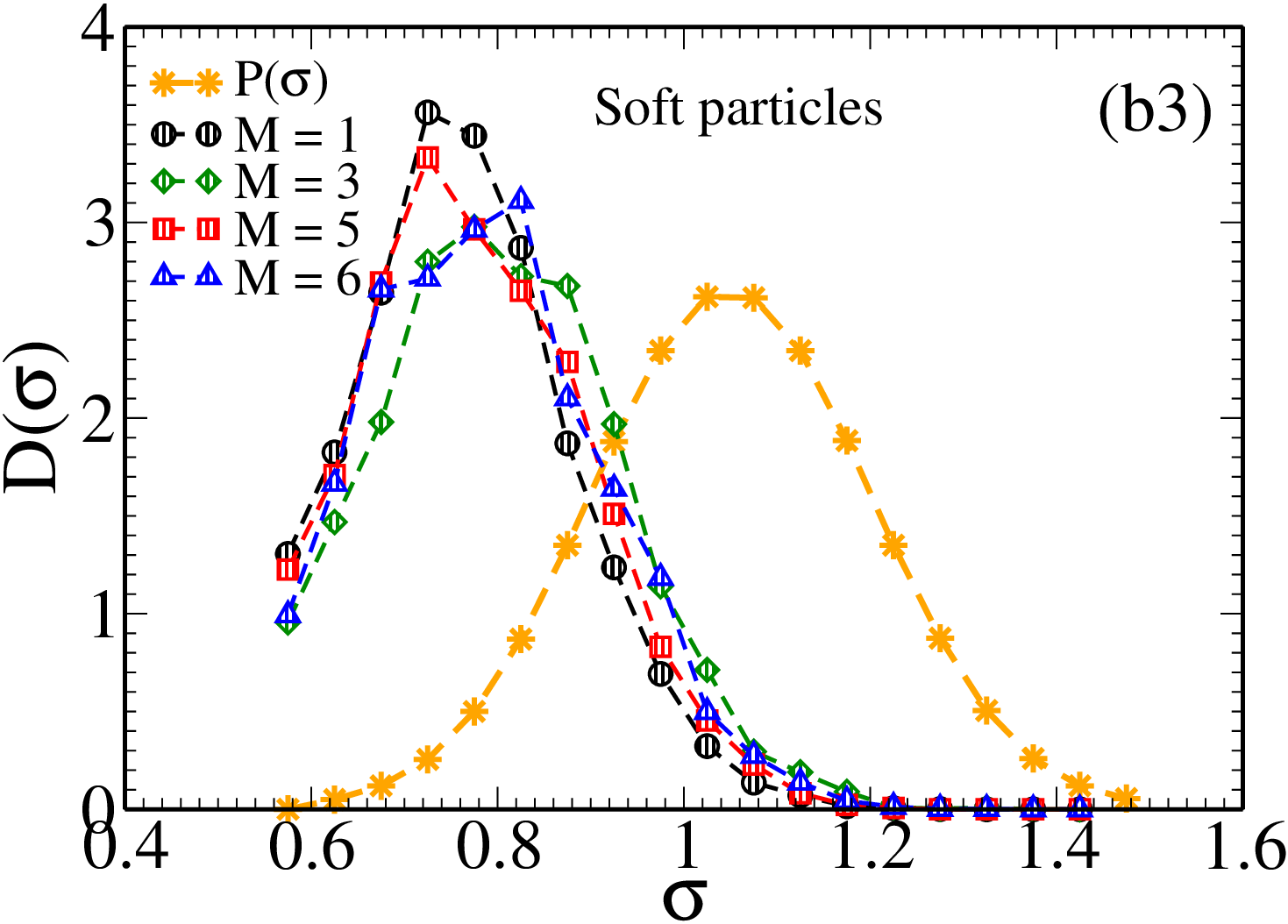}}
   \caption{\emph{ Particle size distribution, $D(\sigma)$, of the hardest and softest particles as a function of $M$ (as defined in Fig. \ref{overlap_hard_soft_region_at_same_t_fig}).Top panel:  (a1)-(a3) Size distribution of all hard particles. Bottom panel:(b1)-(b3) Size distribution of all soft particles. (a1) and (b1) are for 3\% PDI, (a2) and (b2) are for 7\% PDI, and (a3) and (b3) are for 15\% PDI. For comparison, we also plot the size distribution of all the particles, $P(\sigma)$.}}
    \label{sigma_hard_soft_diff_PDI_fig}
\end{figure*}

However, although the structure of a system is not well described for $M=1$, the difference in dynamics between the hardest and softest particles is best determined when we treat the system as monodisperse. This is a contradictory result, and it appears that in these systems, apart from the structure, there can be other parameters that drive the dynamic heterogeneity. To understand the result, in Fig. \ref{sigma_hard_soft_diff_PDI_fig}, we plot the distribution of the particle diameters of the hardest and the softest particles for different values of $M$ for all three systems. We also plot the particle size distribution of the whole system, $P(\sigma)$. When $M=1$, we find that the distribution of the hardest and softest particles is skewed toward the bigger and smaller-sized particles, respectively. This effect is more prominent at higher polydispersity. With an increase in $M$, the distribution of the hardest particles moves toward $P(\sigma)$.  This clearly shows that as we divide particles into species, the cage around smaller particles, which, for $M=1$, is loosely defined, gets better defined at higher $M$. This leads to an increase in the depth of the caging potential and, thus, a decrease in the softness of the potential. The distribution of the diameter of the softer particles also shows some change with $M$, but differently form the hard particles, it always remains skewed toward smaller particles which is similar to that observed for granular systems.\cite{Farhang_PRE} This implies that the cage around the bigger particles is mostly well-defined, and this effect is again more pronounced at higher polydispersity.

\begin{figure}
  \centering
  \subfigure{\includegraphics[width=.4\textwidth]{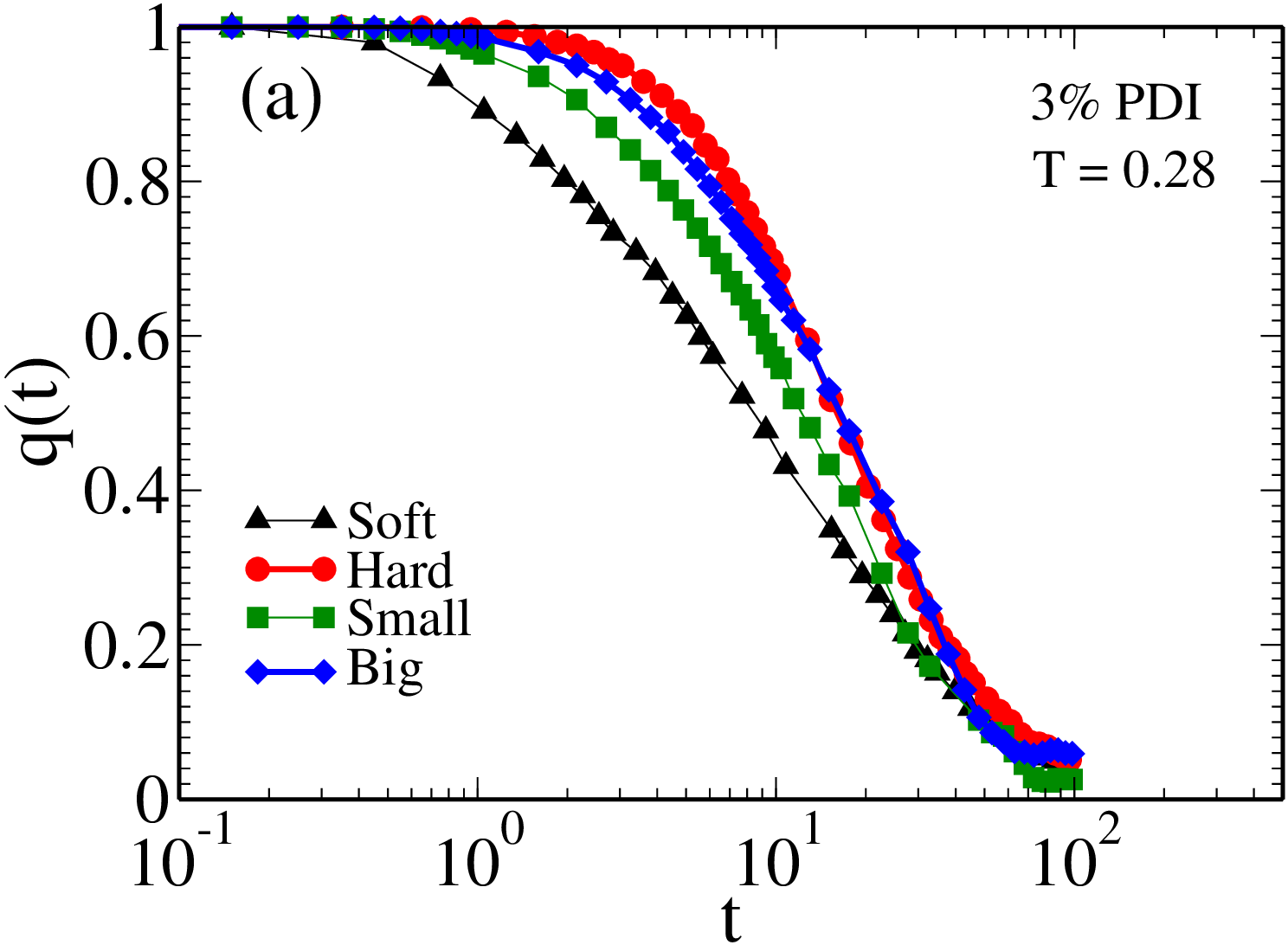}} \\
 \vspace{0.2cm}
  \subfigure{\includegraphics[width=.4\textwidth]{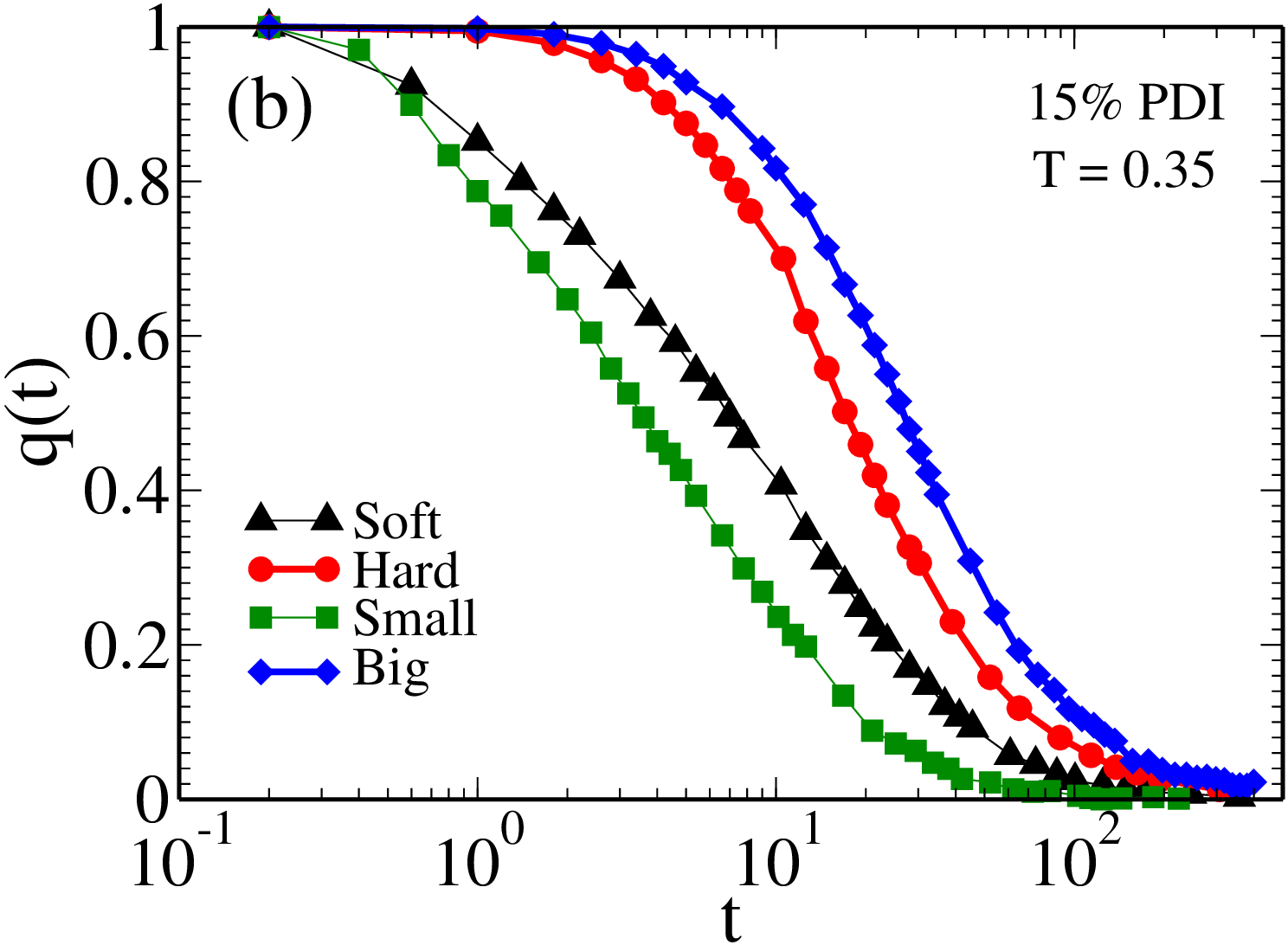}}
    \caption{\emph{Overlap function of 2 hardest particles (red circle), 2 softest particles (black triangle), 2 biggest particles in size (blue diamond), and 2 smallest particles in size (green square) at different PDIs. The structural order parameter is calculated for M=1 (a) 3\% PDI and (b) 15\% PDI.}}
    \label{overlap_hard_soft_region_with_big_n_small_particle_fig}
        \end{figure}

\begin{figure}[h!]
  \centering
      \subfigure{\includegraphics[width=.4\textwidth]{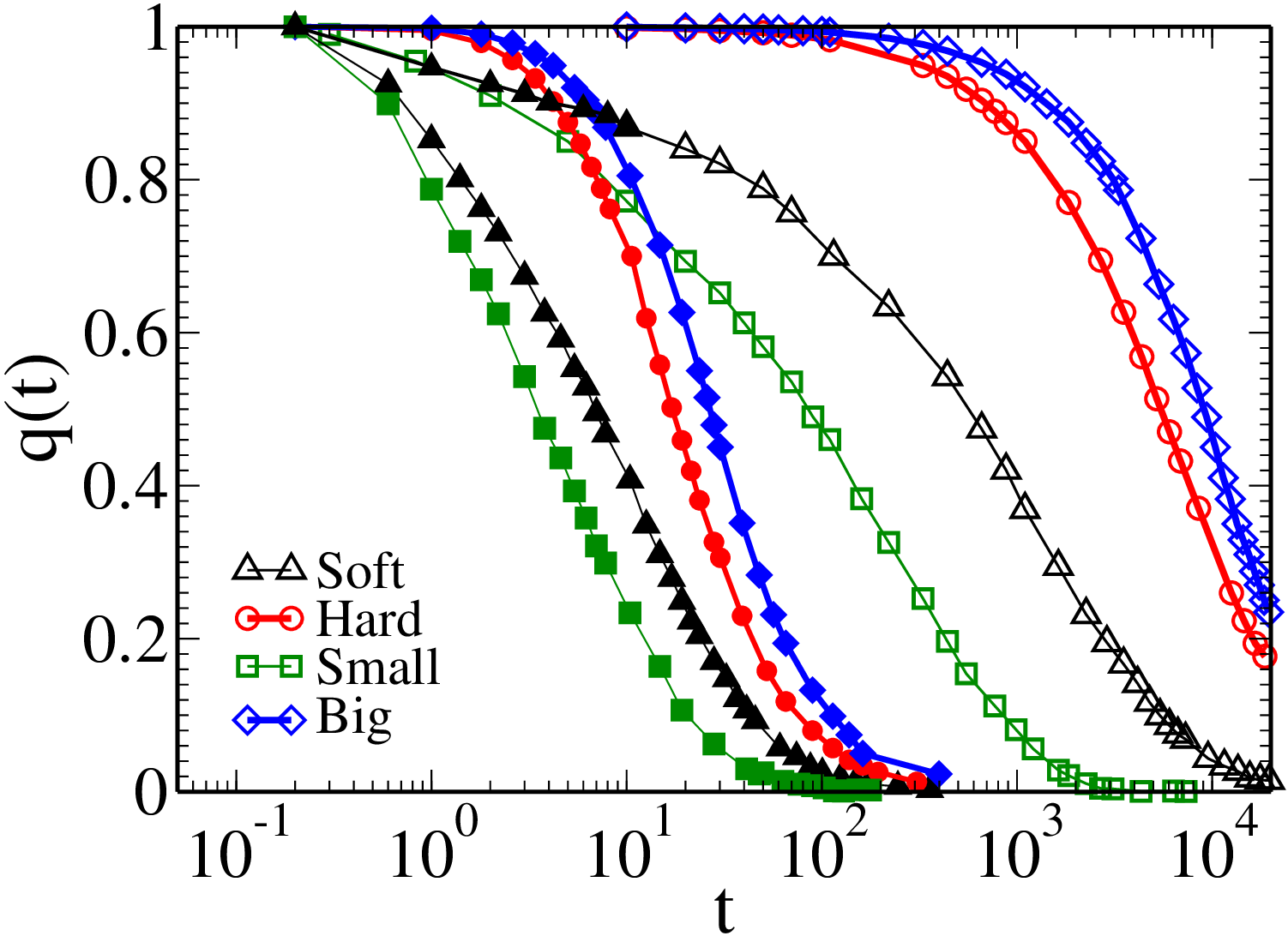}}
    \caption{\emph{Overlap function for 15\% PDI at T = 0.35 (closed symbol), and T = 0.22 (open symbol). Color codes are similar like Fig. \ref{overlap_hard_soft_region_with_big_n_small_particle_fig}.}}
    \label{overlap_PDI_15_diff_T_fig}
        \end{figure}

Notice that the shift in the size distribution of the hardest/softest particles (Fig. \ref{sigma_hard_soft_diff_PDI_fig}) with $M$ is also accompanied by a shift in the overlap function of the hardest/softest particles with $M$ (Fig. \ref{overlap_hard_soft_region_at_same_t_fig}), suggesting that these shifts are correlated. In both cases (particle size distribution and overlap), the shift is more for the harder particles and also increases with polydispersity. This implies that size also plays a role in the dynamics. In Fig. \ref{overlap_hard_soft_region_with_big_n_small_particle_fig}, we plot the dynamics of the two biggest and two smallest particles and compare them with the two hardest and softest particles for M=1. We find that for the 3$\%$ PDI, the difference in dynamics of the biggest and smallest particles is less than that of the softest and hardest particle. This implies that the heterogeneity in the dynamics is primarily determined by the local structural heterogeneity. With an increase in polydispersity, the scenario reverses. For the 15$\%$ PDI system, the difference in dynamics is better described by the size than the local structural order parameter. We know from our earlier study \cite{mohit_PRE} that at lower temperatures, the structure becomes a better predictor of the dynamics. To understand the role of temperature, we choose the 15$\%$ PDI system, where size appears to be dominant, and plot the different overlap functions at two different temperatures (Fig. \ref{overlap_PDI_15_diff_T_fig}). We find that at lower temperatures, although the structure becomes a better predictor of the dynamics, the size still plays a dominant role in determining the dynamics.

\section{COMPARATIVE STUDY OF THE ROLE OF PARTICLE SIZE AND LOCAL STRUCTURE ON THE DYNAMICS}
\label{COMPARATIVE_STUDY_OF_THE_ROLE_OF_PARTICLE_SIZE_AND_LOCAL_STRUCTURE_DYNAMICS}
\begin{figure*}
  \centering
\subfigure{\includegraphics[width=.32\textwidth]{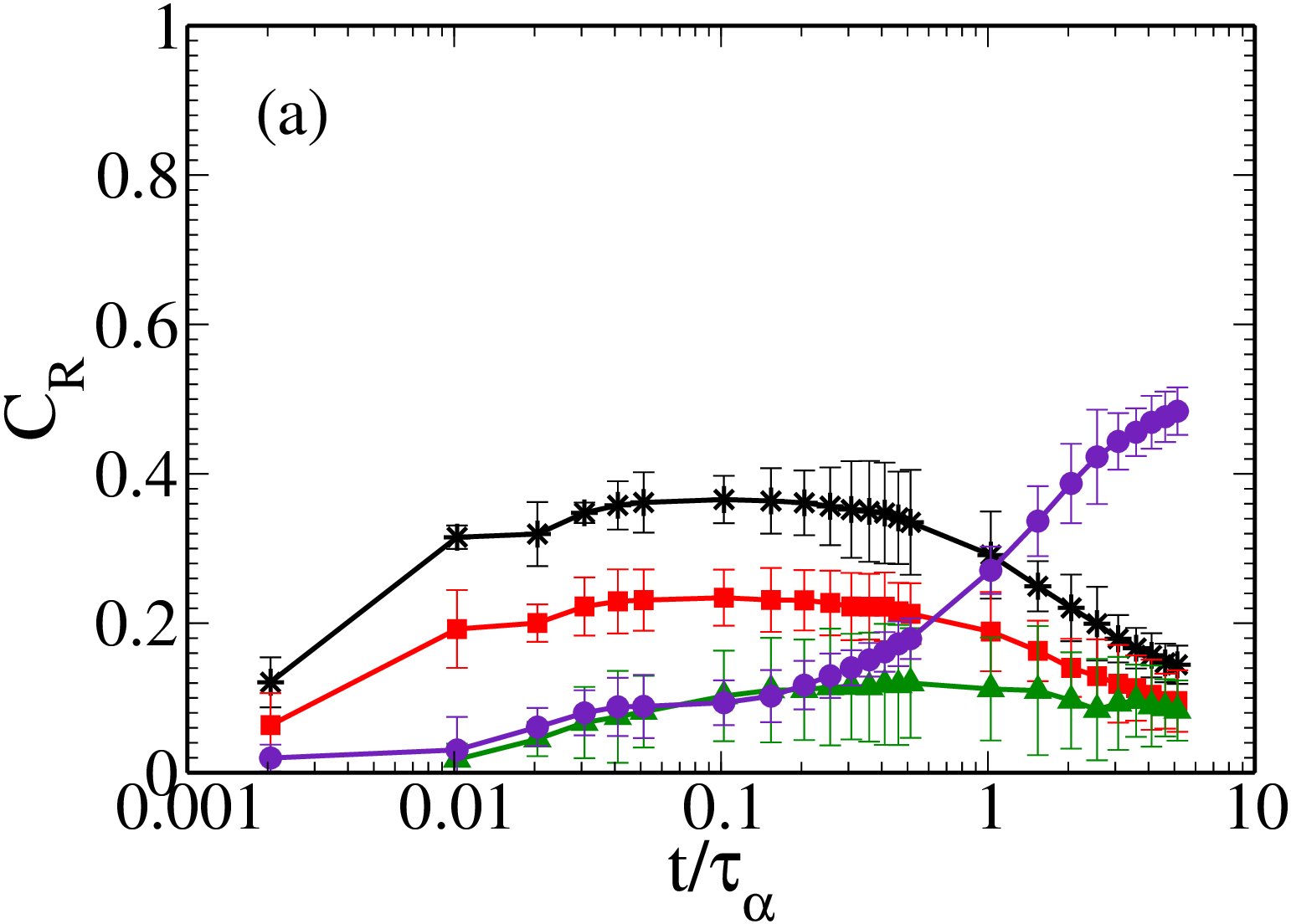}}
\subfigure{\includegraphics[width=.32\textwidth]{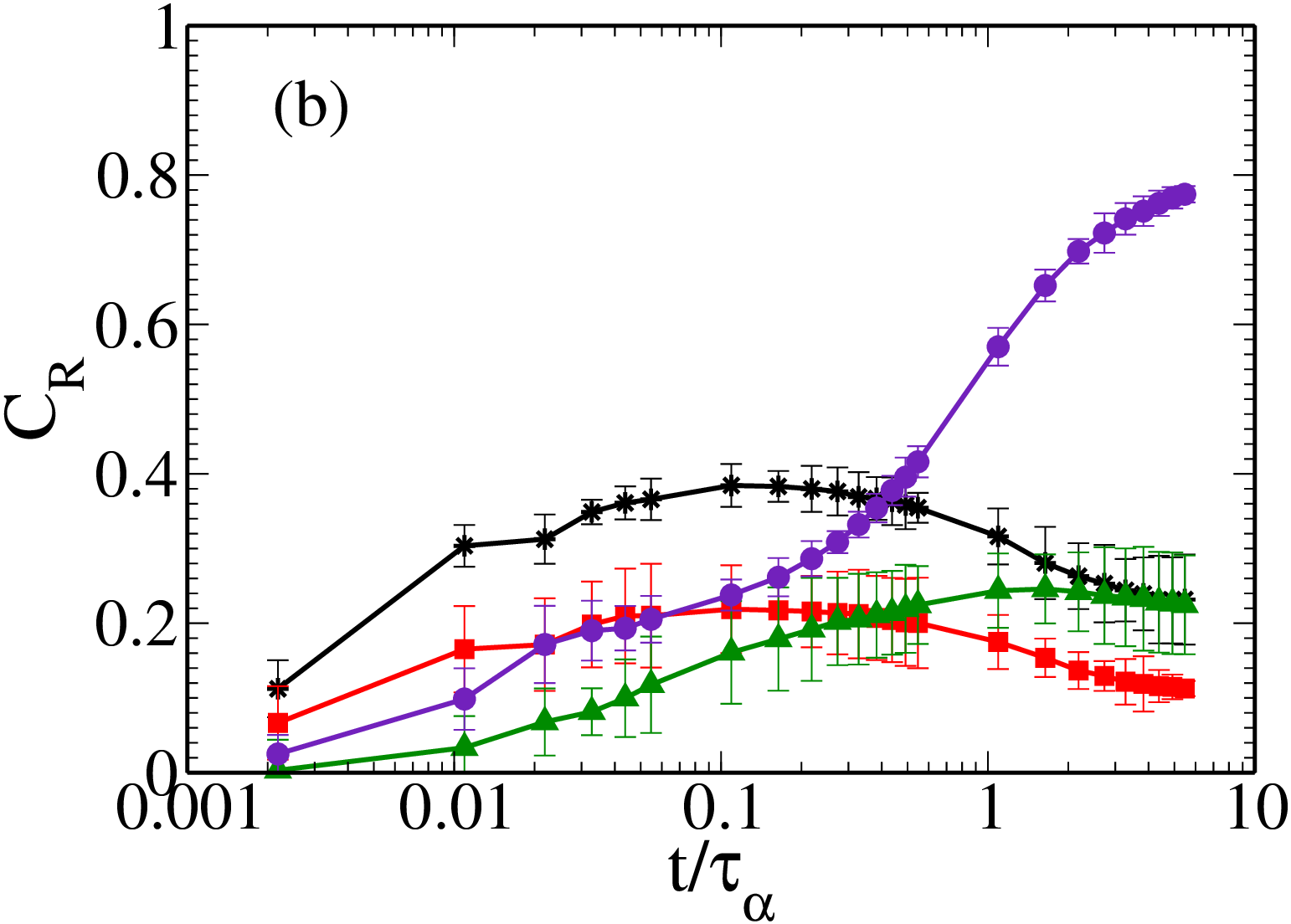}}
\subfigure{\includegraphics[width=.305\textwidth]{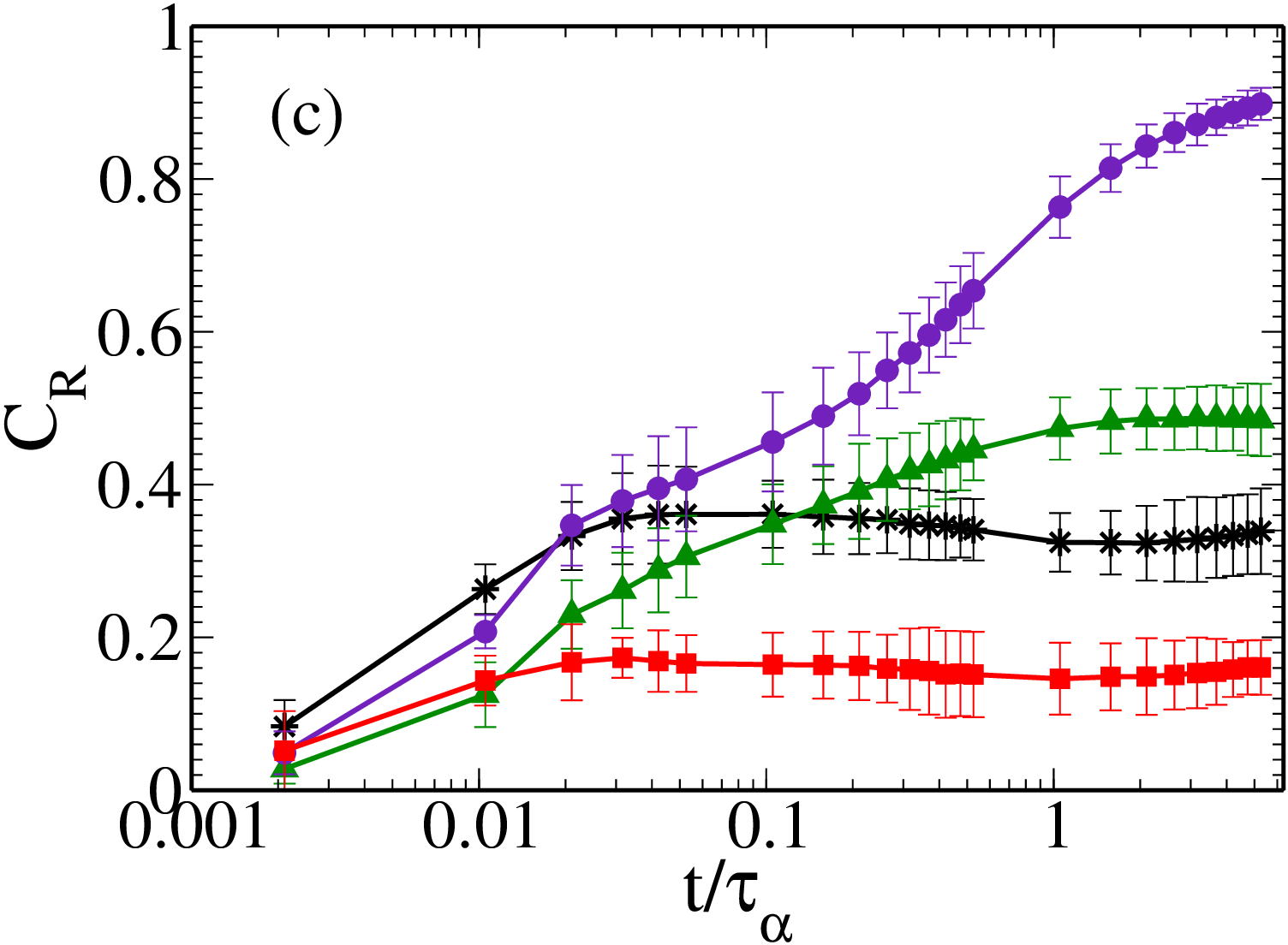}}
    \caption{\emph{Spearman rank correlation $C_R$ between mobility and different order parameters at different PDIs: (a) 3\% PDI; (b) 7\% PDI; and (c) 15\% PDI. (Color code: $C_R (\mu, 1/\beta\Phi_r)$ M = 1 (black star), M = $M_0$ (red square), $C_R (\mu, \Psi)$ (green triangle), and $C_R (\mu, \sigma)$ (violet circle). $M_0$ = 3, 4, and 6 for 3\%, 7\%, and  15\% PDI, respectively.}}
    \label{scattered_plot_diff_PDI_M_M0_fig}
\end{figure*}

The above-mentioned analysis suggests that for polydisperse systems, both size and local structure can play a role in the dynamics. To quantify the dependence of the dynamics on the structure and particle size, we perform isoconfigurational runs (IC). IC is a powerful technique introduced by Harrowell and co-workers to investigate the role of structure in the dynamical heterogeneity of the particles.\cite{cooper_IC_2004,Cooper_IC_2006,Harowell,Bertheir_IC} Among other factors, a particle's displacement can depend on its structure and also its initial momenta. This technique was proposed to remove the uninteresting variation in the particle displacements arising from the choice of initial momenta and extract the role of the initial configuration on the dynamics and its heterogeneity. For each system, five different isoconfigurational runs are carried out for 4000 particles. To ensure that all configurations are different, the configurations are chosen such that the two sets are greater than 100$\tau_{\alpha}$ apart. We run 100 trajectories for each configuration with different starting velocities randomly assigned from the Maxwell-Boltzmann distribution for the corresponding temperatures. Mobility, $\mu$ is the average displacement of each particle over these 100 runs and is calculated as, $ \mu^{j}(t) = \frac{1}{N_{IC}}\sum_{i=1}^{N_{IC}} \sqrt{(r_{i}^{j}(t) - r_{i}^{j}(0))^{2}}$. Here, $\mu^{j}(t)$ is the mobility of the $j^{th}$ particle at time $t$, and $N_{IC}$ is the number of trajectories. 
We calculate the Spearman rank correlation $(C_R)$ between different parameters as a function of time (scaled by the $\alpha$ relaxation time $\tau_{\alpha}$). We plot $C_R(\mu,1/\beta\Phi_r)$ against time for $M=1$ and $M=M_{0}$. We find that $C_R(\mu,1/\beta\Phi_r)$ decreases with an increase in $M$. This result is similar in spirit to that observed for the difference in the overlap functions of the hardest and softest particles (Fig. \ref{overlap_hard_soft_region_at_same_t_fig}).  In Fig. \ref{scattered_plot_diff_PDI_M_M0_fig}, we also plot $C_R(\mu,\sigma)$. We find that for all systems, it grows at longer times, and for systems with higher polydispersity, the correlation is large, even at shorter times. This supports our earlier conclusion that at higher PDI, the size of the particles plays a greater role in describing the dynamic heterogeneity.  

Note that apart from the softness parameter described in this work, other parameters are often used to describe the local static property of a supercooled liquid \cite{Richard_PRM}. 
We check if size plays any role in an order parameter that does not include the radial distribution function. Earlier studies have shown that Vibrality, the local Debye-Waller Facto, \cite{Richard_PRM,Paddy_PRL,Harowell} is a good predictor of dynamics. The analysis is performed on the inherent structure. The Fast Inertial Relaxation Engine (FIRE) algorithm is employed to obtain the inherent structures (IS).\cite{FIRE_method_for_inherent_str} Vibrality is written as, $\Psi (i) =  {\sum_{k=1}^{3N-3}}\frac{{|{\bf{v}_{k}}^{i}|}^{2}}{{\omega_{k}}^2} $, where the sum runs over the entire set of eigenmodes with frequency $\omega_k$. ${\bf{v}_{k}}^{i}$ is a vector that has the three components of the eigenvector ${\overrightarrow{v_{k}}}$ associated with the $i^{th}$ particle. $\Psi $ (i) is the mean square vibrational amplitude of the $i^{th}$ particle, assuming the vibrational energy is equally distributed to all modes. In Fig. \ref{scattered_plot_diff_PDI_M_M0_fig}, we plot $C_R(\mu,\Psi)$ and find that it increases with polydispersity which is similar to $C_{R}(\mu,\sigma)$. It appears that $C_R(\mu,\sigma)$ affects $C_R(\mu,\Psi)$ more compared to $C_R(\mu,1/\beta\Phi_r)$.

To quantify the above-mentioned observations, we now use multiple linear regressions to model mobility in terms of $\Phi$ and  $\sigma$. To evaluate the predictive power of the model, we use the standard 5-fold cross-validation approach, where the data are randomly split into five equal sets, and a model built on four parts is used to predict mobility on the held-out test set. This is performed five times, with each data point tested exactly once. The mean relative error, $MRE =\frac{1}{N}\sum_{j=1}^{N}\frac{|{\mu^{j}}_{p}-{\mu^{j}}_{t}|}{{\mu^{j}}_{t}}$, and the root mean square deviation, $RMSD = \sqrt{\frac{1}{N}{\sum_{j=1}^{N}}({\mu^{j}}_{p}-{\mu^{j}}_{t})^{2}}$, are shown along with the error bar computed from the five test sets. Here, ${\mu^{j}}_p$ and ${\mu^{j}}_t$ are the predicted mobility and true mobility of the $j^{th}$ particle, respectively. The mobility used here is calculated at $t=5$, but the results are independent of $t$. 

\begin{figure}[h!]
  \centering
\subfigure{\includegraphics[width=.4\textwidth]{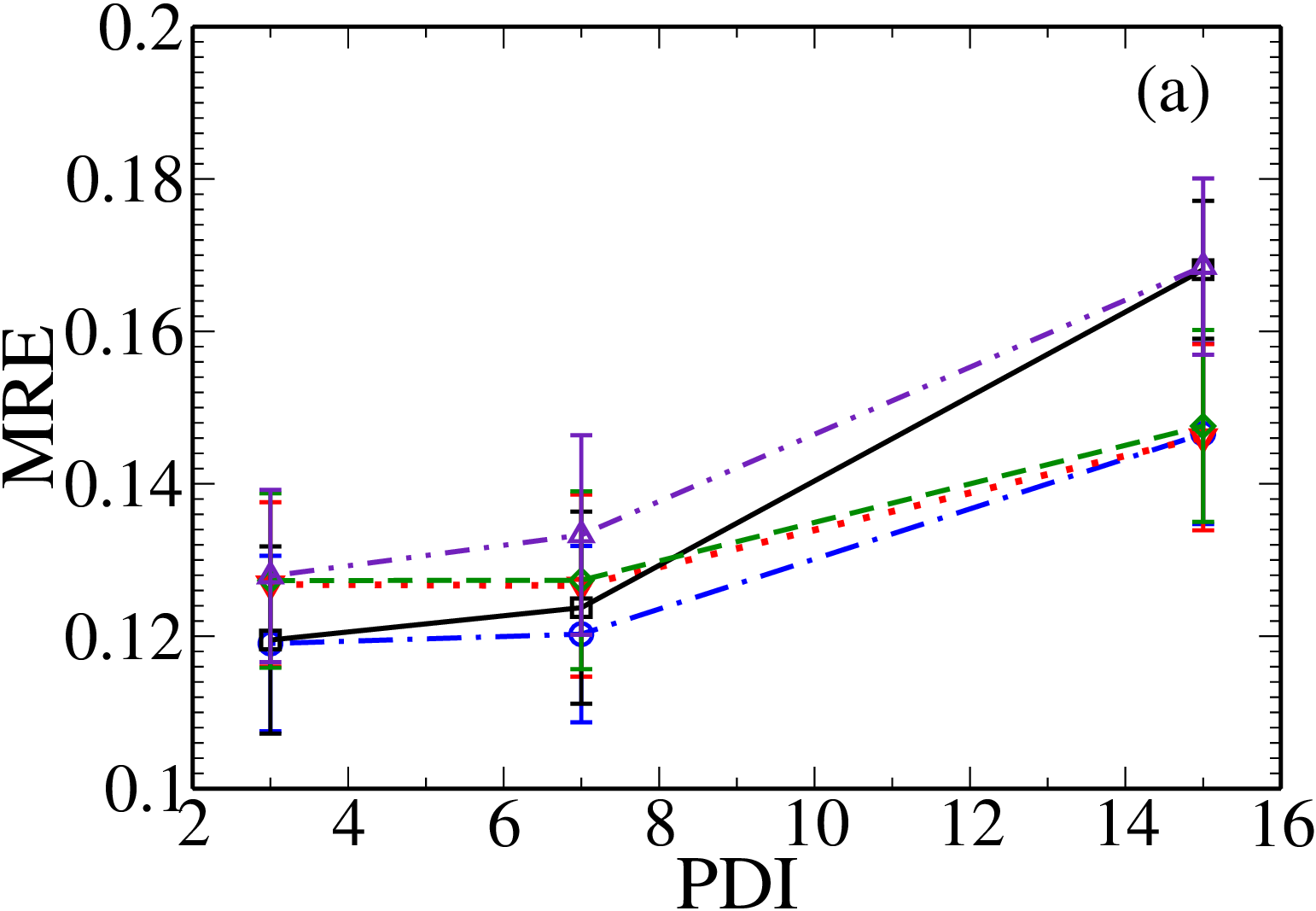}}\\
\vspace{0.2cm}
\subfigure{\includegraphics[width=.4\textwidth]{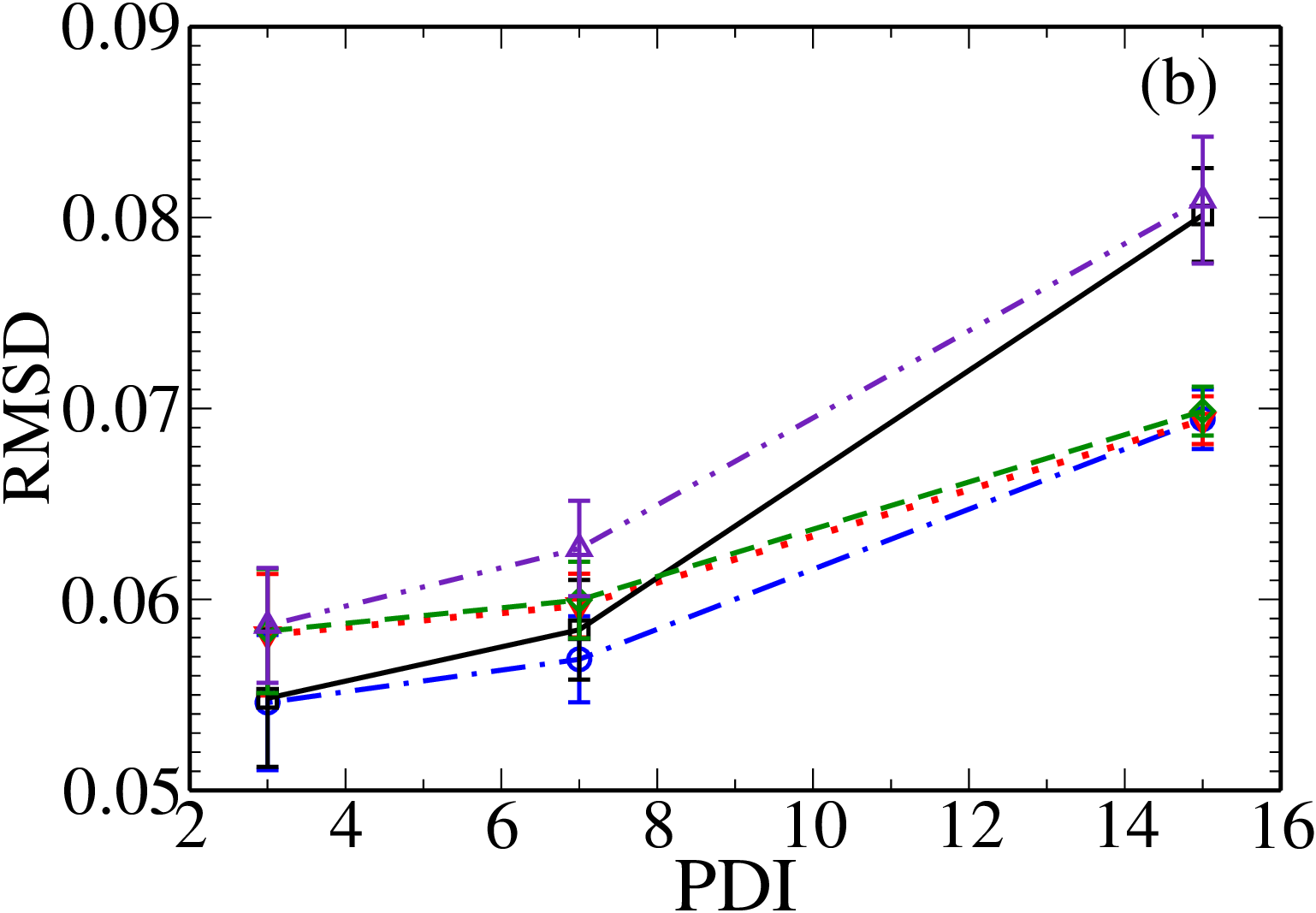}}
    \caption{\emph{Error between predicted and true values of mobility $\mu$. (a) Mean relative error, $MRE =\frac{1}{N} \sum_{j=1}^{N}\frac{|{\mu^{j}}_{p}-{\mu^{j}}_{t}|}{{\mu^{j}}_{t}}$, and (b) root mean square deviation, $RMSD = \sqrt{\frac{1}{N}{\sum_{j=1}^{N}}({\mu^{j}}_{p}-{\mu^{j}}_{t})^{2}}$. Color code: Independent variables are $\Phi_{r}$ and $\sigma$ (blue, dashed-dotted line), independent variable is $\Phi_{r}$ (black, solid line), independent variable is $\sigma$ (green, dashed line), independent variables are $\Psi$ and $\sigma$ (red, dotted line) and independent variable is $\Psi$ (indigo, dashed-double dot line).}}
    \label{RE_mobility_fig}
\end{figure}

We compare results with simple linear regression, also evaluated in the same manner, but using only one of the parameters i.e., either $\Phi$ or $\sigma$. From the analysis of the errors plotted in Fig. \ref{RE_mobility_fig}, we find that for lower PDI, the caging potential is a good parameter to describe mobility. However, with an increase in PDI, size becomes the dominant variable in prediction. We also do a similar analysis using $\Psi$ and $\sigma$ and find that between this pair, size always plays a dominant role for all systems. For smaller PDIs where size does not play a strong role, it appears that among the three variables, SOP is the best predictor of the dynamics. 

Note that in the above-mentioned analysis, although we have treated $\Phi$ and $\Psi$ as independent variables, both have some dependence on the size. The dependence of $\Phi$ on the size can be seen in Fig. \ref{sigma_hard_soft_diff_PDI_fig}, where we find that soft particles are primarily small and hard particles are primarily big in size. The figure also suggests that this dependence increases with PDI and decreases as we increase M. Note that in the figure, we have taken only the hardest and softest particles. To quantify this observation for all particles, in Table \ref{SC_correlation value_table},  we report the Spearman rank correlations between the different parameters, and the correlation values do support the inference drawn from Fig. \ref{sigma_hard_soft_diff_PDI_fig}. For all the systems, the correlation between the local structure and size is more for M=1. Now since dynamics is also correlated with the size of the particles, the local structure appears to be better correlated with dynamics for M=1. This effect increases with polydispersity. We also find that at higher polydispersity, compared to $\Phi$, $\Psi$ is more correlated with $\sigma$. Therefore at higher polydispersity, and longer times, the Vibrality appears to be a better predictor of the dynamics, as seen in Fig. \ref{scattered_plot_diff_PDI_M_M0_fig}. Therefore for systems with a large PDI, any order parameter that is correlated with the size of the particles will appear to be a good predictor of the dynamics.

\begin{table}[h!]
\caption{\emph{Spearman rank correlation, $C_{R}$ between particle size, $\sigma$  and local caging potential, $\beta \Phi_{r}$ when the system is assumed to be monodisperse, $M=1$, and when the system is described in terms of the optimum number of species, $M_{o}$. We also report the Spearman correlation between $\sigma$ and Vibrality, $\Psi$. The systems are polydisperse with a polydispersity index of $3\%$ at T = 0.28, $7\%$ at T = 0.30, and $15\%$ at T = 0.35.}}
\begin{tabular}{|c|c|c|c|}
    \hline
PDI &  $C_R(\sigma, \beta\Phi_r)$ M = 1 & $C_R(\sigma, \beta\Phi_r)$ M = $M_0$ &  $C_R(\sigma,\Psi)$ \\
\cline{1-4}
3\%  & 0.139 & 0.082 & -0.111   \\
\cline{1-4}
7\%  &  0.293 & 0.129 &  -0.253    \\
\cline{1-4}
15\% & 0.464 & 0.195 & -0.516   \\
\cline{1-4}
\end{tabular}
\label{SC_correlation value_table}
\end{table}

\section{CONCLUSION}
\label{CONCLUSION}
In a recent study, we proposed a new structural order parameter that strongly correlates with dynamics.\cite{mohit_PRE} This SOP is the inverse of the depth of the local mean-field caging potential, described in terms of the local liquid structure. We further showed that this correlation between the SOP and dynamics is valid below the onset temperature of the glassy dynamics. Therefore the validity of the theory can be used to determine the onset of glassy dynamics. Since polydisperse systems are good model systems to study supercooled liquid dynamics, in this work, we study the structural order parameter and its correlation with the dynamics of a few polydisperse systems. Note that this SOP needs information on the local structure. It is well known that describing the structure of a polydisperse system is tricky.\cite{main_paper_Truskett,palak_polydisperse} Treating the system as a monodisperse system leads to artificial softening of the structure. In an earlier study, we had shown that for a polydisperse system, the correct structural description is obtained only when the system is expressed in terms of multiple species, $M$.\cite{palak_polydisperse} Here we first show that our earlier method also leads to an increase in mutual information, thus validating the method further. 

We find that the distribution of the particle level SOP, changes with $M$. We also find that this change does not affect all particles similarly. Therefore if we rank particles in terms of the value of the order parameter, then the rank order changes and finally appears to saturate beyond a certain $M$ value.  We also find that the detection of the onset temperature from the correlation of the SOP and the dynamics depends on $M$. The onset temperature first changes with $M$, and at higher values of $M$, it saturates. The saturation of the onset temperature and the rank of the particle order parameter allow us to estimate the optimum number of species needed to describe the system. Like in our earlier study,\cite{palak_polydisperse} the value of $M_{0}$ increases with polydispersity. 

However, the most surprising result is that although the structure is not well defined for $M=1$, the correlation between the structure and dynamics is at its maximum when the system is assumed to be monodisperse. Furthermore, analysis using multiple linear regression shows that although at low polydispersity, the local SOP determines the dynamics, at higher polydispersity, the size of the particle plays a dominant role in the dynamics. We also find that for $M=1$, the bigger particles are primarily well-caged, and the smaller particles appear loosely caged. Therefore, there is a high correlation between the local SOP and the size of the particles. However, with an increase in $M$ and a better description of the structure, the cage is better defined, especially for smaller particles. This reduces the correlation between the SOP and the particle size. Since size plays a dominant role in determining the dynamics, this reduction in the correlation reduces the apparent predictive power of the SOP at higher $M$ values. To test if order parameter-size correlation is present for other order parameters where the local structural information is not needed, we calculate the Vibrality, which is the local Debye-Waller factor, known to be a good predictor of the dynamics.\cite{Richard_PRM,Paddy_PRL} We first show that Vibrality also correlates with size, and this correlation increases sharply with an increase in polydispersity. At lower polydispersity, compared to Vibrality, the SOP is a better predictor of the dynamics. However, at higher polydispersity, the Vibrality performs marginally better. This increase in the predictive power of the Vibrality is due to its stronger coupling with the size of the particle. 

Therefore, our study suggests that for a polydisperse system with a high PDI, any order parameter with a strong coupling with the particle size will appear to be a good predictor of the dynamics. However, this may not reflect the true predictive power of the order parameter. Therefore for a polydisperse system with reasonably high polydispersity, the correlation between dynamics and any static order parameter must be interpreted cautiously, as size can play a role in this correlation and the results may not be generic. 

In this paper, we have studied the structure-dynamics correlation at a single particle level which is an acceptable practice.\cite{Richard_PRM,Andrea_liu_nature,Andrea_liu_PRE,Harowell,cooper_IC_2004} However, the correlation between structure and dynamics is weak when we use single particle information.\cite{Bertheir_IC,tanka_PRX,Tanaka_PRL_2020,Tanaka_nature,Cooper_IC_2006} On the other hand, the correlation improves when we consider the collective dynamical property over a certain region \cite{Paddy_nature_MI,Bertheir_IC,Cooper_IC_2006} or correlate the coarse grained structural property with longtime dynamics \cite{Tanaka_PRL_2020,Tanaka_nature,tanka_PRX,coslovich_JCP_2020}. In a polydisperse system, this coarse graining of the SOP over a static length reduces the coupling between the order parameter and particle size. It can thus be a useful way to study the real correlation between the order parameter and the dynamics.\\

\begin{figure}[h!]
  \centering
  \subfigure{\includegraphics[width=.23\textwidth]{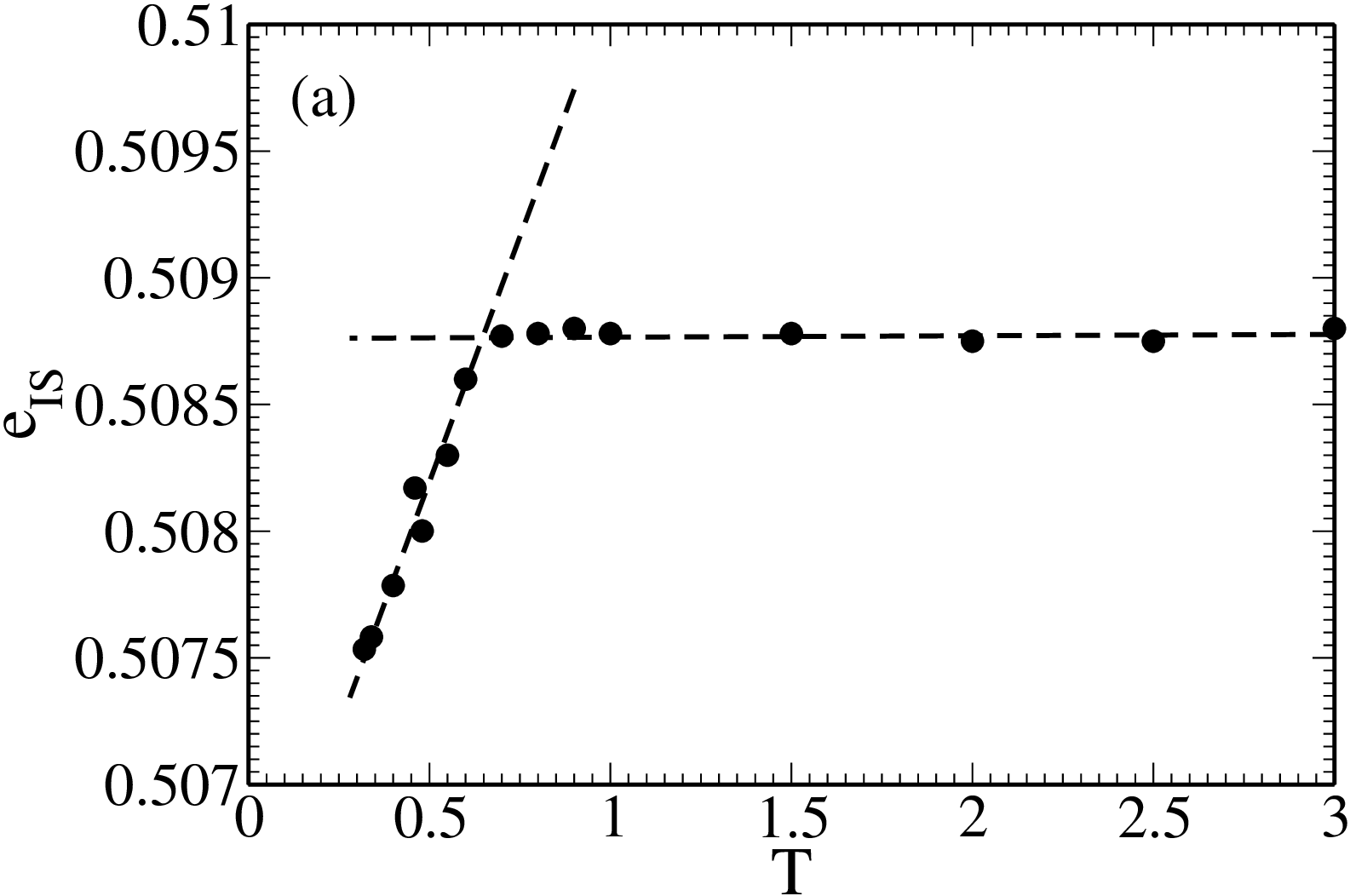}}
  \subfigure{\includegraphics[width=.23\textwidth]{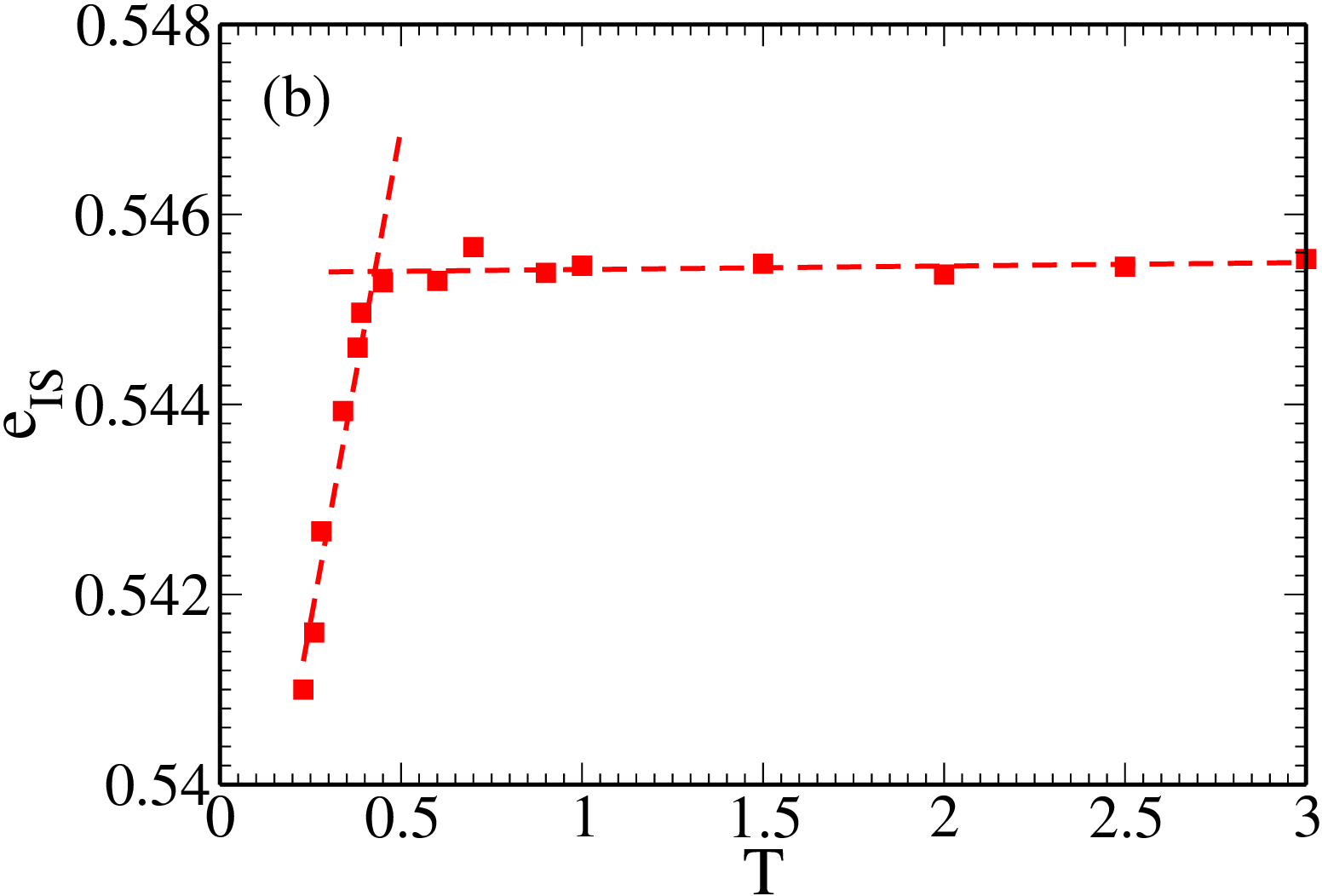}}
  \subfigure{\includegraphics[width=.23\textwidth]{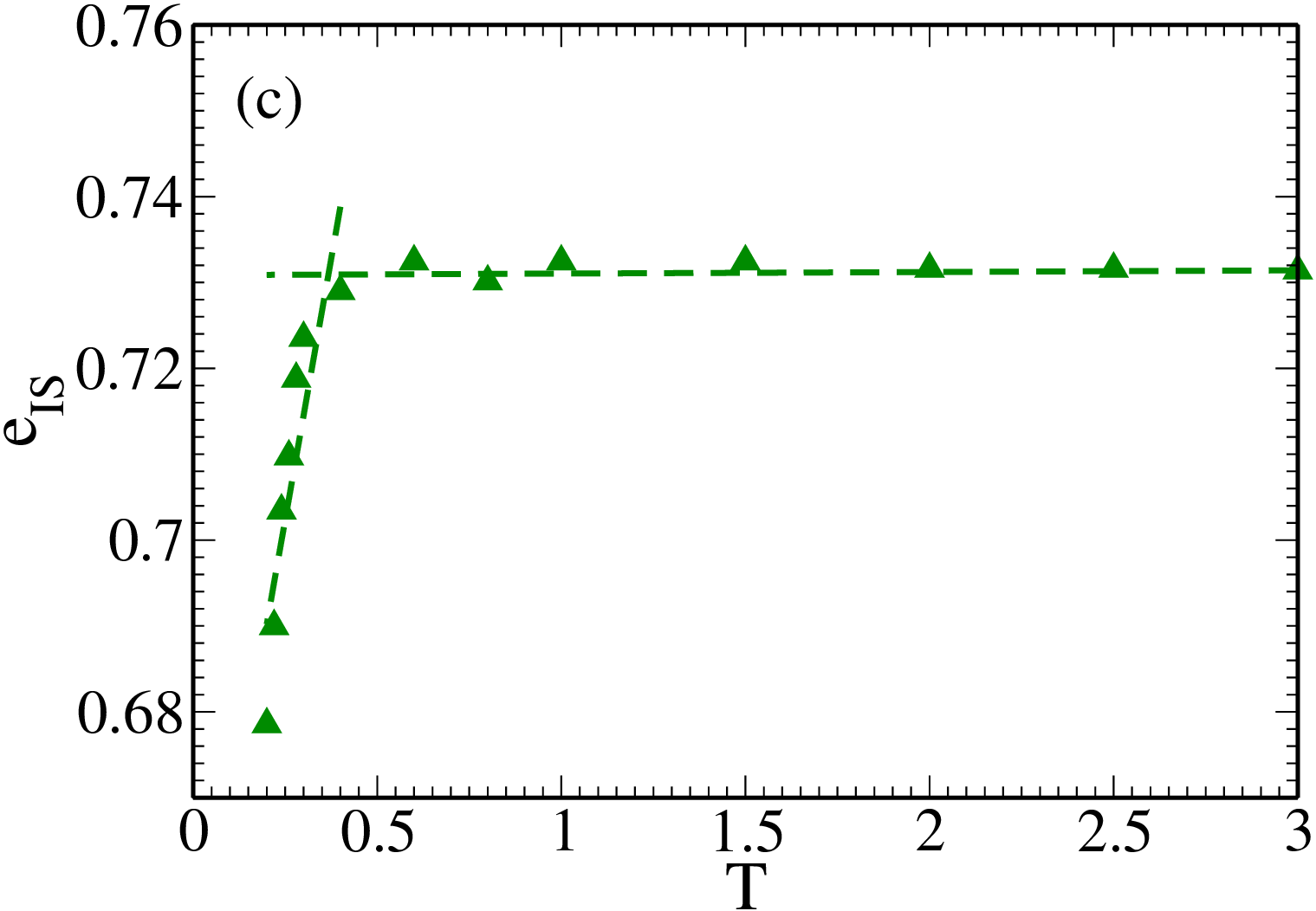}}
    \caption{\emph{Inherent structure energy ($e_{IS}$) as a function of temperature. (a) 3\% PDI; (b)7\% PDI; (c) 15\% PDI. Near the onset temperature, the value of IS starts deviating from its high temperature value thus allowing us to predict the onset.}}
    \label{inherent_structure}
\end{figure}

\begin{figure}[h!]
  \centering
  \subfigure{\includegraphics[width=.3\textwidth]{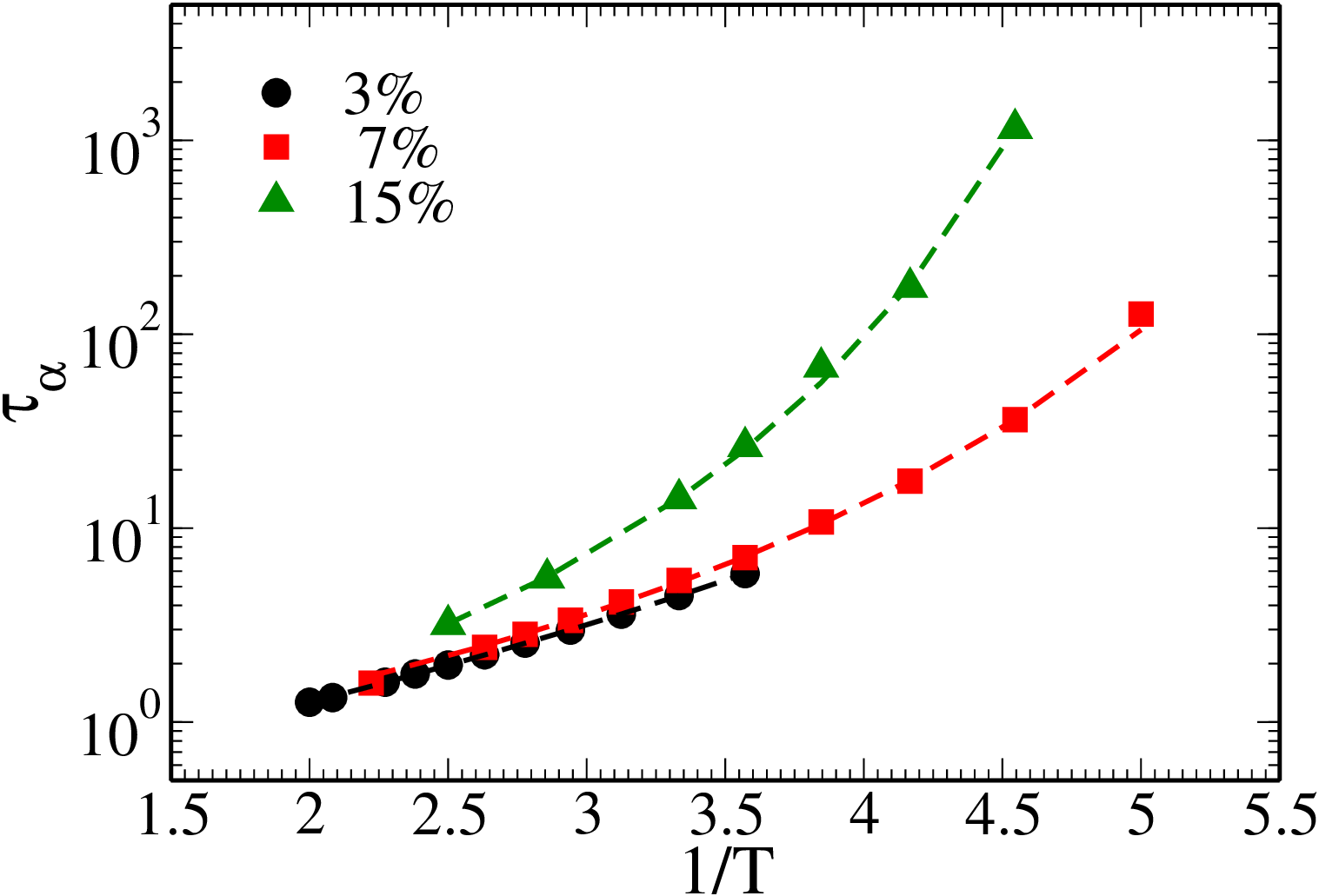}}
    \caption{\emph{The temperature dependence of the $\alpha$ relaxation time at different PDIs. The dashed lines are the VFT fits.}}
    \label{system_T_range}
\end{figure}

\textbf {Appendix A: DYNAMICS AND EXCESS ENTROPY}\\
To elucidate the temperature range of the system, we first obtain the onset temperature of the glassy dynamics for the systems by analyzing the temperature dependence of their inherent structures (IS)\cite{sastry_inherent_1998} (Fig. \ref{inherent_structure}). The IS is obtained using the FIRE algorithm \cite{FIRE_method_for_inherent_str}. For PDI 3\%, 7\%, and 15\% the onset temperatures are 0.64, 0.43, and 0.37, respectively. We calculate the relaxation time $\tau_\alpha$ by examining the overlap function [see Eq. \ref{overlap_equation}] decay to 1/e = 0.367. The relaxation time vs temperature below the onset temperature is plotted in Fig. \ref{system_T_range}. The temperature dependence of the relaxation time is fitted to the well known Vogel-Fulcher-Tammann (VFT) equation,\cite{VFT} and the resulting VFT temperatures for the different systems are as follows: 3\% - 0.073, 7\% - 0.117, and 15\% - 0.154. However, as mentioned in the main text the system with $3\%$ PDI crystallizes at a reasonably high temperature (below $T = 0.28$) compared to its VFT temperature.

\begin{figure}[h!]
  \centering
    \subfigure{\includegraphics[width=.38\textwidth]{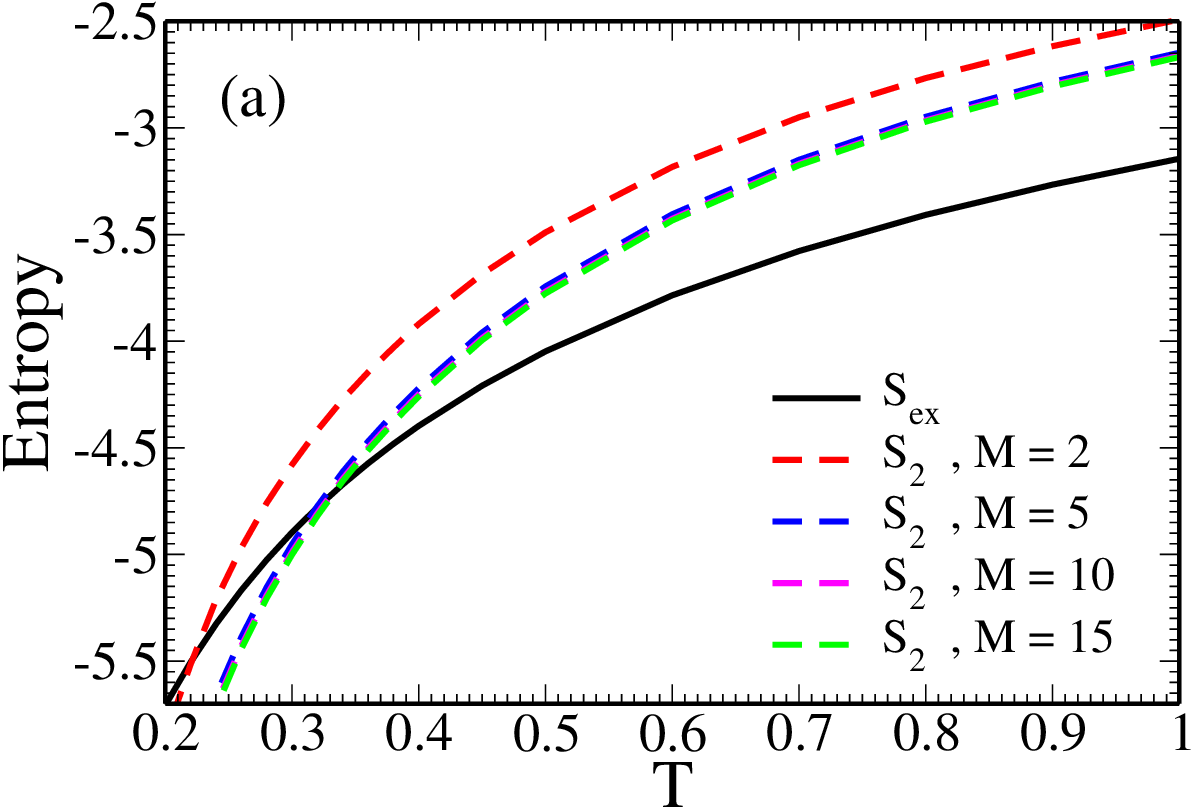}}
        \subfigure{\includegraphics[width=.38\textwidth]{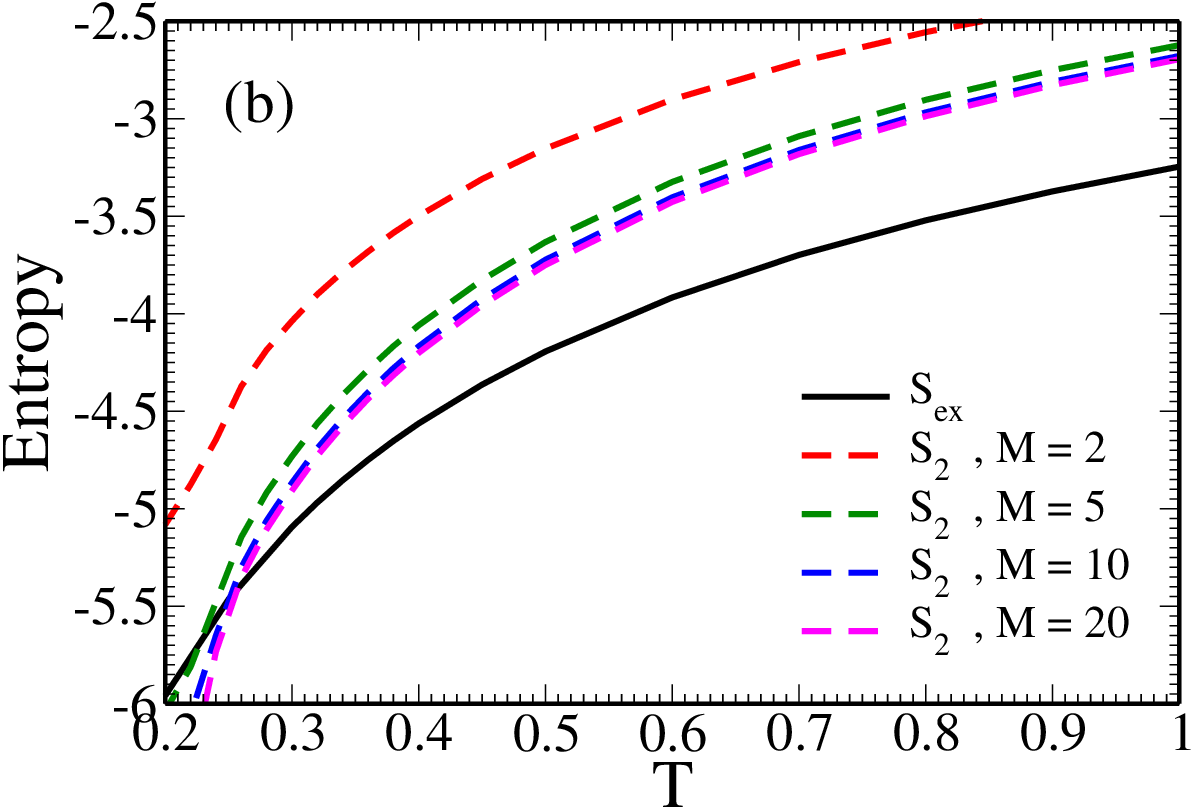}}
    \caption{\emph{Temperature dependence of $S_2$ changes with M and this also changes the $T_{cross}$ values (a) 7\% PDI and (b) 15\% PDI}}
    \label{S2_Sexc_Cross_point_fig}
\end{figure}

Excess entropy $S_{ex}$, is a loss of entropy due to the interaction between particles. Excess entropy is calculated via the temperature integration (TI) method.\cite{Frankel_n_smith,palak_ujjwal_JCP} As discussed in the main text pair excess entropy, $S_{2}$ takes into account the excess entropy due to the two-body correlation. $S_2$ and $S_{ex}$ cross each other at a temperature $T_{cross}$ which is similar to the onset temperature.\cite{Atreyee_2017} In Fig. \ref{S2_Sexc_Cross_point_fig} we plot  the temperature dependence of $S_{ex}$ and $S_{2}$ for 7$\%$ and 15$\%$ PDI where $S_{2}$ is calculated at different values of $M$. Both $S_{2}$ and $T_{cross}$ change with $M$. Initially, they vary strongly, and then the variation is weaker with M.

\begin{figure}[h!]
  \centering    \subfigure{\includegraphics[width=.4\textwidth]{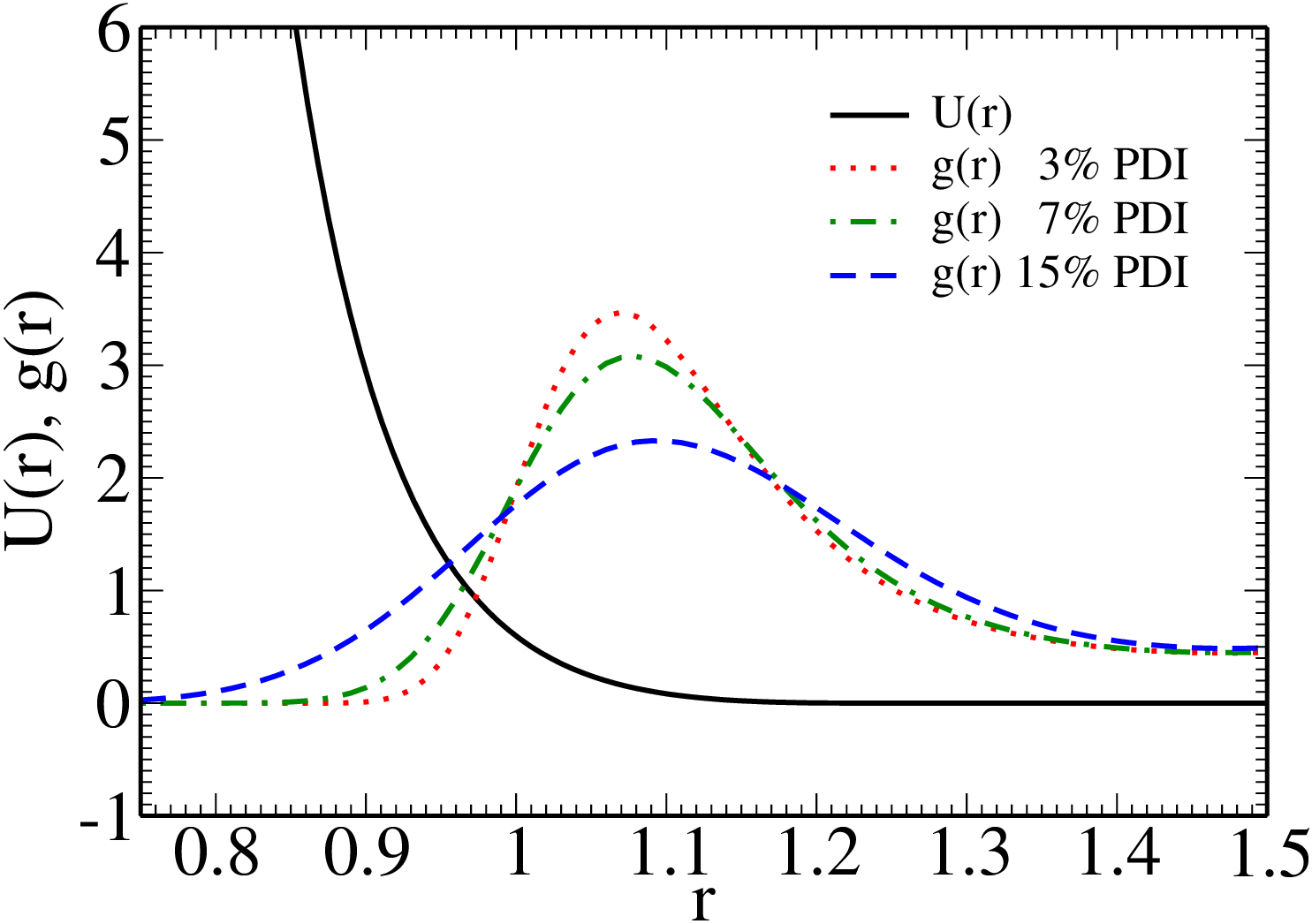}}
    \caption{\emph{Inverse power law potential, U, and radial distribution function, g(r), for different PDIs at M=1.}}
\label{justification_of_approximation_fig}
\end{figure}

\begin{figure*}
  \centering
    \subfigure{\includegraphics[width=.37\textwidth]{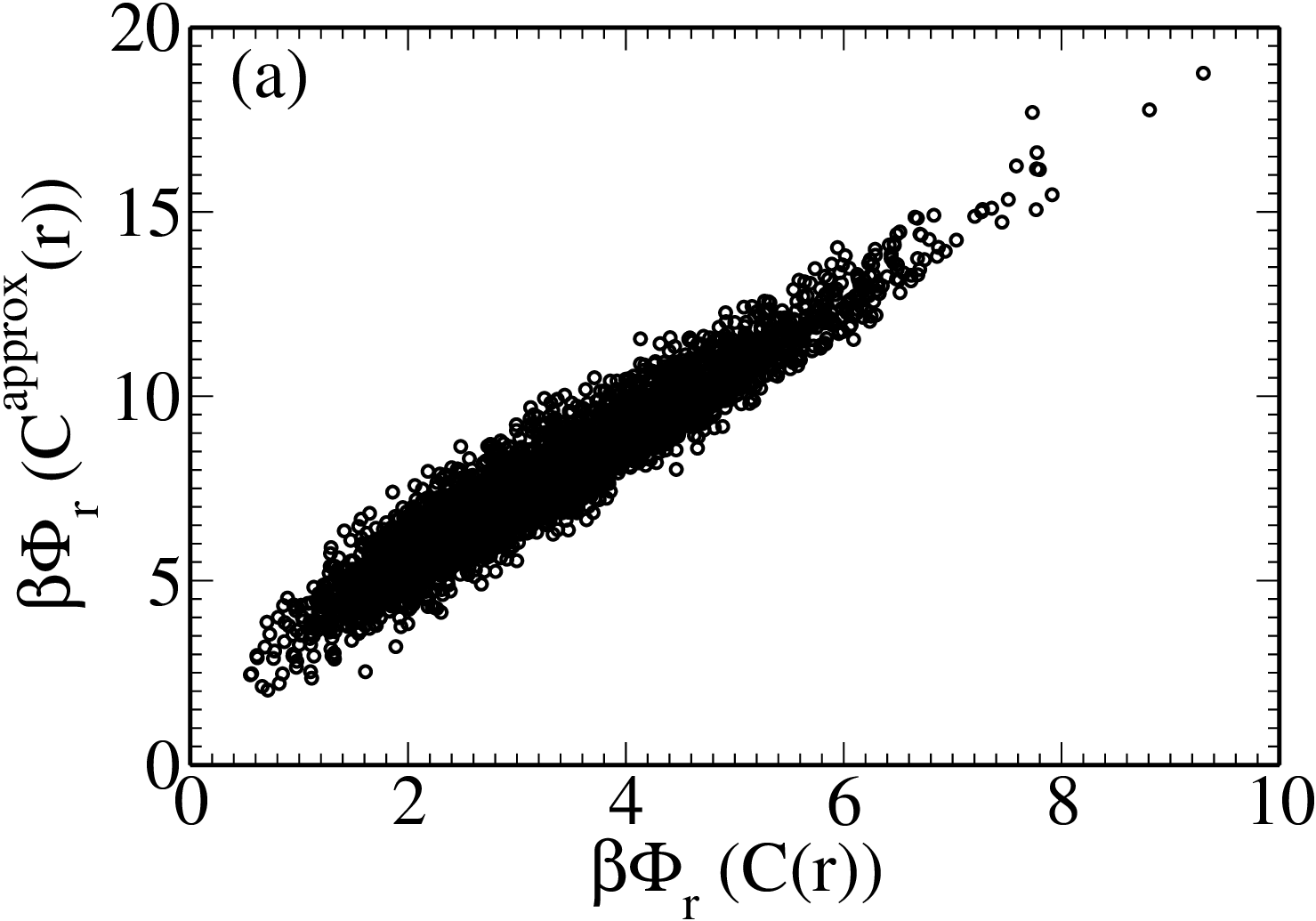}}
    \hspace{0.2cm}     \vspace{0.3cm}\subfigure{\includegraphics[width=.36\textwidth]{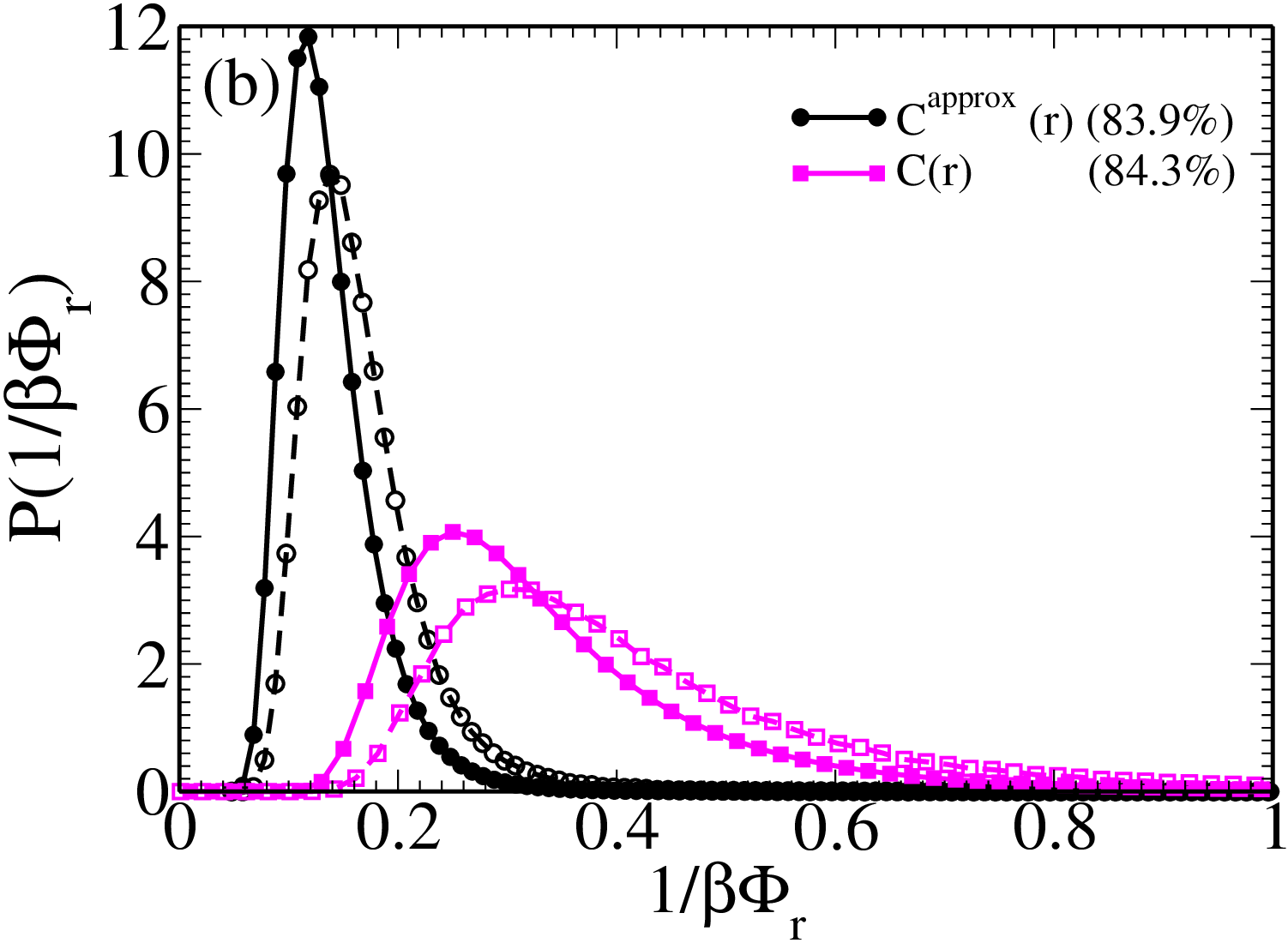}}
    \hspace{0.2cm}
         \vspace{0.3cm}
 \subfigure{\includegraphics[width=.37\textwidth]{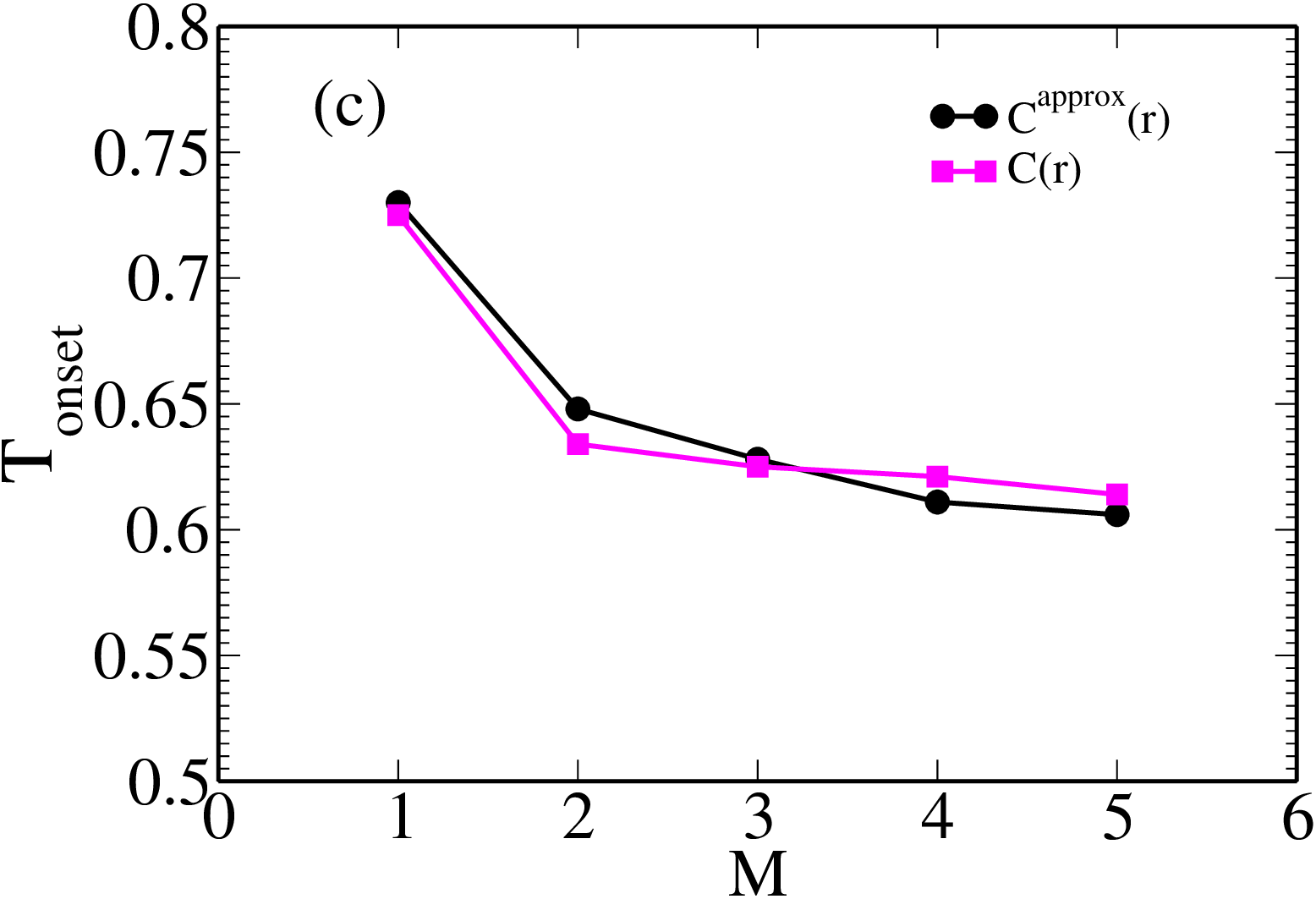}}
  \hspace{0.2cm}
  \subfigure{\includegraphics[width=.36\textwidth]{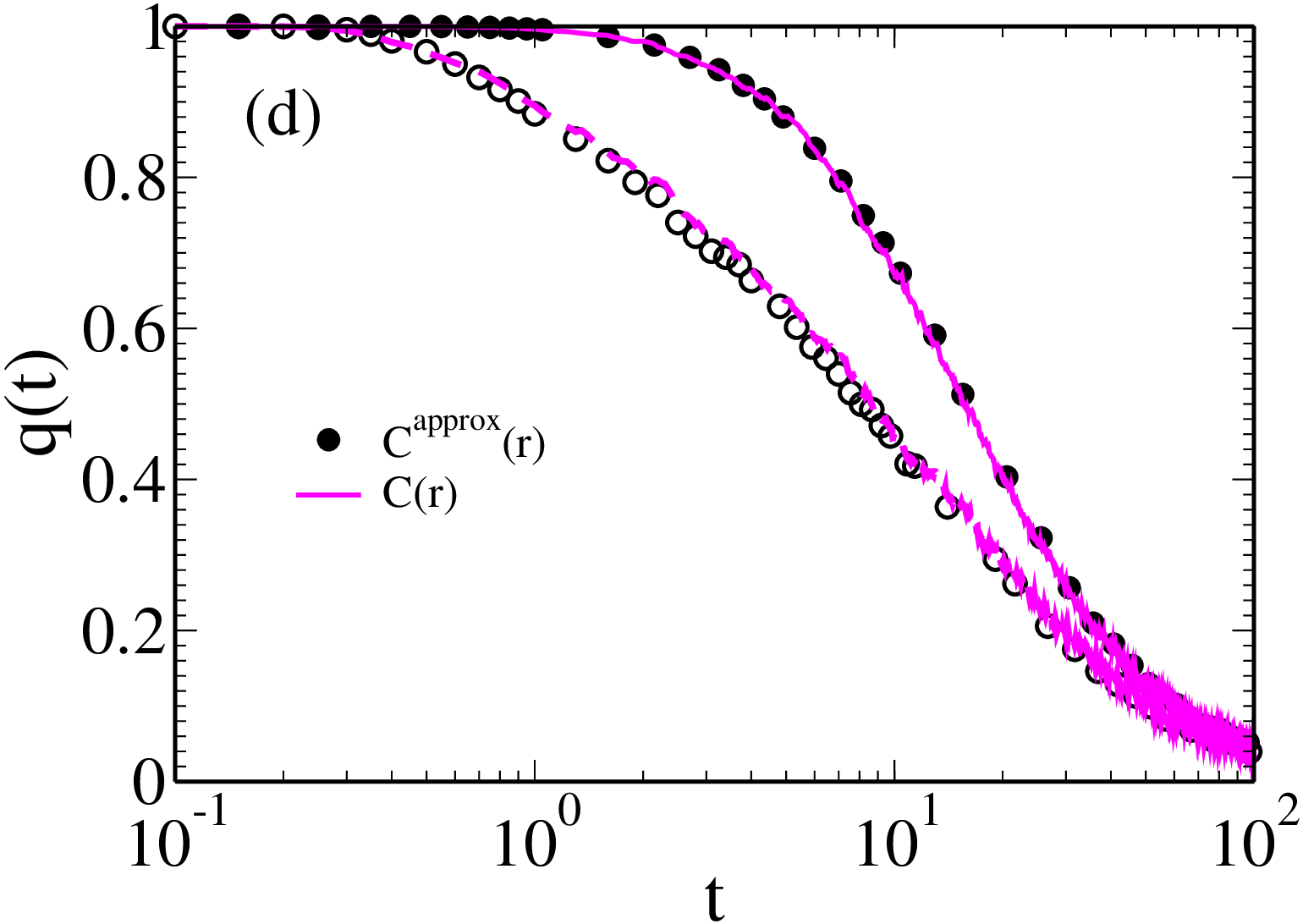}}
    \caption{\emph{Different parameters for justification of $1/\beta \Phi_r$ can be calculated from $C_{uv}(r)$ and ${C^{approx}}_{uv}(r)$ both. This analysis is performed at 3\% PDI. (a) Single frame scatter plot from $C_{uv}(r)$ and ${C^{approx}}_{uv}(r)$ at T =0.28 (b) Distribution of the inverse of $\beta \Phi_r$ of all particles P($1/\beta \Phi_r$) in the system (solid line with filled symbol) and of those which are about to rearrange (dotted line with open symbol) at T = 0.28, from C(r) (black) and $C^{approx}(r)$ (magenta). (c) Onset temperature obtained from $C(r)$ (square) and $C^{approx}(r)$ (circle) (d) Overlap function of few particles that has the highest \{solid line (C(r)) or closed symbol ($C^{approx}(r)$)\} and lowest value of $\beta\Phi(r)$ \{Dotted line (C(r)) or open symbol ($C^{approx}(r)$)\} at T = 0.28.}}
    \label{justification_of_one_and_all_term_fig}
\end{figure*}

\textbf{Appendix B:  CALCULATION OF LOCAL CAGING POTENTIAL USING $C_{uv}$ AND ${C_{uv}}^{approx}$}\\
\\
The potential energy depth calculation using this direct correlation function is given in Eq. \ref{potential_at_zero_displacement_eq}. The expression of $C_{uv}(r)$ according to the Hypernetted chain (HNC) approximation\cite{Hansen_and_McDonald} is given in Eq. \ref{c_mu_nu__with_all_term_equation}.
At higher PDI, when the system is described by one species, the rdf shows a large softening and is non-zero at very small values of 'r' compared to the interaction potential (as shown in Fig. \ref{justification_of_approximation_fig}. In experimental systems where the interaction potential is not known, it is often assumed that the potential of mean force is the same as the interaction potential i.e., $-\beta U_{uv}(r) = \ln[g_{uv}(r)]$. Under this assumption, the expression of the direct correlation function, $C^{approx}_{uv}(r) \approx [g_{uv}(r)-1]$. Here we present an analysis that shows that using $C(r)$ and $C^{approx}(r)$ primarily shifts the distribution of the potential energy depth but does not affect the correlation between the structural order parameter and the dynamics. 
In Fig. \ref{justification_of_one_and_all_term_fig} (a), we show a scatter plot of $\beta\Phi(r)$ calculated using $C_{uv}(r)$ vs that using ${C_{uv}}^{approx}$ and find that they are strongly correlated. Averaged over 1000 frames, the Spearman rank correlation between $(C_{uv},{C_{uv}^{approx}})$ = 0.948 and the Pearson correlation is = 0.955. This confirms that the use of the approximate direct correlation function primarily shifts the distribution of $\beta\Phi(r)$ as shown in Fig. \ref{justification_of_one_and_all_term_fig} (b). We also plot the distribution of the softness for the fast particles, and it shows that in both cases, more than $80\%$ of the fast particles have a softness value higher than the average softness. In Fig. \ref{justification_of_one_and_all_term_fig} (c), we plot the onset temperature obtained when $\beta\Phi(r)$ is calculated using $C(r)$ and $C^{approx}(r)$ (details of onset temperature calculation are given in Section. \ref{onset_caluclation_PRS}), and interestingly, both results are identical. The dynamics of a of the few hardest and a few of the softest particles are plotted in Fig. \ref{justification_of_one_and_all_term_fig} (d). It clearly shows that using an approximate direct correlation function does not reduce the predictive power of the structural order parameter.\\

\textbf {Appendix C: IDENTIFICATION OF FAST PARTICLES}\\
\\
There are many ways of identifying fast particles \cite{walter_PRL_1997,walter_JCP_2002,Widmer-Cooper_2005,Candelier_PRL,smessaert_PRE}. Here we use the method proposed by Candelier et al.\cite{Candelier_PRL,smessaert_PRE}. In that method, for each particle in a certain time window W = {$t_1,t_2$}, they calculated the quantity $p_{hop}(i,j)$. When the average position of a particle changes rapidly, a cage jump happens. Expression for $p_{hop}(i,j)$ is,

\begin{equation}
    p_{hop}(i,t) = \sqrt{\big<(r_i - {\big<r_i\big>}_U)^2\big>_V   \big<(r_i - {\big<r_i\big>}_V)^2\big>_U} ,
\end{equation}
\noindent
 where $\Delta t$ timescale over which the particles can rearrange U = [t - $\Delta$ t/2,t] and V = [t,t + $\Delta$ t/2]. For a time window W, a small value of $p_{hop}$ means the particle is within the same cage, and a large $p_{hop}$ means that within that time window, the particle has moved between two distinct cages. With the help of $p_{hop}$, fast particles are defined in this work. If $p_{hop}$ is greater than $p_c$, then we identify that as a fast particle\cite{Andrea_liu_PRE,mohit_PRE}. $p_c$ is the value of the mean square displacement at time $t_{max}$, where $t_{max}$ is the time at which the non-Gaussian parameter is at its maximum. Although we are working with a polydisperse system we have kept the $p_{c}$ value fixed for all particles. For more details refer to Reference. \cite{mohit_PRE}\\

{\bf ACKNOWLEDGMENT}\\
P.~P. acknowledges the CSIR for the research fellowships. M.~S. acknowledge SERB for funding, S.~M.~B. acknowledge SERB for the funding and S.~M.~B. acknowledge, Chandan Dasgupta for the discussion. The authors would like to acknowledge Leelavati Narlikar for the helpful discussions. \\[4mm]

{\bf AVAILABILITY OF DATA}\\
The data that support the findings of this study are available from the corresponding author upon reasonable request.\\[3mm]

\section{REFERENCES}

\bibliographystyle{h-physrev}
\bibliography{Reference}

\end{document}